%% file: HLLHC_PDFs.tex
\documentclass[11pt,a4paper]{article}

\usepackage[colorlinks=true, linkcolor=black!50!blue, urlcolor=blue, citecolor=blue, anchorcolor=blue]{hyperref}
\usepackage[font=small,labelfont=bf,margin=0mm,labelsep=period,tableposition=top]{caption}
\usepackage[a4paper,top=4cm,bottom=2.5cm,left=4cm,right=2.5cm,bindingoffset=0mm]{geometry}

\usepackage{graphicx}
\usepackage{float}
\usepackage{afterpage}
\usepackage{epsfig,cite}
\usepackage{amssymb}
\usepackage{amsmath}
\usepackage{dsfont}
\usepackage{multirow}
\usepackage{url}
\usepackage{xcolor}
\usepackage{float}
\usepackage{afterpage}

\usepackage{url}
\usepackage{hyperref}

\usepackage{multirow,booktabs,multirow}

\newcommand{\be}{\begin{equation}}
\newcommand{\ee}{\end{equation}}
\newcommand{\bea}{\begin{eqnarray}}
\newcommand{\eea}{\end{eqnarray}}
\newcommand{\bi}{\begin{itemize}}
\newcommand{\ei}{\end{itemize}}
\newcommand{\ben}{\begin{enumerate}}
\newcommand{\een}{\end{enumerate}}

\newcommand{\lp}{\left(}
\newcommand{\rp}{\right)}

\def\frac#1#2{{{#1}\over {#2}}}
\def\gsim{\mathrel{\rlap{\lower4pt\hbox{\hskip1pt$\sim$}}
    \raise1pt\hbox{$>$}}}         
\def\lsim{\mathrel{\rlap{\lower4pt\hbox{\hskip1pt$\sim$}}
    \raise1pt\hbox{$<$}}}         

\newcommand{\draft}[1]{}

\def\beq{\begin{equation}}
\def\eeq{\end{equation}}

\def\lapprox{\lower .7ex\hbox{$\;\stackrel{\textstyle <}{\sim}\;$}}
\def\gapprox{\lower .7ex\hbox{$\;\stackrel{\textstyle >}{\sim}\;$}}

\numberwithin{equation}{section}
\numberwithin{figure}{section}
\numberwithin{table}{section}

\usepackage{tabularx}
\newcolumntype{C}[1]{>{\centering\arraybackslash}p{#1}}

\begin{document}
\newgeometry{top=1.5cm,bottom=1.5cm,left=2.5cm,right=2.5cm,bindingoffset=0mm}
\vspace{-2.0cm}
\begin{flushright}
Nikhef/2018-041\\
\end{flushright}
\vspace{2cm}

\begin{center}
  {\Large \bf
Towards Ultimate Parton Distributions\\[0.20cm] at the High-Luminosity LHC
  }
\vspace{1.4cm}

  Rabah Abdul Khalek$^1$, Shaun Bailey$^2$, Jun Gao$^3$, Lucian Harland-Lang$^2$, 
  and Juan Rojo$^1$.\\

\vspace{0.7cm}
       {\it

             $^1$ Department of Physics and Astronomy, Vrije Universiteit
  Amsterdam, 1081 HV
  Amsterdam, \\
  Nikhef Theory Group, Science Park 105, 1098 XG Amsterdam, The
  Netherlands.\\
 $^2$ Rudolf Peierls Centre for Theoretical Physics, University of Oxford,\\ 
 Clarendon Laboratory, Parks Road, Oxford OX1 3PU, United Kingdom.\\
$^3$ Institute of Nuclear and Particle Physics,\\
Shanghai Key Laboratory for Particle Physics and Cosmology,\\
School of Physics and Astronomy, Shanghai Jiao Tong University, Shanghai, China.

}

\vspace{1.0cm}

{\bf \large Abstract}

\end{center}
Since its start of data taking, the LHC has provided an impressive
wealth of information
on the quark and gluon structure of the proton.
Indeed, modern global analyses of
parton distribution functions (PDFs) include a wide range of LHC measurements of
processes such as the production of jets, electroweak gauge bosons, and top quark pairs.
In this work, we assess the ultimate constraining power of LHC
data on the PDFs that can be expected from the complete dataset,
in particular after the High-Luminosity (HL) phase, starting in around 2025.
The huge statistics of the HL-LHC, delivering 
$\mathcal{L}=3$ ab$^{-1}$ to ATLAS and CMS and $\mathcal{L}=0.3$ ab$^{-1}$
to LHCb, will lead to an extension
of the kinematic coverage of PDF-sensitive measurements
as well as to an improvement in their statistical and systematic uncertainties.
Here we generate HL-LHC pseudo-data for different projections
of the experimental uncertainties, and then quantify the resulting
constraints on the PDF4LHC15 set by means of the Hessian profiling method.
We find that HL-LHC measurements can reduce
PDF uncertainties by up to a factor of 2 to 4 in comparison to state-of-the-art fits,
leading to few-percent uncertainties for important observables
such as the Higgs boson transverse momentum distribution via gluon-fusion.
Our results illustrate the significant improvement in
the precision of PDF fits achievable from hadron collider
data alone, and motivate the continuation of the ongoing successful program
of PDF-sensitive measurements by the LHC collaborations.

\clearpage

\tableofcontents

\input{sec-introduction}
\input{sec-pseudodata}

\input{sec-resultsprocs}

\input{sec-ultimatepdfs}
\input{sec-summary}

\bibliography{HLLHC_PDFs}

\end{document}

%% file: sec-introduction.tex
\section{Introduction}

A detailed understanding of the quark and gluon
structure of the proton~\cite{Gao:2017yyd,Rojo:2015acz,Forte:2013wc}
is an essential ingredient of theoretical predictions for hadron colliders
such as the LHC.
This is quantified by the parton distribution functions (PDFs),
which determine how the proton's momentum is shared among its constituents
in a hard--scattering collision.
PDF uncertainties represent one of the dominant theoretical systematic
errors in many important LHC processes,
including the profiling of the Higgs boson sector~\cite{deFlorian:2016spz};
direct searches for new heavy beyond the Standard
Model (BSM) particles~\cite{Beenakker:2015rna}; indirect BSM searches
by means of the SM Effective Field Theory (SMEFT)~\cite{Alioli:2017jdo};
as well as in
the measurement of fundamental SM parameters such as the $W$ boson mass~\cite{Aaboud:2017svj},
the Weinberg mixing angle~\cite{Aaltonen:2018dxj}, and the strong coupling
constant~\cite{Ball:2018iqk} and its running~\cite{Becciolini:2014lya}.

Since the start of data taking in 2009, the LHC has provided an impressive
wealth of information on the proton's PDFs.
Indeed, modern global PDF fits~\cite{Ball:2017nwa,Dulat:2015mca,
  Harland-Lang:2014zoa,Alekhin:2017kpj}
include a wide range of LHC measurements in
processes such as the production of jets, weak gauge bosons,
and top quark pairs.
Crucially, the recent breakthroughs in the calculation
of NNLO QCD and NLO QED and electroweak corrections (including
photon--induced ones) to
most PDF--sensitive processes have been instrumental in
allowing for the full exploitation of the information provided by the LHC measurements.
The impact of high precision LHC data combined with state--of--the art
perturbative calculations has been quantified for many of the processes
of interest, such as top quark pair production~\cite{Czakon:2016olj,Guzzi:2014wia},
the transverse momentum spectrum of $Z$ bosons~\cite{Boughezal:2017nla}, 
direct photon production~\cite{d'Enterria:2012yj,Campbell:2018wfu},
$D$ meson production in the forward region~\cite{Zenaiev:2015rfa,Gauld:2016kpd}, $W$ production
in association with charm quarks~\cite{Aad:2014xca,Chatrchyan:2013mza,CMS-PAS-SMP-17-014},
and inclusive jet production~\cite{Currie:2016bfm,Rojo:2014kta}.
See the reviews~\cite{Rojo:2015acz,Gao:2017yyd} for a more extensive list
of references.

With experimentalists warming up to analyse the complete Run II dataset, the high energy physics community is already busy looking ahead to the future.
Following Run III, around 2023, a major upgrade of the
LHC accelerator and detector systems will make the start
of its High Luminosity (HL) operation phase possible.
The ten--fold increase in its instantaneous luminosity will lead
to the collection of huge datasets, with the HL--LHC
expected to deliver around
$\mathcal{L}=3$ ab$^{-1}$ to ATLAS and CMS and around $\mathcal{L}=0.3$ ab$^{-1}$
to LHCb.
This unprecedented dataset will open new exciting physics opportunities,
such as the measurement of the Higgs boson couplings to second generation
fermions as well as of its self--interactions.
These opportunities will be summarised in
a CERN Yellow Report~\cite{yellowreport} to be presented before the end of 2018,
in order to contribute to
the update of the European Strategy for Particle Physics (ESPP)\footnote{See
  also~\cite{Bediaga:2018lhg} for a recent review of the potential of the
  LHCb experiment in the HL--LHC era.}.

From the point of view of PDF determinations, the availability of these
immense data samples 
will permit a significant extension
of the kinematic coverage of PDF--sensitive measurements
as well as a marked improvement in their statistical and systematic uncertainties.
With this motivation, the goal of this work is to quantify the impact of the future HL--LHC
measurements on the proton PDFs.
In other words, we aim to assess  the ultimate constraining power
of hadron collider data on the PDFs.
In turn, the resulting projections
for the expected PDF uncertainties will feed into other related
projections for HL--LHC processes, which will benefit from the associated
reduction of theoretical errors.

It is important to emphasise here that
while this type of study has
previously been carried out in the context of
lepton--hadron colliders such as the Large Hadron electron Collider (LHeC)
and the Electron Ion Collider (EIC)~\cite{AbelleiraFernandez:2012cc,AbelleiraFernandez:2012ty,Paukkunen:2017phq,Ball:2013tyh,Marquet:2017bga,Aschenauer:2017oxs,Aschenauer:2015ata,Boer:2011fh,Cooper-Sarkar:2016udp}, 
this is the first time that such a
systematic effort has been devoted to determine
the PDF--constraining potential of a future hadron collider.
Clearly, being able to compare the information on PDFs that will be provided by the HL--LHC
with that from proposed electron--proton colliders such as the LHeC represents
an important input to inform the upcoming ESPP update.

Our analysis has been carried out as follows.
First, we have generated HL--LHC pseudo--data for a number of PDF--sensitive
processes: Drell--Yan production (both at high dilepton invariant mass and in the forward rapidity
regions); $W$ production in association with charm quarks (central and forward regions);
inclusive jet and prompt photon production; the transverse momentum of $Z$ bosons;
and differential distributions in top quark pair production.
We have selected those processes that should benefit more directly
from the increased statistics available at the HL--LHC. We done consider measurements such as inclusive $W,Z$ production in the
central region, which are already
completely limited by systematic uncertainties~\cite{Aaboud:2016btc,Khachatryan:2016pev}, with no significant improvement anticipated from increased statistics alone.
For each process, the binning and kinematic cuts applied to the pseudo--data is constructed
from a suitable extension of reference measurements
at $\sqrt{s}=8$ and 13 TeV.
We consider different scenarios for the expected  systematic uncertainties,
from a conservative one with approximately the same systematics
as the corresponding baseline measurements from Run I
and a factor of 2 reduction for those from Run II,
to an optimistic one with
a reduction by a factor 2.5 (5) as compared to Run I~(II). 

Subsequently, we quantify the 
constraints of the HL--LHC pseudo--data
on the PDF4LHC15\_100 set~\cite{Butterworth:2015oua,Gao:2013bia,Carrazza:2015hva,Carrazza:2015aoa}
by means
of the Hessian Profiling method~\cite{Paukkunen:2014zia} (see also~\cite{Schmidt:2018hvu}).
We have chosen the PDF4LHC15 set since it broadly represents
the state--of--the--art understanding of
the proton structure. 
While it is beyond the scope of this work to construct forecasts of the experimental correlation
models, we account for their effective impact by using available Run I and II
measurements as benchmarks.
The resulting profiled sets are then implemented in the {\tt LHAPDF6} interface~\cite{Buckley:2014ana}, 
thus being readily available
for phenomenological applications.

By performing this analysis, we find that the legacy HL--LHC measurements can reduce the
uncertainties in the PDF luminosities by a factor between 2 and 5 in comparison to state--of--the--art fits,
depending on the specific flavour combination of the initial state and the invariant mass
of the produced final state.
We also show that our projections for the  PDF error reduction,
which are predominantly driven by the increased statistics of the HL--LHC data sample,
depend only moderately on the specific scenario
adopted for the reduction of the experimental systematic errors.

We then explore the implications of the profiled PDFs for representative
LHC cross sections at $\sqrt{s}=14$ TeV,
both within the Standard Model (SM) and beyond it.
Our analysis highlights how $\mathcal{O}\lp 1\%\rp$ PDF uncertainties
are within the reach of the HL--LHC for key observables such as
the transverse momentum distribution in Higgs production from gluon fusion.
Therefore, our study illustrates the significant improvement
in the precision of PDF determinations achievable from hadron collider
data alone, and motivates the continuation of the ongoing successful program
of PDF--sensitive measurements at the LHC.

The outline of the paper is the following.
First, in Sect.~\ref{sec:pseudodata} we describe the features of the
PDF--sensitive processes used to generate the HL--LHC pseudo--data.
Then in Sect.~\ref{sec:resultsindividual} we quantify the constraints
on the PDFs of individual processes using the Hessian
profiling method.
The full set of HL--LHC pseudo--data is combined in Sect.~\ref{sec:ultimatepdfs}
to construct the ultimate HL--LHC parton distributions, which is
then used to assess their phenomenological
implications for different processes both
in the SM and beyond it.
Finally, in Sect.~\ref{sec:summary} we summarise our results and indicate
how they are made publicly available.

%% file: sec-pseudodata.tex
\section{Pseudo--data generation}
\label{sec:pseudodata}

In this section we present the PDF--sensitive processes
for which HL--LHC pseudo--data have been generated, provide
details about the binning and kinematic cuts, and
also describe the baseline Run I and II measurements that are used to model
the experimental systematic uncertainties expected in the HL--LHC era.

\subsection{PDF--sensitive processes}
\label{sec:pdfsensprocs}

\begin{figure}[t]
\centering
\epsfig{width=0.90\textwidth,figure=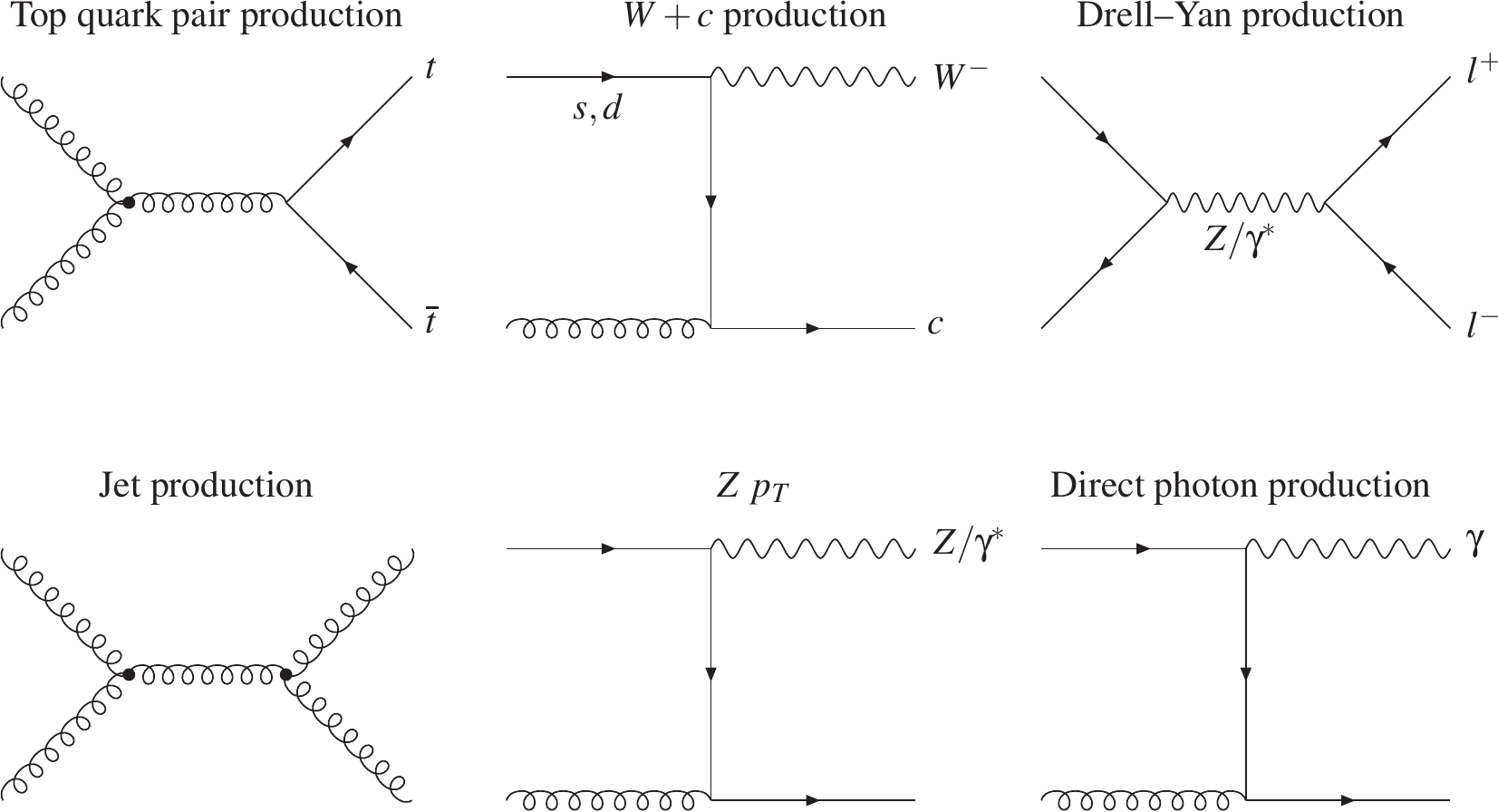}
\caption{\small Representative Feynman diagrams at the Born level
  of the six types of collider processes for which HL--LHC pseudo--data
  has been generated in this analysis: the production of top quark pairs, $W$
  bosons in association with charm quarks, and the neutral and charged
  current Drell--Yan processes; the production of inclusive jets, $Z$ bosons
  at finite transverse momentum, and direct photons.
 \label{fig:feynman}} 
\end{figure}

We start by describing the  PDF--sensitive processes
that will be considered in this study to generate
HL--LHC pseudo--data.
Our analysis is based on six different types of processes:
the production of top quark pairs, jets,
  direct photons, and $W$ bosons in association with charm quarks, the
  transverse momentum of $Z$ bosons, and the neutral and charged
  current Drell--Yan processes.
  In Fig.~\ref{fig:feynman} we show representative Feynman diagrams at the Born level
  for all of these processes, in order to illustrate their sensitivity to the different
  partonic initial states.
  For instance, we see that jets, photon, and top quark pair production
  are directly dependent on the gluon content of the proton, while $W$+charm
  is sensitive to strangeness, and the Drell--Yan process to the quark--antiquark
  luminosity.

This choice of input processes is driven by the fact that some types
of hard--scattering reactions should benefit more directly
from the increased statistics offered by the HL--LHC than others.
Indeed, some of the existing LHC measurements, such as inclusive $W,Z$ production in the
central region~\cite{Aaboud:2016btc,Khachatryan:2016pev}, are already 
limited by systematic uncertainties, and therefore are unlikely to
improve significantly at higher luminosities.
On the other hand, our selection of processes will greatly benefit from
the huge HL--LHC dataset either because they are relatively rare,
such as $W$+charm, or because their kinematic coverage can be extended
to regions of large invariant masses and transverse momentum or
forward rapidities where event rates exhibit a steep fall--off. While these pseudo--data sets do include some regions
which are currently systematics dominated, i.e. towards central rapidity and lower mass/transverse momentum, as we will see the dominant PDF impact comes from the regions which are not, where the existing data are less constraining and the contributing PDFs are currently less well determined.

In more detail, the specific processes for
which HL--LHC pseudo--data have been generated are the following:
\begin{itemize}

\item High--mass Drell--Yan, specifically the dilepton
  invariant mass differential distribution $d\sigma(pp\to ll)/dm_{ll}$
  for $m_{ll}\gtrsim 110$ GeV for a central rapidity acceptance, $|\eta_{l}|\le 2.4$.
  This process is particularly useful for quark flavour separation, specifically
  to constrain the poorly known large--$x$ sea quarks.

  Here the ATLAS 8 TeV measurement of differential Drell--Yan
  cross sections~\cite{Aad:2016zzw} is taken as reference, with
  additional bins in the high $m_{ll}$ region included to benefit from
  the enhanced kinematic coverage.

\item The differential distributions for on--peak $W$ and $Z$ boson
  production in the forward region, $2.0 \le \eta_{l}
  \le 4.5$, covered
  by the LHCb experiment.
  These measurements constrain quark flavour separation,
  including the strange and charm content of the proton,
  in the large and small $x$ region~\cite{Rojo:2017xpe}, complementary
  to the data from the central region.

  The reference analysis is the LHCb measurement of the rapidity distributions
  of $W$ and $Z$ bosons in the muon final state at 8 TeV~\cite{Aaij:2015zlq}.
  In comparison to the reference measurement, a finer binning by a factor of 2 to 5 has been
  adopted as allowed by the increased event rates.
  Events are selected if $p_T^l \ge 20$ GeV, the lepton rapidities
  fall in the LHCb acceptance, and, in the case of $Z$ production,
  there is the additional requirement that $60~{\rm GeV}\le m_{ll} \le 120~{\rm GeV}$.
  
\item   Differential distributions in top quark pair production,
  providing direct information on the large $x$ gluon~\cite{Czakon:2016olj}.
  Specifically, we consider the top quark
  transverse momentum $p_T^t$ and rapidity $y_t$, and
  the top quark pair rapidity $y_{t\bar{t}}$ and invariant mass $m_{t\bar{t}}$.

  The reference measurements here are  the ATLAS 8 TeV differential
  distributions in the lepton+jets final state~\cite{Aad:2015mbv}.
  We assume that the statistical correlations between different
  distributions will be available, as is the case for the 8 TeV
  data~\cite{ATL-PHYS-PUB-2018-017},
  and therefore include the four distributions simultaneously in the fit.
  To account for the increased statistics of the HL--LHC, the number
  of bins in the rapidity distributions is doubled, while
  the $p_T^t$ and $m_{t\bar{t}}$ distributions are extended to higher values
  in the
  TeV region.

\item The transverse momentum distribution of the $Z$ bosons in the
  dilepton final state, $20 \,{\rm GeV}<p_T^{ll}<3.5\,{\rm TeV}$ region for central rapidities $|\eta_{Z}|\le 2.4$
  and different bins of the dilepton invariant mass $m_{ll}$.
  This process is relevant to constrain the gluon and the antiquarks
  at intermediate values of $x$~\cite{Boughezal:2017nla}.

  For the reference analysis, we take the ATLAS measurements of
  the transverse momentum of lepton pairs at 8 TeV~\cite{Aad:2015auj}.
  The pseudo--data is generated for six different bins of the
  dilepton invariant mass $m_{ll}$, with boundaries 12, 20, 30, 40, 66, 116, and 150 GeV
  respectively.
  In each of the invariant mass $m_{ll}$ bins,
  additional bins are added to the $p_T^{ll}$ distribution
  to exploit the improved coverage of the large
  transverse momentum region.

\item The production of $W$ bosons in association with charm quarks.
  This process provides a sensitive handle on the strangeness content
  of the proton~\cite{Stirling:2012vh,Chatrchyan:2013mza},
  which is the least well known of
  the light quark PDFs.
  The pseudo--data for this process has been generated as a function of the lepton
  psuedorapidity $\eta_l$ from the $W$ boson decay, and is inclusive over the kinematics
  of the charm quark provided it satisfies the selection cuts.

  For this process, pseudo--data have been generated both for the central rapidity
  region relevant for ATLAS and CMS as well as for the forward region covered by
  LHCb.
  In the central rapidity region, $|\eta^l|\le 2.4$, the reference measurement is the CMS analysis
  at 13 TeV~\cite{CMS-PAS-SMP-17-014}, where events are selected provided that $p_T^c \ge 5 $ GeV
  and $p_T^l \ge 26$ GeV with $l$ indicating the result of the $W\to l\nu$ decay.
  At forward rapidities, $2\le \eta^l \le 4.5$, we use a dedicated selection strategy
  with $2.2 \le \eta^c \le 4.2$, $p_T^\mu \ge 20$ GeV, $p_T^c \ge 20$ GeV, and
  $p_T^{\mu+c}\ge 20$ GeV~\cite{LHCbprivcomm}. We take the acceptance to be 30\% due to $c$--jet tagging and an overall normalisation error of 5\%.
 
\item Prompt isolated photon production represents
  a complementary probe of the gluon PDF at intermediate
  values of $x$~\cite{Campbell:2018wfu}.
  Here the pseudo--data has been generated as differential distributions
  in the photon transverse momentum $p_T^\gamma$ for different bins
  in the photon rapidity $\eta^\gamma$.

  The reference measurements here is the ATLAS 13 TeV analysis of~\cite{Aaboud:2017cbm}, where
  additional bins have been added to the
  $p_T^\gamma$ distribution in each rapidity bins
  to benefit from the improved coverage of the large $p_T^\gamma$ region.

\item Finally, we consider the inclusive production of hadronic jets in different bins
  of their rapidity up to $|y_{\rm jet}|\le 3$ as a function of their $p_T^{\rm jet}$.
  This process
  provides direct information on the gluon
  and the valence quarks at large--$x$.
  Here the jets are reconstructed using the anti--$k_T$  algorithm
  with $R=0.4$ as radius parameter.

  The reference measurement here is the 13 TeV ATLAS analysis of inclusive jet
  and dijet production based on
  a luminosity $\mathcal{L}=3.2$ fb$^{-1}$ from the 2015 data--taking period.
  The coverage of the high--$p_T$ region has
  been extended to the $p_T^{\rm jet}\simeq 2-3$ TeV in comparison to these.
    
\end{itemize}

It is important to emphasise that the list of processes
considered in this work is by no means  exhaustive.
Clearly, there are other important
processes that will provide useful information on the proton PDF
in the HL--LHC era.
Among these, one could consider dijet production~\cite{Currie:2017eqf} and
single top quark production~\cite{Berger:2017zof}, providing
information on the gluon and on the quark flavour separation respectively.
In both cases, the NNLO corrections for differential distributions
are available, as well as reference LHC measurements at 8 and 13 TeV.
Our choices of processes are in addition generally geared towards the high and intermediate $x$ region. Other PDF--sensitive processes, such as inclusive $D$ meson production, can play a role at lower $x$. Although in this case it is unlikely to benefit from the high--luminosity phase, as it already
occurs at very high rates, this may not be true for other rarer processes sensitive to this region.

In addition, one should take into account that progress from both the experimental and theoretical
sides could lead to novel processes being added to the PDF fitting toolbox,
for instance more exclusive processes or processes for which the standard
DGLAP description breaks down.
With these caveats, the set of processes adopted in this work
is representative enough to provide a reasonable snapshot of the PDF--constraining
potential of the HL--LHC.

It is also important to mention that the HL--LHC projections presented
in this work are  based on pseudo--data  generated
specifically for this study, and that they are not endorsed by the LHC collaborations.
However, we have taken into account all the feedback and suggestions received from the
ATLAS, CMS, and LHCb contacts involved in the Yellow Report studies.

\subsection{Theory calculations and pseudo--data generation}
\label{sec:pseudodatagen}

For the various processes described above, we have
generated pseudo--data for a centre--of--mass
energy of $\sqrt{s}=14$ TeV assuming a total integrated
luminosity of $\mathcal{L}=3$ ab$^{-1}$ for the CMS and ATLAS experiments, and of
$\mathcal{L}=0.3$ ab$^{-1}$ for the LHCb experiment. Note that in the former case we explicitly include pseudo--data from both experiments.
Statistical uncertainties are evaluated
from the expected number of events per bin,
taking into account branching ratios and
acceptance corrections determined
from the corresponding reference analysis.
Systematic uncertainties are taken to be those of the
13 or 8 TeV baseline analyses and then rescaled appropriately.
We consider various scenarios for the reduction of systematic errors,
from a more conservative one to a more optimistic one.

Theoretical predictions are computed
at next--to--leading order (NLO) in the QCD
expansion using {\rm MCFM}~\cite{Boughezal:2016wmq} interfaced to
{\tt APPLgrid}~\cite{Carli:2010rw} to produce
the corresponding fast grids.
The only exception is inclusive jet production, for which the NLO
calculation is obtained from the {\tt NLOJET++} program~\cite{Nagy:2003tz}.
The central value of the pseudo--data initially coincides with
the corresponding prediction obtained
using this NLO calculation with the PDF4LHC15 NNLO set as input.
Subsequently, this central value is fluctuated
according to the corresponding experimental uncertainties.
This implies that, by construction, one should find $\chi^2/N_{\rm dat}\simeq 1$
from the fit to the pseudo--data.

Specifically, if $\sigma_i^{\rm th}$ is the theoretical cross section
for  bin $i$ of a given process,
then the central value of
the HL--LHC pseudo--data $\sigma_i^{\rm exp}$ is constructed by means of
\be
\label{eq:pseudodataGen}
\sigma_i^{\rm exp} = \sigma_i^{\rm th} \times \lp 1 + r_i\cdot \delta^{\rm exp}_{{\rm tot},i}
 + \lambda\cdot \delta^{\rm exp}_{\mathcal L}+ s\cdot \delta^{\rm exp}_{\mathcal N}\rp \, ,
\ee
where $r_i$, $\lambda$, and $s$ are univariate Gaussian random numbers, $\delta^{\rm exp}_{{\rm tot},i}$ is the total
(relative) experimental uncertainty corresponding to this specific bin
(excluding the luminosity and normalization uncertainties), and $\delta^{\rm exp}_{\mathcal L}$
is the luminosity uncertainty, which is fully
correlated among all the pseudo--data bins of the same experiment
(but uncorrelated among different experiments).
We take this luminosity uncertainty to be $\delta^{\rm exp}_{\mathcal L}=1.5\%$
for the three LHC experiments.
$\delta^{\rm exp}_{\mathcal N}$
are possible additional normalization uncertainties as in the case of
$W$ boson production in association with charm quarks that will be explained later.

In Eq.~(\ref{eq:pseudodataGen}), the total experimental
uncertainty $\delta^{\rm exp}_{{\rm tot},i}$ is defined as
\be
\label{eq:totalExpError}
\delta^{\rm exp}_{{\rm tot},i} \equiv \lp \lp \delta^{\rm exp}_{{\rm stat},i}\rp ^2 +
\lp f_{\rm corr}\times f_{\rm red}\times
\delta^{\rm exp}_{{\rm sys},i}\rp^2 \rp^{1/2} \, .
\ee
In this expression, the relative statistical error $\delta^{\rm exp}_{{\rm stat},i}$ is
computed as
\be
\label{eq:acceptance}
\delta^{\rm exp}_{{\rm stat},i} = \lp f_{\rm acc} \times N_{{\rm ev},i}\rp^{-1/2} \, ,
  \ee
  where $N_{{\rm ev},i}=\sigma_i^{\rm th} \times \mathcal{L}$ is the expected number
  of events in bin $i$ at the HL--LHC with $\mathcal{L}=3~(0.3)$ ab$^{-1}$.
  In Eq.~(\ref{eq:acceptance}),
  $f_{\rm acc}\le 1$
  is an acceptance correction which accounts for the fact that, for some of the processes
  considered, such as top quark pair production, there is a finite experimental acceptance
  for the final state products
  and/or one needs to include the effects of branching fractions.
  The value of
  $f_{\rm acc}$ is determined by extrapolation using the reference dataset,
  except for forward $W$+charm production (where there is no baseline measurement)
  where the acceptance is set to $f_{\rm acc}=0.3$, due dominantly to the $c$--jet tagging efficiency.  
  
  In Eq.~(\ref{eq:totalExpError}),   $\delta^{\rm exp}_{{\rm sys},i}$ indicates the total
  systematic error of bin $i$ taken from the reference LHC measurement at either 8 TeV
  or 13 TeV, while $f_{\rm red}\le 1$ is a correction factor that accounts for the fact that on average systematic
  uncertainties will decrease at the HL--LHC in comparison to Run II due to both detector
  improvements and the enlarged dataset for calibration.
  Finally,  $f_{\rm corr}$ represents
  an effective correction factor that accounts
  for the fact that data with correlated systematics may be more constraining than the same
  data where each source of error is simply added in quadrature,
  as we do in this analysis.
  We discuss below in Sect.~\ref{sec:corrimpact}
  how the value of  $f_{\rm corr}$ can be
  determined by means of available LHC measurements
  for which the full information on correlated systematics
  is available.
  
\begin{table}[t]
  \centering
  \renewcommand{\arraystretch}{1.50}
  \small
  \begin{tabular}{c|c|c|c|c|c}
    Process    &   Kinematics  &   $N_{\rm dat}$  &  
      $f_{\rm corr}$  &  $f_{\rm red}$ &  Baseline  \\
\toprule
\multirow{3}{*}{$Z$ $p_T$}  &    $20\,{\rm GeV}\le p_T^{ll} \le 3.5$ TeV           &
\multirow{3}{*}{338} &   \multirow{3}{*}{0.5}  & \multirow{3}{*}{$\lp 0.4, 1\rp$}
& \multirow{3}{*}{\cite{Aad:2015auj} (8 TeV)} \\
  &    $12\,{\rm GeV}\le m_{ll} \le 150$ GeV           &              & & \\
   &    $|y_{ll}|\le 2.4$            &                    &   &     \\
\midrule
\multirow{2}{*}{high-mass Drell-Yan}  &   $p_T^{l1(2)}\ge 40(30)\,{\rm GeV}$           &  \multirow{2}{*}{32}         &        \multirow{2}{*}{0.5}      
&       \multirow{2}{*}{$\lp 0.4, 1\rp$}        &    \multirow{2}{*}{\cite{Aad:2016zzw} (8 TeV)}     \\
&  $|\eta^l|\le 2.5$, $m_{ll}\ge 116\,{\rm GeV}$   & & & &  \\
\midrule
top quark pair  &     $m_{t\bar{t}}\simeq 5$ TeV, $|y_t|\le 2.5$          &       110                   & 0.5
&         $\lp 0.4, 1\rp$      &   \cite{Aad:2015mbv} (8 TeV)    \\
\midrule
\multirow{2}{*}{$W$+charm (central)}  &        $p_T^\mu \ge26\,{\rm GeV}$, $p_T^c \ge5\,{\rm GeV}$     &  \multirow{2}{*}{12}         &        \multirow{2}{*}{0.5}       
&       \multirow{2}{*}{$\lp 0.2, 0.5\rp$}       &    \multirow{2}{*}{\cite{CMS-PAS-SMP-17-014} (13 TeV)}     \\
&  $|\eta^\mu|\le 2.4$    & & & &\\
\midrule
\multirow{3}{*}{$W$+charm (forward)}  &      $p_T^\mu \ge20\,{\rm GeV}$, $p_T^c \ge20\,{\rm GeV}$           &          \multirow{3}{*}{10}     &        \multirow{3}{*}{0.5}   
&      \multirow{3}{*}{$\lp 0.4, 1\rp$}            &   \multirow{3}{*}{LHCb projection}         \\
&   $p_T^{\mu+c} \ge20\,{\rm GeV}$    & & & & \\
&  $2\le \eta^\mu \le 4.5$, $2.2\le \eta^c \le 4.2$     & & & & \\
\midrule
Direct photon  &     $E_T^\gamma \lsim 3$ TeV, $|\eta_{\gamma}|\le 2.5$          & 118              &      0.5        
&    \multirow{1}{*}{$\lp 0.2, 0.5\rp$}           &   \cite{Aaboud:2017cbm} (13 TeV)      \\
\midrule
\multirow{2}{*}{Forward $W,Z$}  &  $p_T^{l}\ge 20\,{\rm GeV}$, $2.0\le \eta^l\le 4.5$           &  \multirow{2}{*}{90}         &        \multirow{2}{*}{0.5}     
&       \multirow{2}{*}{$\lp 0.4, 1\rp$}        &    \multirow{2}{*}{\cite{Aaij:2015zlq} (8 TeV)}     \\
&    $60\,{\rm GeV}\le m_{ll}\le 120\,{\rm GeV}$  & & & & \\
\midrule
Inclusive jets  &       $|y| \le 3$, $R = 0.4$      &       58        &      0.5        &
\multirow{1}{*}{$\lp 0.2, 0.5\rp$} 
&   \cite{Aaboud:2017wsi}    (13 TeV)               \\
\bottomrule
Total   &    &   768 &   &   &   \\
\bottomrule
  \end{tabular}
  \vspace{0.3cm}
  \caption{\small \label{tab:PseudoData}
    Summary of the features of the HL--LHC pseudo--data generated for the present
    study.
    For each process we indicate the kinematic coverage, the number of pseudo--data
    points used across all detectors $N_{\rm dat}$, the values of the correction factors
    $f_{\rm corr}$ and $f_{\rm red}$; and finally the reference from the 8 TeV or
    13 TeV measurement used as baseline to define the binning and the systematic
    uncertainties of the HL--LHC pseudo--data, as discussed in the text.
  }
\end{table}

Concerning the theoretical calculations adopted here,  
since the present study relies on pseudo--data,
it is not necessary to account for higher--order QCD effects
or electroweak corrections.
Indeed, by far the dominant contribution to the PDF sensitivity of hadron
collider processes is contained within the NLO calculation.
As in the case of PDF closure tests~\cite{Ball:2014uwa},
here we are only interested in the relative
reduction of the PDF uncertainties once the HL--LHC data is included
in the fit, while the central value itself will be essentially
unaffected.
Note that this also holds for the contribution of
photon--initiated (PI) processes, since
the photon PDF is very well
know~\cite{Manohar:2017eqh,Bertone:2017bme,Nathvani:2018pys}.
Therefore, PI processes effectively induce
an overall rescaling of the cross section which becomes irrelevant
when generating pseudo--data.

  In Table~\ref{tab:PseudoData} we present
  the summary of the main features of the HL--LHC pseudo--data generated for
  the present study.
    For each process, we indicate the kinematic coverage, the number of pseudo--data
    points used  $N_{\rm dat}$, the values of the correction factors $f_{\rm acc}$,
    $f_{\rm corr}$, and $f_{\rm red}$; and finally the reference for the 8 TeV or
    13 TeV measurement used as baseline to define the binning and the systematic
    uncertainties of the HL--LHC pseudo--data.
    A total of around $N_{\rm dat}= 768$ pseudo--data points are then used in the PDF profiling.
    The values of the reduction factor for the systematic errors
    $f_{\rm red}$ is varied between 1 (0.5) and 0.4 (0.2) in the conservative
    and optimistic scenarios for a 8 TeV (13 TeV) baseline measurement.
    This different treatment is motivated by the fact that available 13 TeV measurements
    are based on a smaller dataset and therefore
    tend to have larger systematic errors in comparison to the 8 TeV case. Thus we can expect some improvement here at the
    HL--LHC even in the most conservative scenario;
    Run II measurements based on the complete integrated luminosity will
    certainly benefit from reduced systematics.

\begin{figure}[t]
\centering
\epsfig{width=0.98\textwidth,figure=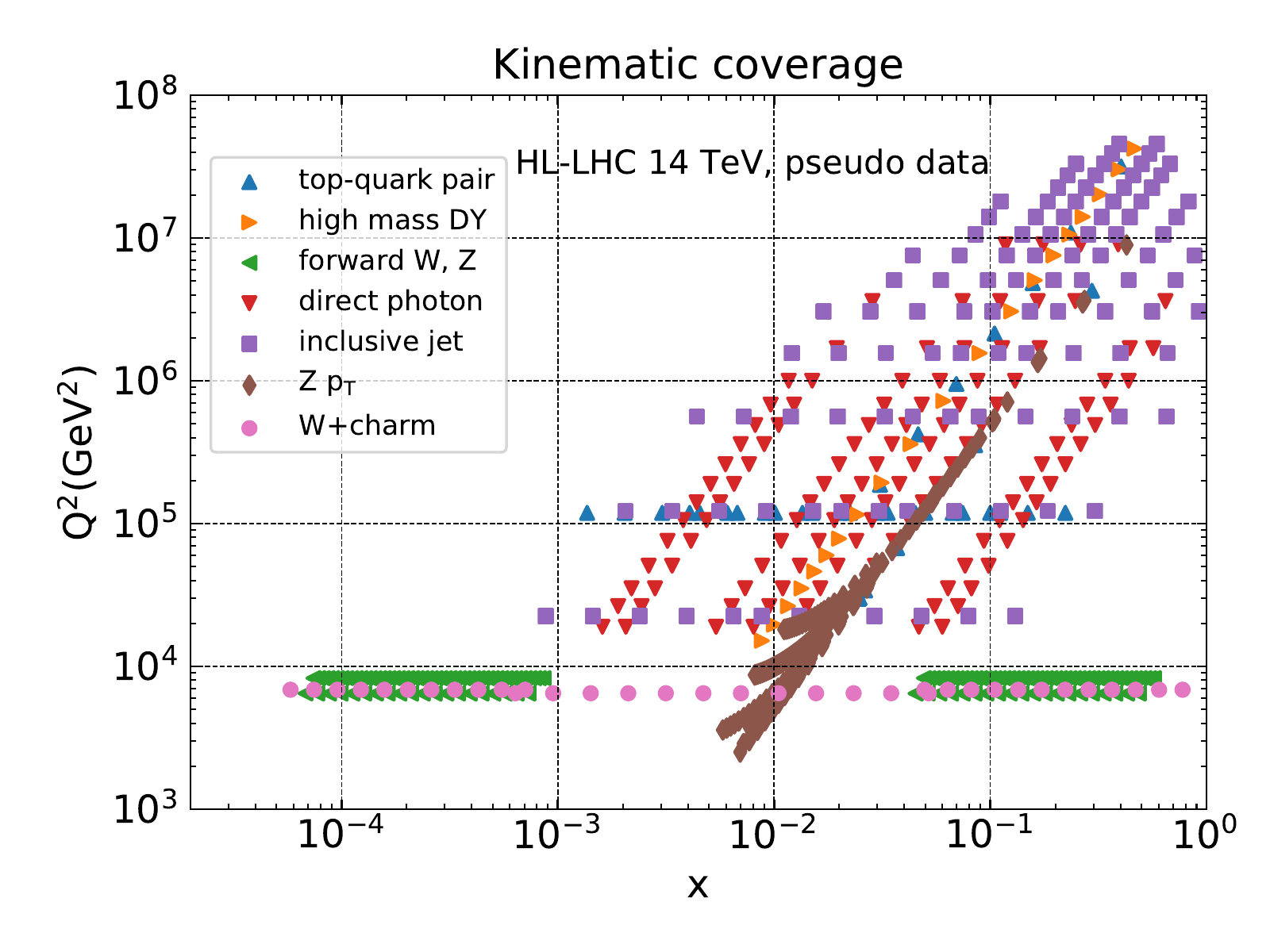}
\caption{\small 
  The kinematical coverage in the $(x,Q^2)$ plane of the
  HL--LHC pseudo--data included in this analysis.
  For each data point, the values of $(x_1,Q^2)$ and $(x_2,Q^2)$
  corresponding to each of the two colliding partons are determined
  approximatly from the corresponding leading--order kinematics.
  We assume $x_1=x_2$ if rapidities are not specified for the
  final states.
  The HL--LHC pseudo--data therefore spans a wide region in the kinematic
  plane, namely $6\times 10^{-5} \lsim x \lsim 0.7$ and
  $40~{\rm GeV} \lsim Q \lsim 7~{\rm TeV}$. 
 \label{fig:kinHLLHC}} 
\end{figure}

In Fig.~\ref{fig:kinHLLHC} we show  the kinematical coverage in
the $(x,Q^2)$ plane of the
  HL--LHC pseudo--data included in this analysis.
  For each data point, the values of $(x_1,Q)$ and $(x_2,Q)$
  corresponding to the two colliding partons are determined
  approximately from leading--order kinematics, which is sufficient
  for illustration purposes.
  We assume $x_1=x_2$ if rapidities are not specified for the
  final states. 
  We see that the HL--LHC pseudo--data covers a wide kinematic region,
  including the large momentum transfers up to $Q\simeq 6$ TeV, as well
  as the large-$x$ region, with several different processes.
  Specifically, the input pseudo--data  spans
  the range $6\times 10^{-5} \lsim x \lsim 0.7$ and
  $40~{\rm GeV} \lsim Q \lsim 7~{\rm TeV}$ in
  the $(x,Q)$ kinematic plane.
  Note that the LHCb measurements are instrumental to constrain
  the small--$x$ region, $6\times 10^{-5} \lsim x \lsim 10^{-3}$, beyond
  the acceptance of ATLAS and CMS.
  
\subsection{Impact of correlating uncertainties}
\label{sec:corrimpact}

As we will also discuss in Sect.~\ref{sec:resultsindividual},
 when constructing the $\chi^2$ estimator for the HL--LHC pseudo--data
 we will not explicitly include the correlations between the systematic errors.
 Instead, we add statistical and systematic uncertainties
 in quadrature as indicated in  Eq.~(\ref{eq:totalExpError}).
  This choice is motivated by the fact that
  it is already challenging to estimate how specific systematic uncertainties
  will be reduced at the HL--LHC, let alone how their mutual correlations
  will be modified.
  Note that
  even restricting ourselves to Run I measurements, the determination of the experimental
  correlation model is a delicate problem, and can in some cases
  complicate the PDF interpretation of measurements such
  as inclusive jet production~\cite{Harland-Lang:2017ytb}.

  On the other hand, completely neglecting the effects of the experimental
  correlations may artificially reduce the
  impact of the pseudo--data into the fit.
  Precisely for this reason, we have introduced
  the correction factor $f_{\rm corr}$ in Eq.~(\ref{eq:totalExpError}).
  Its value has been tuned to the LHC measurements
  of the top quark pair differential
  distributions~\cite{Khachatryan:2015oqa,Aad:2015mbv} at $\sqrt{s}=8$ TeV and
  of the central $W+$charm rapidity distribution ~\cite{CMS-PAS-SMP-17-014}
  at $\sqrt{s}=13$ TeV,
  for which the full breakdowns of systematic errors are available. Although not shown explicitly here, in the latter case we also check against the published 7 TeV data~\cite{Chatrchyan:2013uja}, finding similar results to the preliminary 13 TeV.
  In the following comparison, we use a value for
  the tolerance of $T=1$ (defined in the next section)
  to exaggerate
  the effect of $f_{\rm corr}$ due to the new data having a more dominate
  role in the $\chi^2$, enabling an easier determination of the optimal value.
  
  \begin{figure}[t]
    \begin{center}
  \includegraphics[width=0.49\linewidth]{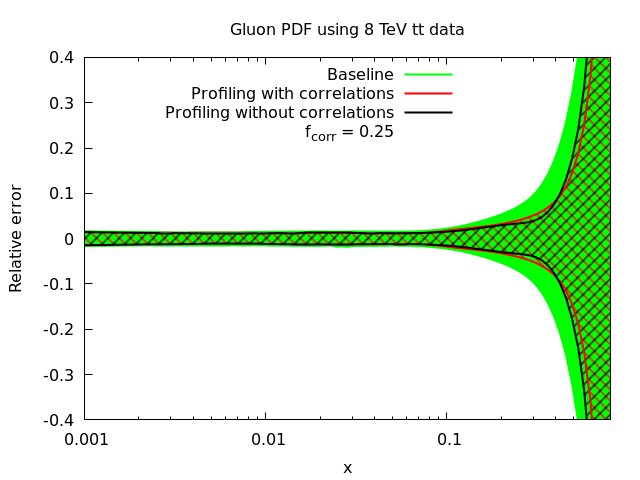}
  \includegraphics[width=0.49\linewidth]{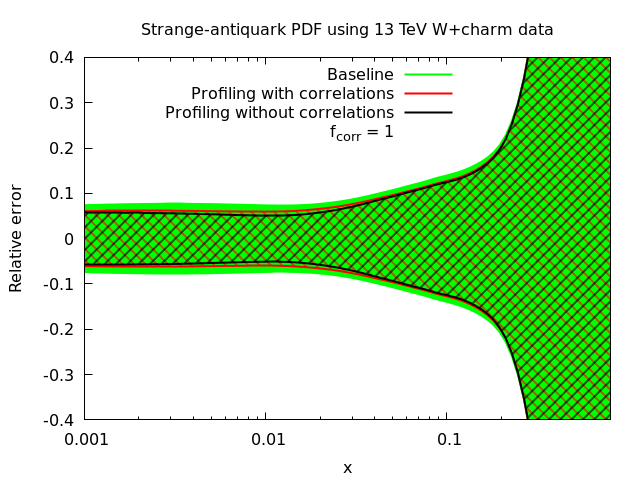}
  \caption{\small Comparison between
    the baseline PDF4LHC15 set and the sets profiled with the LHC
    data, either with or without the correlations between
    the experimental systematic uncertainties accounted
    for.
    In the latter case, the $f_{\rm corr}$ factor is chosen to reproduce
    the results of the profiling when the correlations are included, see text.
    We show here the results of profiling with the
    top differential distributions at $\sqrt{s}=8$ TeV
    with $f_{\rm corr}= 0.25$ (left) and the $W+$charm rapidity distribution
    at $\sqrt{s}=13$ TeV with $f_{\rm corr}= 1$ (right plot).
    A tolerance factor of $T=1$ has been used for this
    specific comparison.
       \label{fig:f_corr} }
    \end{center}
  \end{figure}

  In Fig.~\ref{fig:f_corr} we compare
  the baseline PDF4LHC15 set and the sets profiled with these two
  LHC datasets, with or without the correlations between
    the experimental systematic uncertainties accounted
    for.
    In the latter case, the $f_{\rm corr}$ factor is chosen to reproduce
    the results of the profiling when the correlations are included.
    We can see that for the two considered datasets, rather different values of $f_{\rm corr}$ are preferred; for the top data, we require $f_{\rm corr}\sim 0.25$ while for the $W+$charm data we require instead $f_{\rm corr}\sim 1$.
    Clearly the precise value of this correction therefore appears to depend quite sensitively on the considered datasets, in terms of the corresponding breakdown of systematic uncertainties and overall PDF impact.

    The results of Fig.~\ref{fig:f_corr} might suggest that,
    for projections which are dominantly driven by the potential improvement in systematic uncertainties, our  approach could be questionable
    and require a more complete treatment of experimental correlations.
    However, here we have explicitly chosen our input dataset to be
    composed of those processes
    for which the PDF impact will be driven instead by the improvement in the statistics and extension to unconstrained kinematic regions.
  Indeed, we will see later on that  the specific value of this parameter does not have a large impact on the final results, and we will simply take $f_{\rm corr}=0.5$ in what follows as an average, somewhat weighted towards the value required by the top quark differential data, as this shows a larger PDF impact and would therefore be more important to account for accurately.

%% file: sec-resultsprocs.tex
\section{HL--LHC constraints from individual processes}
\label{sec:resultsindividual}

In this section, we study the constraints on the PDFs
that are expected from individual HL--LHC measurements listed
in Table~\ref{tab:PseudoData}.
First of all, we review the formulation of the Hessian profiling
used in this work to quantify the PDF constraints.
Then we present the results for the
various HL--LHC processes and study how the description
of the pseudo--data is affected.
The complete set of processes is  combined together
into a single profiled PDF set in the next section.

\subsection{The Hessian profiling method}

Quantifying the impact of new experimental data into a Hessian
PDF set such as PDF4LHC15\_100 can be efficiently carried out by means of the
Hessian Profiling technique~\cite{Paukkunen:2014zia,Schmidt:2018hvu}.
This approach is based on
the minimization of the following figure of merit:
\bea
\nonumber
\chi^2\lp {\rm \beta_{exp}},{\rm \beta_{th}}\rp
&=&\sum_{i=1}^{N_{\rm dat}}\frac{1}{\lp \delta^{\rm exp}_{{\rm tot},i}\sigma^{\rm th}_i\rp^2}\lp \sigma_i^{\rm exp}
+\sum_j\Gamma_{ij}^{\rm exp}\beta_{j,\rm exp}
 -\sigma_i^{\rm th}
 +\sum_k\Gamma_{ik}^{\rm th}\,\beta_{k,\rm th} \rp^2 \\ \label{eq:hessianchi2}&&
 +\sum_j \beta_{j,\rm exp}^2+T^2\sum_k \beta_{k,\rm th}^2 \; ,
 \eea
 where $\sigma_i^{\rm exp}~(\sigma_i^{\rm th})$ are the central values
 of a given experimental measurement (theory prediction), see Eq.~(\ref{eq:pseudodataGen}),
 $\beta_{j,\rm exp}$ are the nuisance parameters corresponding
 to the set of fully correlated experimental systematic
 uncertainties, $\beta_{k,\rm th}$ are the nuisance parameters
 corresponding to the PDF Hessian eigenvectors, $N_{\rm dat}$ is the number of data
 points and $T$ is the tolerance factor.
 The matrices $\Gamma_{ij}^{\rm exp}$ and
 $\Gamma_{ik}^{\rm th}$ encode the effects of the corresponding
 nuisance parameters on the experimental data and on the
 theory predictions, respectively.

The minimisation of Eq.~(\ref{eq:hessianchi2}) produces
 approximately equivalent results to carrying out the corresponding Hessian fit
 from scratch, provided settings such as the input PDF parameterisations,
 the tolerance factor $T$, and the theoretical calculations are unchanged.
 An advantage of the Hessian profiling method in comparison
 to related techniques such as the Bayesian
 reweighting method~\cite{Ball:2011gg,Ball:2010gb}, relevant
 for Monte Carlo PDF sets, is that there is no information loss
 even when the added measurements provide
 significant new information.
 This property is crucial in the present analysis, since the HL--LHC
 pseudo--data induces significant constraints on the PDFs.

 As mentioned in Sect.~\ref{sec:pseudodatagen}, in this study we add in quadrature statistical
 and experimental uncertainties (except for the luminosity, which is
 kept fully correlated), and then account for the effects of the missing
 correlations by means of the factor $f_{\rm corr}$.
 For this reason, we only consider nuisance parameters for the luminosity
 errors, as well as for an overall normalization uncertainty of 5\% in forward
 $W$+charm production, arising from charm--jet tagging.
 If we then minimise Eq.~(\ref{eq:hessianchi2}) with respect to these
 experimental nuisance parameters we obtain
  \be
\label{eq:hessianchi2rev}
\chi^2\lp {\rm \beta_{th}}\rp
=\sum_{i,j=1}^{N_{\rm dat}}\lp \sigma_i^{\rm exp}
 -\sigma_i^{\rm th}
 +\sum_k\Gamma_{ik}^{\rm th}\,\beta_{k,\rm th}\rp \lp \text{cov} \rp_{ij}^{-1}\lp \sigma_j^{\rm exp}
  -\sigma_j^{\rm th}
  +\sum_m\Gamma_{jm}^{\rm th}\,\beta_{m,\rm th}\rp  +T^2\sum_k \beta_{k,\rm th}^2 \; ,
 \ee
 where we have defined the experimental covariance matrix as follows:
 \be
 \label{eq:covariancematrix}
	 \lp \text{cov} \rp_{ij} = \delta_{ij}\lp \delta^{\rm exp}_{{\rm tot},i}\sigma^{\rm th}_i\rp^2 + 
	 \Gamma_{i,\text{lumi}}^{\rm exp}\Gamma_{j,\text{lumi}}^{\rm exp} +
            \Gamma_{i,\text{norm}}^{\rm exp}\Gamma_{j,\text{norm}}^{\rm exp} \, .
	    \ee
            Note that since Eq.~(\ref{eq:covariancematrix}) is defined in terms of a fixed
            theoretical prediction (rather than of the fit output itself), our
            results are resilient with respect to 
            the D'Agostini bias~\cite{Ball:2009qv,Ball:2012wy}.

At this point, the minimisation of Eq.~(\ref{eq:hessianchi2rev})
         with respect to the Hessian
 PDF nuisance parameters $\beta_{k,\rm th}$  can
  be interpreted as leading
 to PDFs that have been optimized 
 to describe this new specific measurement.
 The resulting Hessian matrix in the $\beta_{k,\rm th}$ parameter
 space at the minimum can be
diagonalized to construct the new eigenvector directions, and
PDF uncertainties are determined from the $\Delta\chi^2=T^2$ criteria.
In the studies presented here, we use $T=3$, which roughly corresponds to the average
tolerance determined dynamically in the CT14 and MMHT14 analyses.
The resulting profiled PDF set\footnote{Note that sometimes in this paper we will for brevity use the shorthand `fit', but it always understood that a profiling has been performed rather than a full refit.} can be straightforwardly used
for phenomenology using the uncertainty prescription
of symmetric Hessian sets, and the default output format
is compliant with the {\tt LHAPDF} interface.

\subsection{Inclusive gauge boson production}

 We now  present results for the Hessian profiling of the
 PDF4LHC15 set after the inclusion of HL--LHC pseudo--data
 from individual processes.
 Then in Sect.~\ref{sec:ultimatepdfs} we will
 consider the results of the combination for all the 
 processes together.
We begin with the inclusive gauge boson production processes listed
 in Table~\ref{tab:PseudoData} and described
 in Sect.~\ref{sec:pdfsensprocs}.
 We consider two sets of pseudo--data: inclusive $\gamma^*/Z$ production
 in the central rapidity  $|\eta_{ll}|\le 2.4$
 and high invariant mass $m_{ll}\ge 116$ GeV regions,
 and inclusive $W^+,W^-,\gamma^*/Z$ production in the forward
 region, $2.0\le \eta^l\le 4.5$.

 In this section, we will use the same structure to discuss the
 impact on the PDFs of the individual HL--LHC processes that
 are being considered.
 First, we will display representative examples of
 the correlations between the PDFs and the pseudo--data, to
 illustrate the sensitivity of the latter.
 Second, we will show how the description of the HL--LHC pseudo--data
 is modified once it is included in the PDF4LHC15 set by means of profiling.
 Finally, we will assess its impact on the PDFs in a specific
 scenario for the projections of the experimental systematic errors.
In particular, we adopt the `optimistic' choice of Table~\ref{tab:Scenarios}, i.e. $F\equiv f_{\rm corr}\cdot f_{\rm red}=0.2$,
 which corresponds to a value $f_{\rm red}=0.4$ for the
 reduction of the systematic uncertainties compared to
 the 8 TeV baseline measurements. As discussed above, for 13 TeV baselines, in this scenario we take a lower value of 
  $f_{\rm red}=0.2$, to account for the smaller 13 TeV datasets these are based on.

\begin{figure}[t]
  \begin{center}
\includegraphics[width=0.49\linewidth]{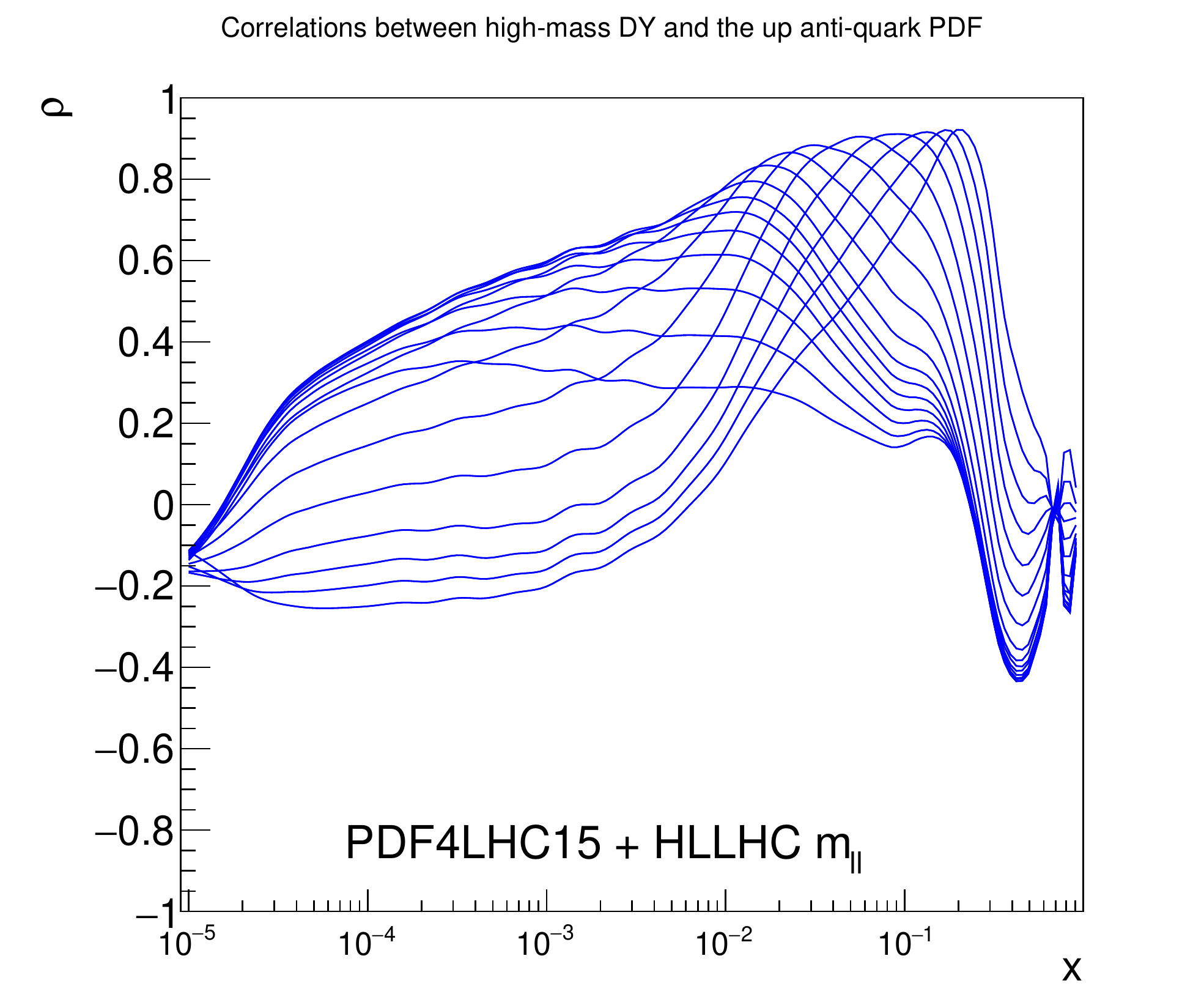}
\includegraphics[width=0.49\linewidth]{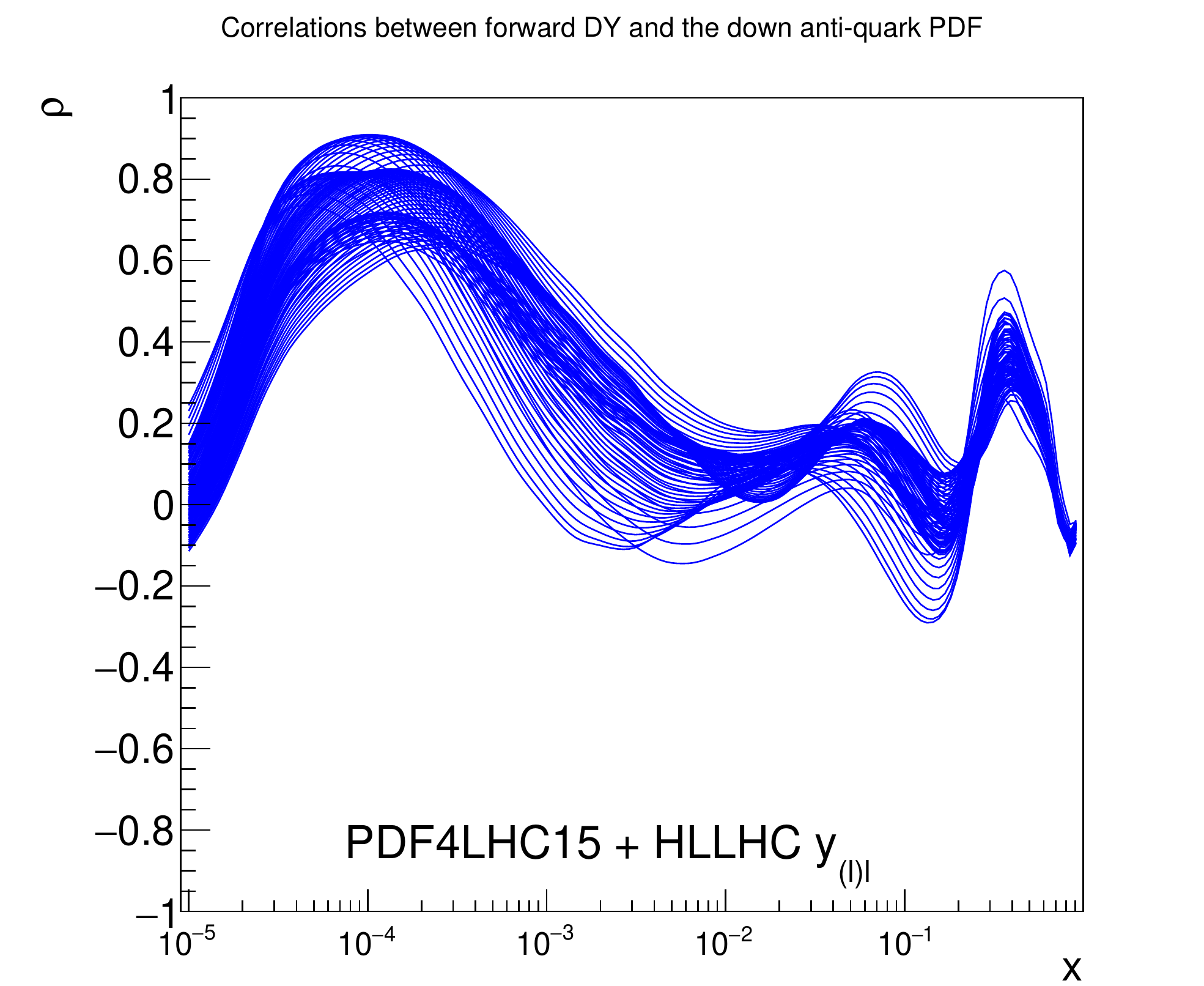}
\caption{\small The correlation coefficients $\rho$ between the
  PDFs and the HL--LHC pseudo--data.
  Left: the correlation between the anti--up quark
  and the high--mass Drell--Yan pseudo--data as a function
  of $x$ for $Q=100$ GeV.
  Right: the correlation between the anti--down quark and the inclusive
  $W,Z$ production process in the forward region.
  In each plot, the different curves correspond to each
  of the bins of the pseudo--data used in the fit.
     \label{fig:correlations_DY} }
  \end{center}
\end{figure}

 We start by discussing the correlations.
 In Fig.~\ref{fig:correlations_DY} we show
 the correlation coefficients $\rho$ between the
 PDFs and the HL--LHC pseudo--data on the Drell--Yan process.
 The left (right) plot displays the correlation
 between the anti--up (anti--down) quark
  as a function
  of $x$ for $Q=100$ GeV
  for the high--mass (forward) Drell--Yan pseudo--data.
  A value of $\rho$ close to 1 (-1) in a given
  region of $x$ indicates that this process is strongly
  (anti--)correlated with the input PDFs in this same region,
  and thus that could potentionally be used to reduce
  PDF uncertainties there.

As we can see from Fig.~\ref{fig:correlations_DY}, in the case of
high--mass Drell--Yan we have $\rho \ge 0.9$ for
$0.05 \lsim x \lsim 0.5$, indicating that this process can provide
information on the large--$x$ antiquarks.
In the case of the forward $W,Z$ production measurements
the correlation coefficient
for the $\bar{d}$ PDF peaks at $x\simeq 10^{-4}$, highlighting
that the forward kinematic coverage of LHCb allows
the quark flavour separation to be pinned down to small values of $x$.

\begin{figure}[t]
  \begin{center}
\includegraphics[width=0.49\linewidth]{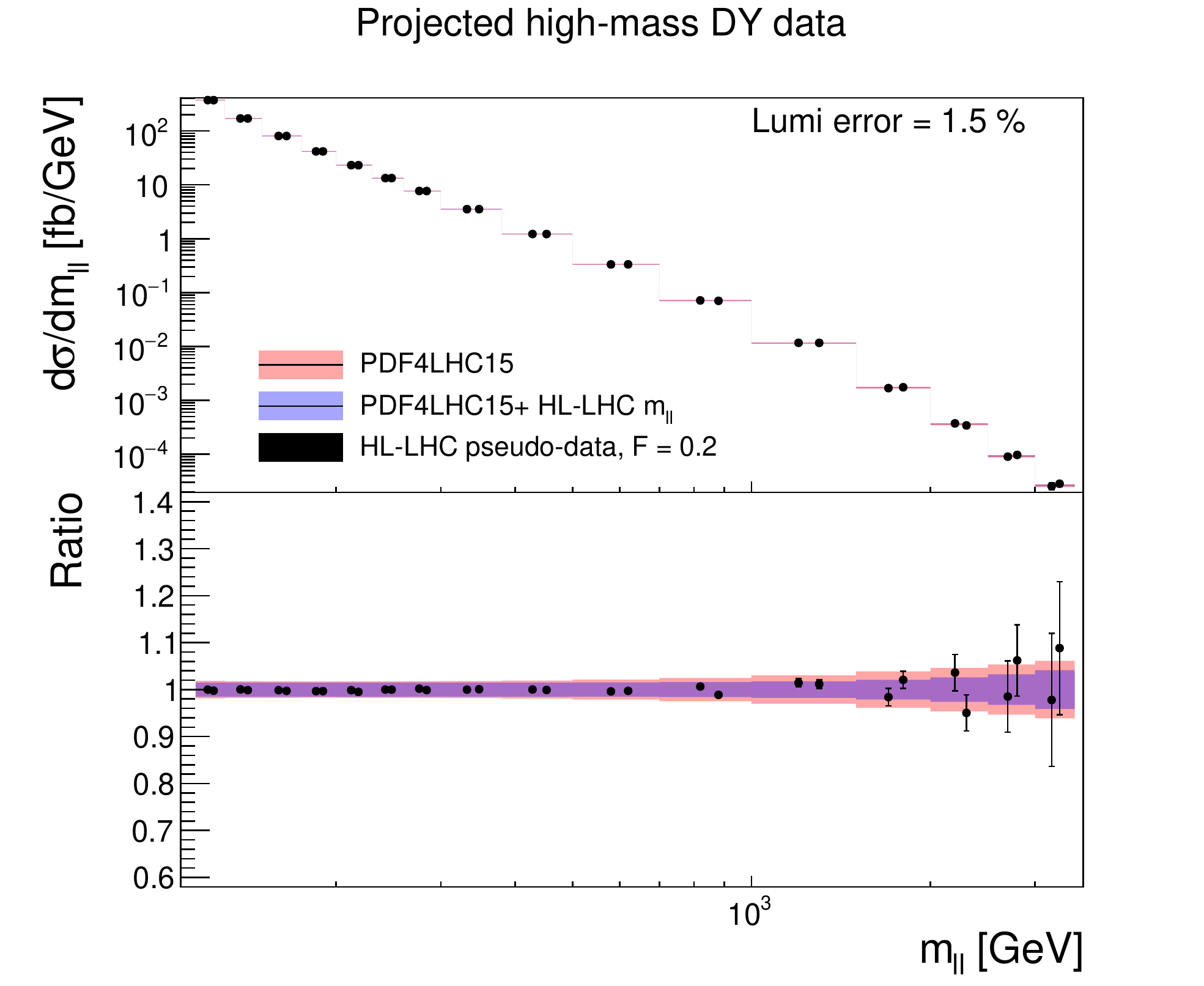}
\includegraphics[width=0.49\linewidth]{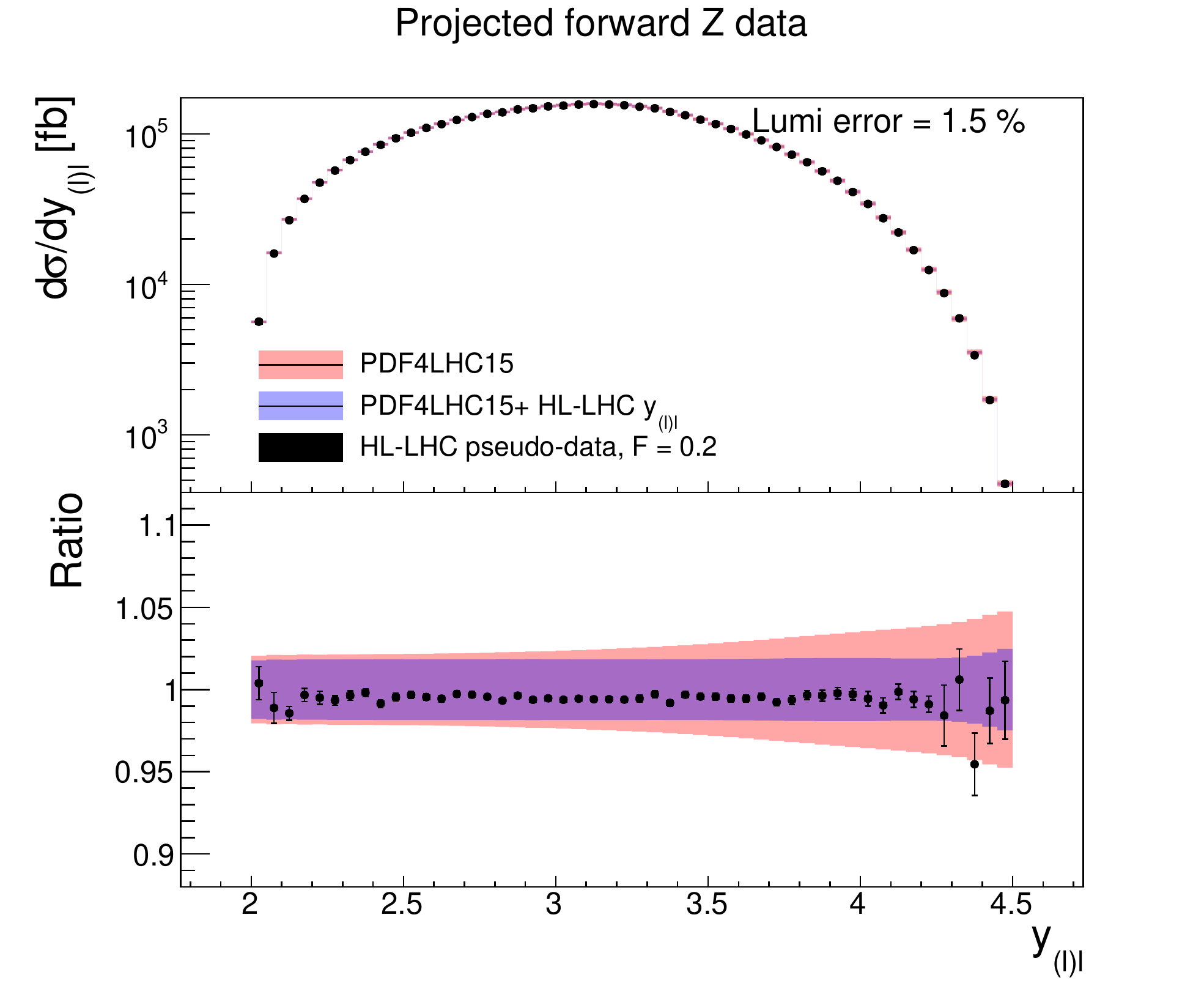}
\caption{\small Comparison between the HL--LHC pseudo--data
  and the theoretical predictions for high--mass (left)
  and forward (right) Drell--Yan production.
  The theory calculations are shown both before (PDF4LHC15)
  and after profiling.
  Luminosity uncertainties are not shown in the experimental
  errors.
  In the bottom panel, we show the same results normalised
  to the central value of the original theory calculation.
  Note in the right plot the comparison are only made for
  forward $Z$ data though both $W$ and $Z$ data are included
  in the profiling.
     \label{fig:datatheory_DY} }
  \end{center}
\end{figure}

We next assess the impact of inclusive gauge
boson production, after profiling.
In Fig.~\ref{fig:datatheory_DY} we show the
comparison between the HL--LHC pseudo--data
  and the theoretical predictions for high--mass (left)
  and forward (right) Drell--Yan production.
  Note in the right plot the comparison is only made for
  forward $Z$ data, but both $W$ and $Z$ data are included
  in the profiling. In addition, in the left plot, in each bin there are two experimental pseudo--data
points, corresponding to the ATLAS and CMS projections; this is true for all central rapidity pseudo--datasets which follow.
  The theory calculations are shown both before (that is,
  using the PDF4LHC15 set)
  and after profiling.
  Luminosity uncertainties are not shown in the experimental
  errors, but are included in the profiling.
  In the bottom panel, we show the same results normalised
  to the central value of the original theory calculation.

From these comparisons, we see that the impact of the high--mass
Drell--Yan pseudo--data on the PDFs is rather moderate, presumably because
even at the HL--LHC the expected precision of the measurements
is comparable or larger than current PDF uncertainties, in particular
in the high $m_{ll}$ range.
On the other hand, for the $W,Z$ measurements that will
be carried out by LHCb we can observe a marked error reduction
of up to a factor two, highlighting the usefulness of
the forward kinematic coverage.
Note that in both cases the central values of the theoretical predictions
are relatively unaffected, with the dominant impact being on the uncertainties. This is expected, as by construction we assume 
the datasets are consistent with the underlying theory and PDFs.

\begin{figure}[t]
  \begin{center}
\includegraphics[width=0.49\linewidth]{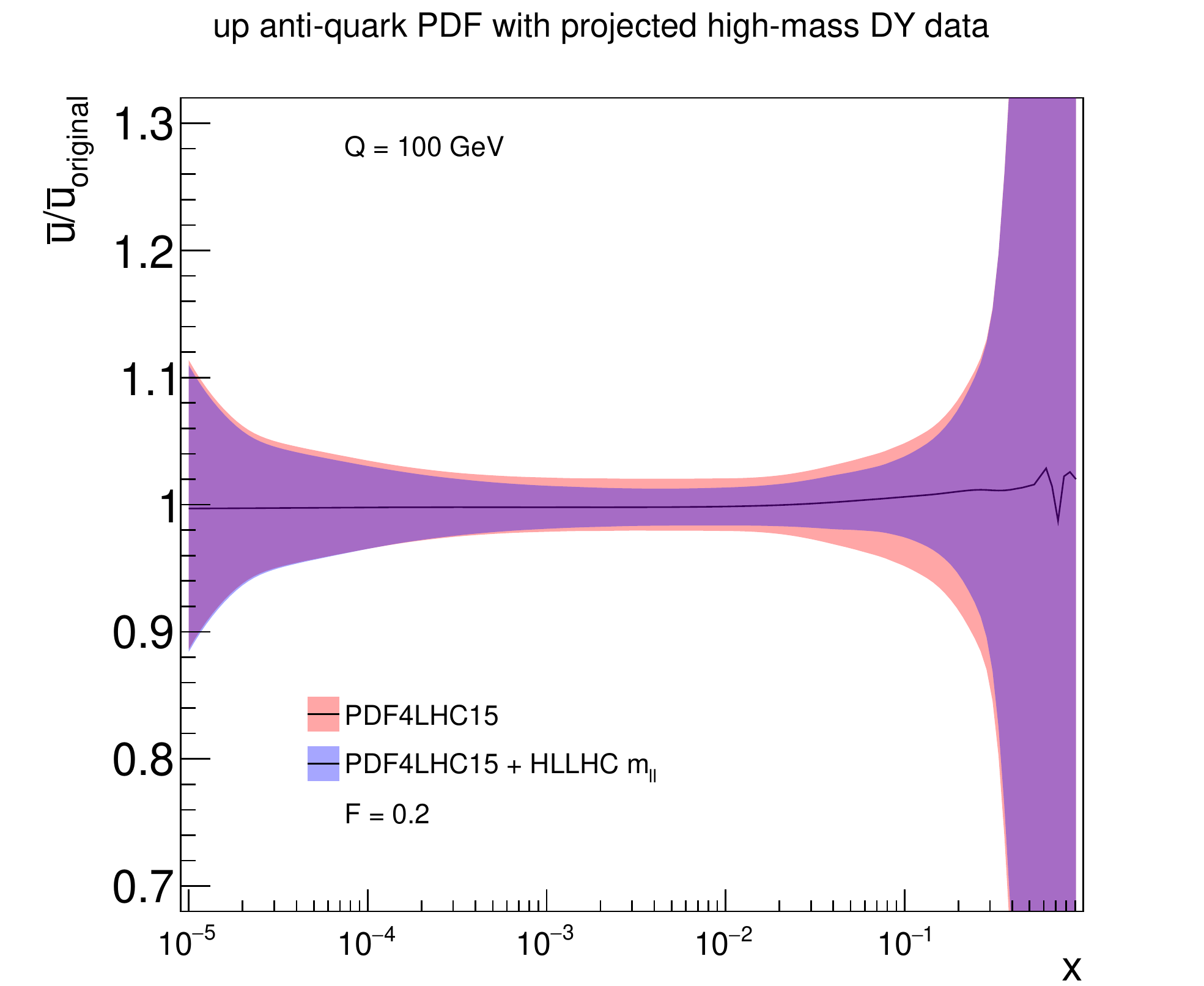}
\includegraphics[width=0.49\linewidth]{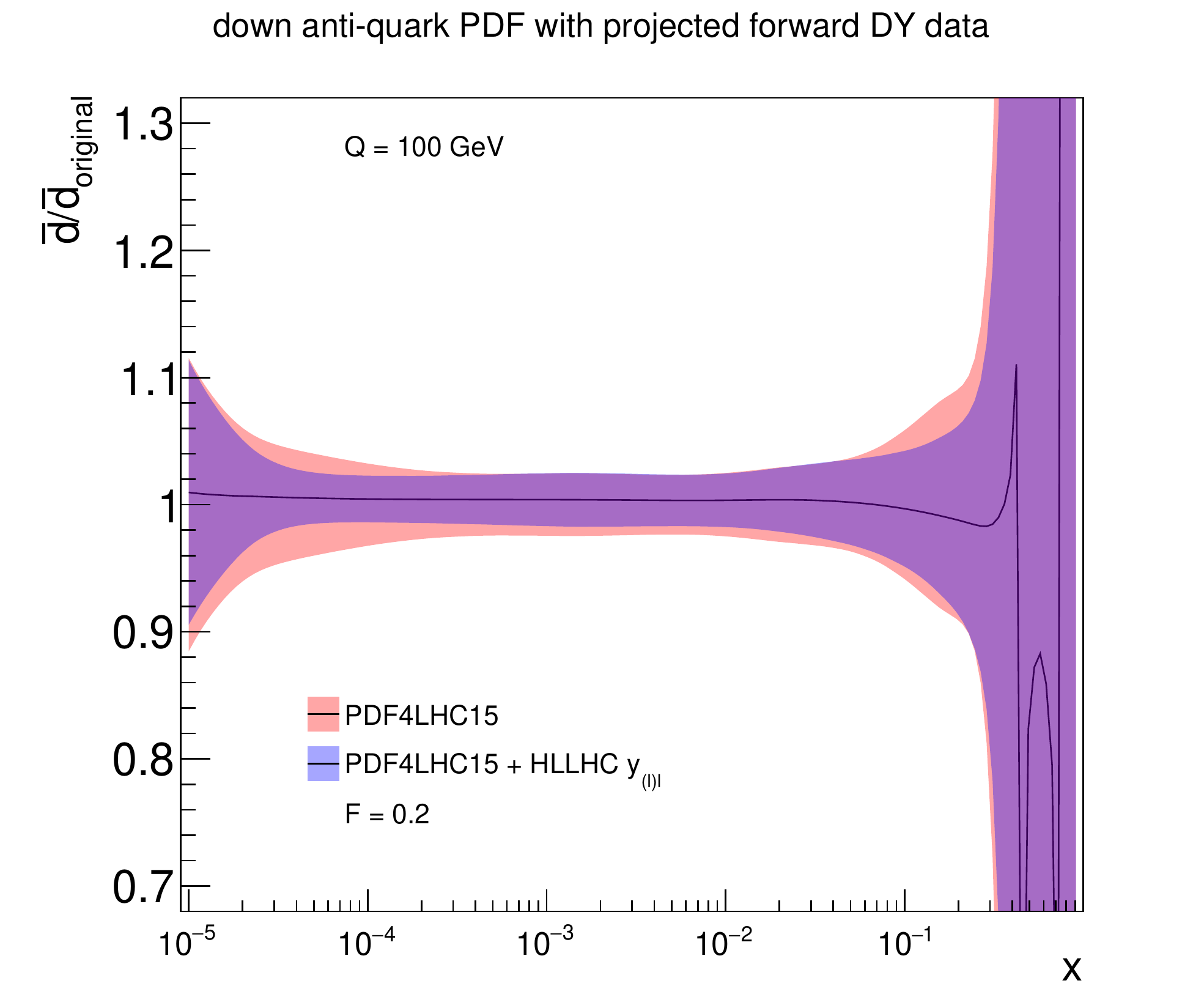}
\caption{\small The impact of the HL--LHC pseudo--data on the PDFs
  at $Q=100$ GeV.
  Left: impact of high--mass Drell--Yan production on the up antiquark.
  Right: impact of  forward $W,Z$ process on the down antiquark.
     \label{fig:PDFimpact_DY} }
  \end{center}
\end{figure}

Concerning the corresponding impact of the HL--LHC pseudo--data on the PDFs,
in Fig.~\ref{fig:PDFimpact_DY} we show the reduction of
the PDF uncertainties found upon the inclusion of the
high--mass Drell--Yan (left) and the forward $W,Z$ (right) pseudo--data
on the PDF4LHC15 set.
We display the same PDF flavours as those used in the
calculation of the
correlation coefficients in Fig.~\ref{fig:correlations_DY},
namely the up and down antiquarks respectively.
What we find is consistent with Fig.~\ref{fig:datatheory_DY}:
a rather moderate effects on the up antiquark from the high--mass Drell--Yan process,
while a more marked effect on the down antiquark from
the forward $W,Z$ process specially in the small--$x$ region.

\subsection{Top quark pair production}

\begin{figure}[t]
  \begin{center}
\includegraphics[width=0.49\linewidth]{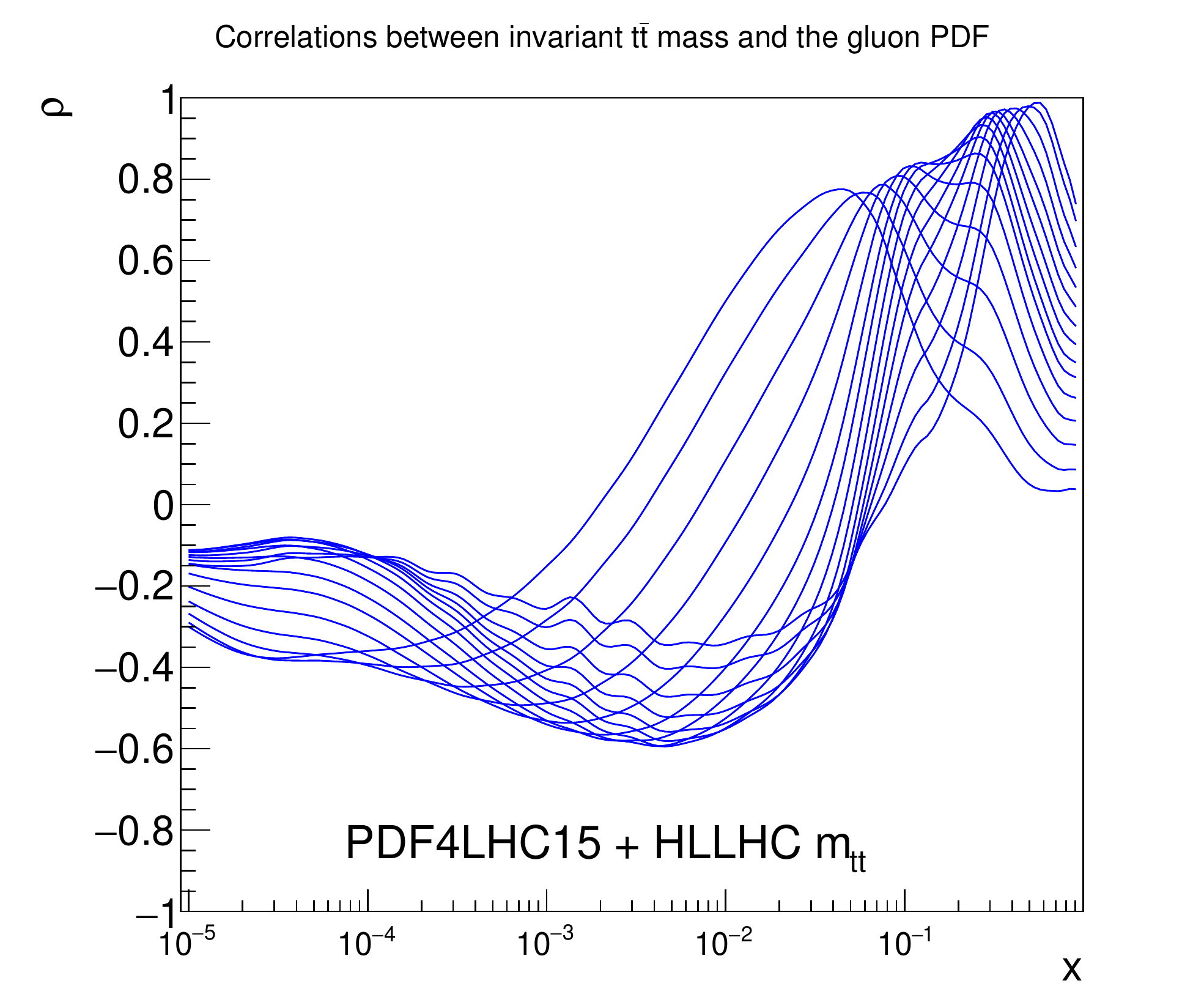}
\caption{\small As in Fig.~\ref{fig:correlations_DY}, now for
  the correlation coefficient between the gluon PDF and the various
  bins of $m_{t\bar{t}}$, the invariant mass 
  of the top quark pair.
     \label{fig:correlations_g} }
  \end{center}
\end{figure}

Here we will focus on the gluon PDF, given that at the LHC
top quark pairs are mostly produced via gluon fusion.
As explained in Sect.~\ref{sec:pdfsensprocs}, we include four different distributions simultaneously: $p_T^t$, $y_t$, $m_{t\bar{t}}$, and $y_{t\bar{t}}$,
assuming that the statistical correlations among them will be available.
First, in Fig.~\ref{fig:correlations_g} we show
the same correlation coefficients as in Fig.~\ref{fig:correlations_DY}
now  between the gluon PDF and the various
  bins of $m_{t\bar{t}}$, the invariant mass distribution
  of top quark pairs.
  The fact that $\rho$ peaks in the large--$x$ region indicates that
  adding the $t\bar{t}$ distributions will directly constrain
  the gluon PDF here.

\begin{figure}[t]
  \begin{center}
\includegraphics[width=0.49\linewidth]{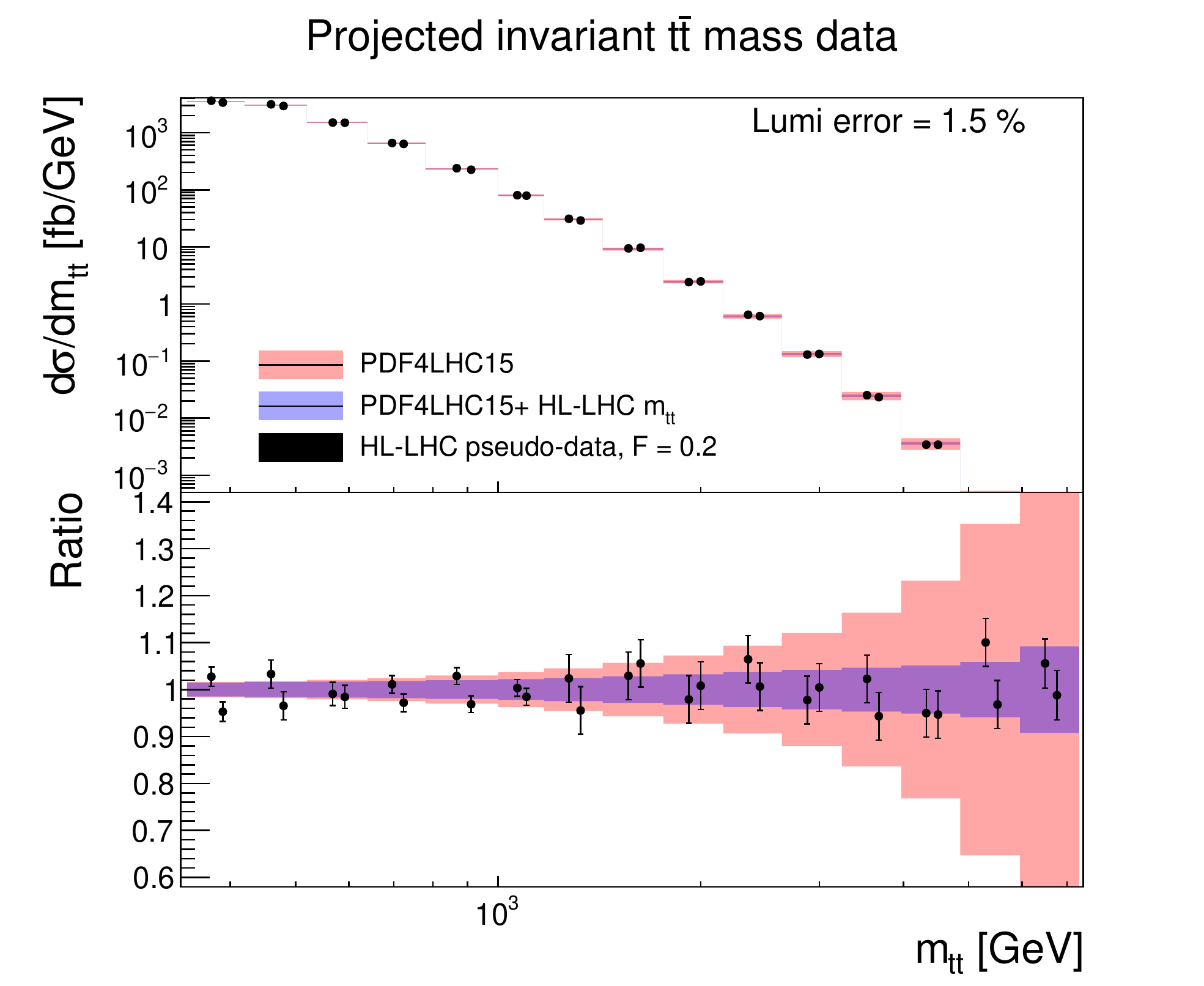}
\includegraphics[width=0.49\linewidth]{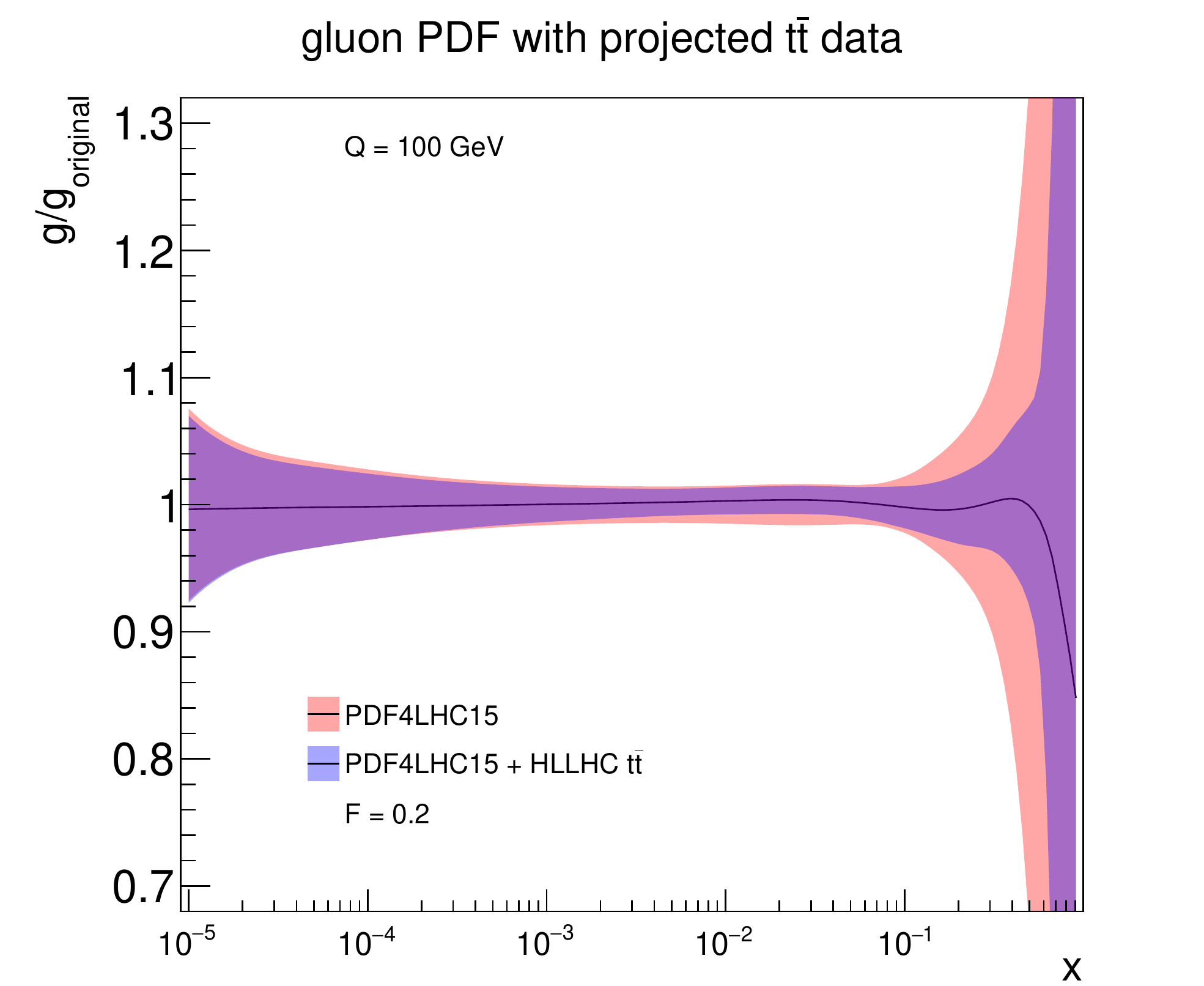}
\caption{\small Left: As in Fig.~\ref{fig:datatheory_DY}, now
  for the $m_{t\bar{t}}$ distribution in top quark
  pair production.
  Right: As in Fig.~\ref{fig:PDFimpact_DY}, now for the gluon
  PDF after including the HL--LHC $t\bar{t}$ pseudo--data
  in the fit.
     \label{fig:mtt_output_F_0_1} }
  \end{center}
\end{figure}

Next, in Fig.~\ref{fig:mtt_output_F_0_1} (left) we show
the same comparison as in Fig.~\ref{fig:datatheory_DY} now
for the  $m_{t\bar{t}}$
distribution.
We can observe a very marked PDF uncertainty reduction at large
values of the invariant mass.
As expected, we find in Fig.~\ref{fig:mtt_output_F_0_1} (right) that
the addition of the HL--LHC $t\bar{t}$ pseudo--data
leads to a significant reduction in the PDF uncertainties
in the gluon PDF at large--$x$, highlighting the good constraining
power of this type of measurements.

\subsection{Jet and photon production}

\begin{figure}[t]
  \begin{center}
    \includegraphics[width=0.49\linewidth]{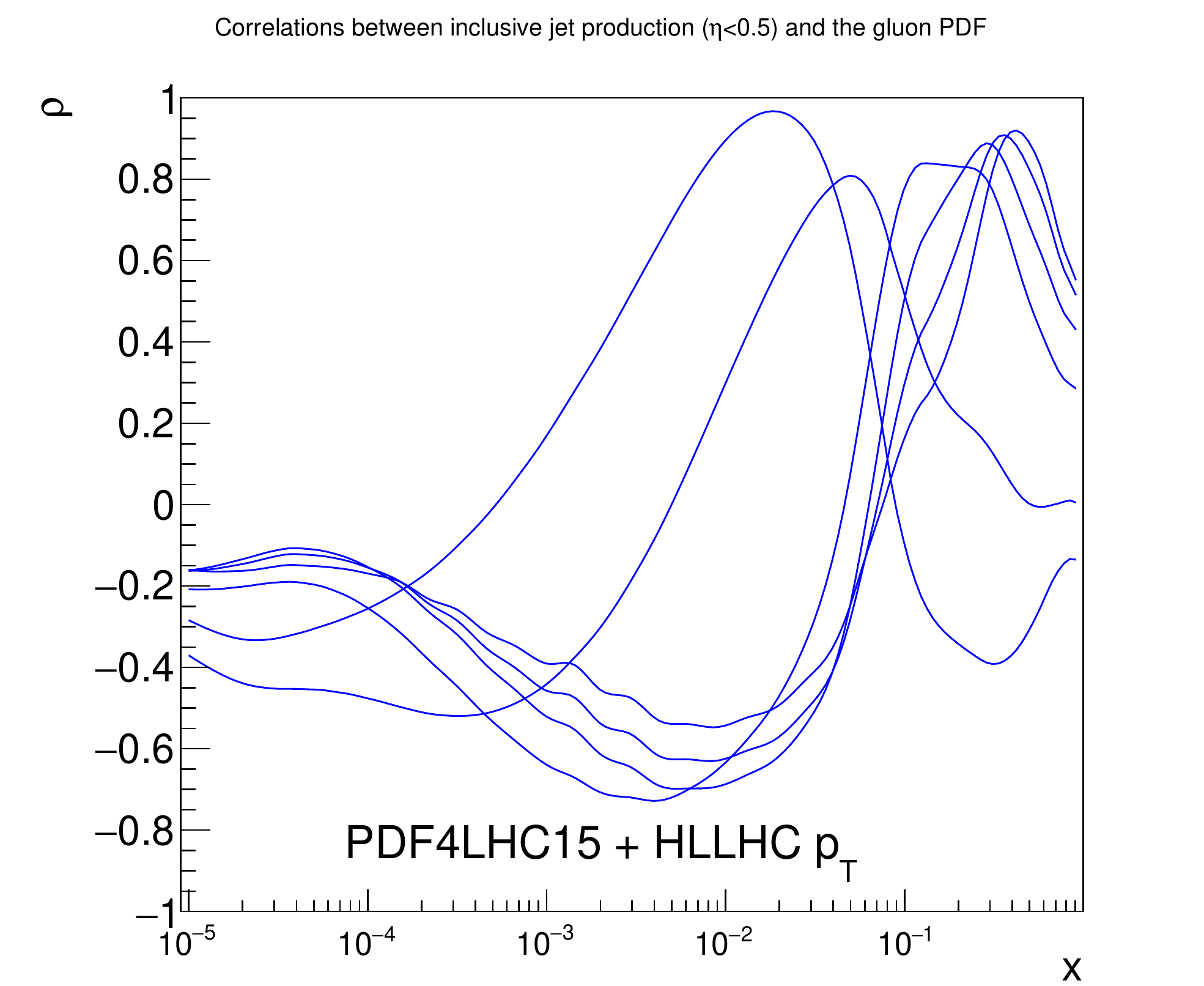}
    \includegraphics[width=0.49\linewidth]{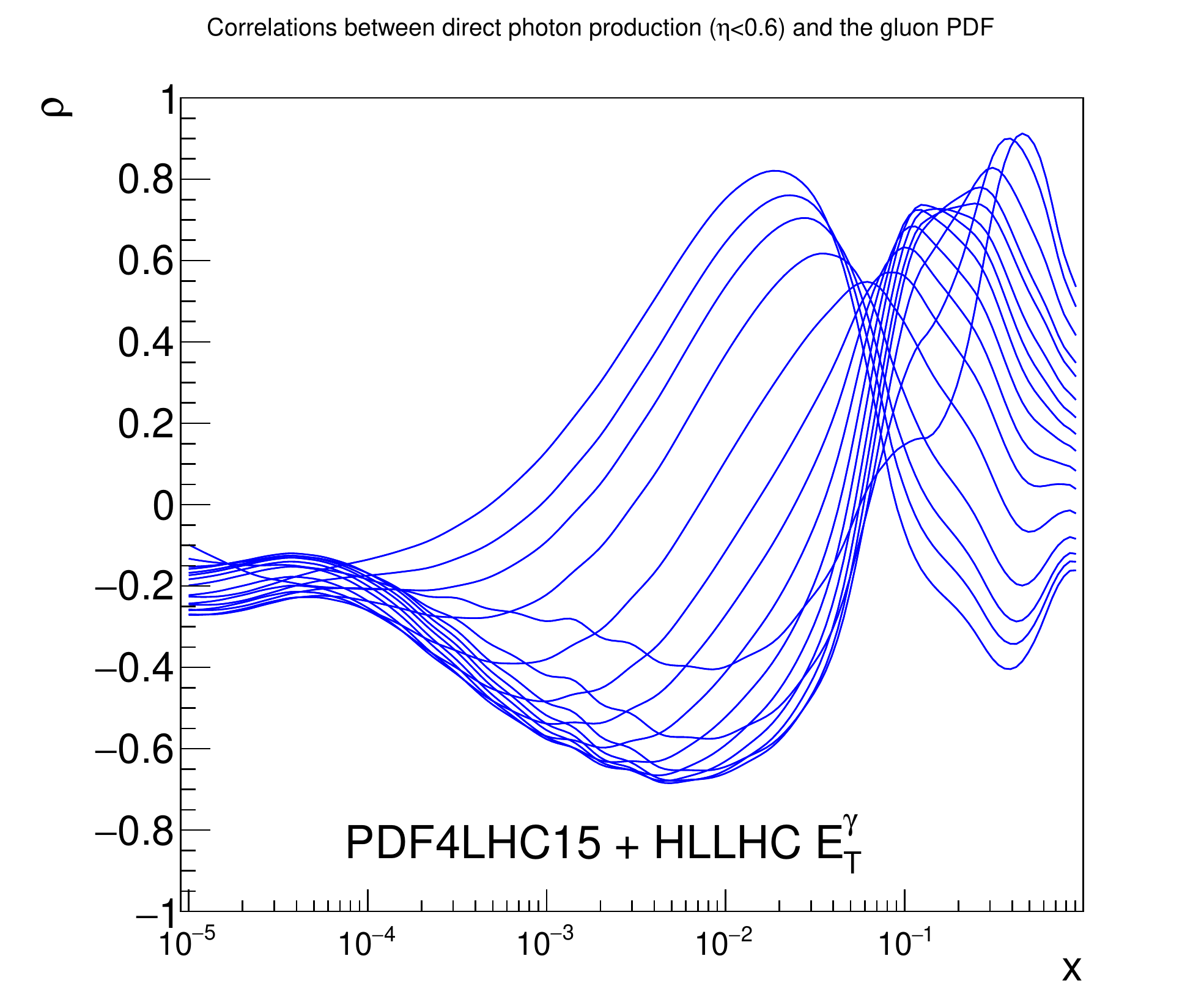}
\caption{\small As in Fig.~\ref{fig:correlations_DY}, now for
  the correlation coefficient between the gluon PDF and
  the central rapidity bin of the inclusive jet (left)
  and direct photon (right) pseudo--data.
  \label{fig:correlations_Jets} }
  \end{center}
\end{figure}

We now turn to consider two of the processes that can be used to provide information on the gluon: inclusive jet
production and direct photon production.
Note that these measurements also provide a handle on the valence
quark distributions, due to the significant fraction of events that
originate from quark--gluon scattering.
In Fig.~\ref{fig:correlations_Jets} we display the correlation coefficient between
the gluon PDF and
  the central rapidity bin of the inclusive jet (left)
  and direct photon (right) pseudo--data.
  From this comparison we see that the correlation profiles
  are similar for the two processes.
  In both cases the correlation coefficient
  is significant around $x\simeq 10^{-2}$,
  is then reduced a bit, and then becomes large again, peaking
  at $x\simeq 0.5$.
  One can verify that the value of $\rho$ decreases as we move to
  more forward rapidities, due to the enhanced contribution
  from quark--initiated diagrams.
  
\begin{figure}[t]
  \begin{center}
\includegraphics[width=0.49\linewidth]{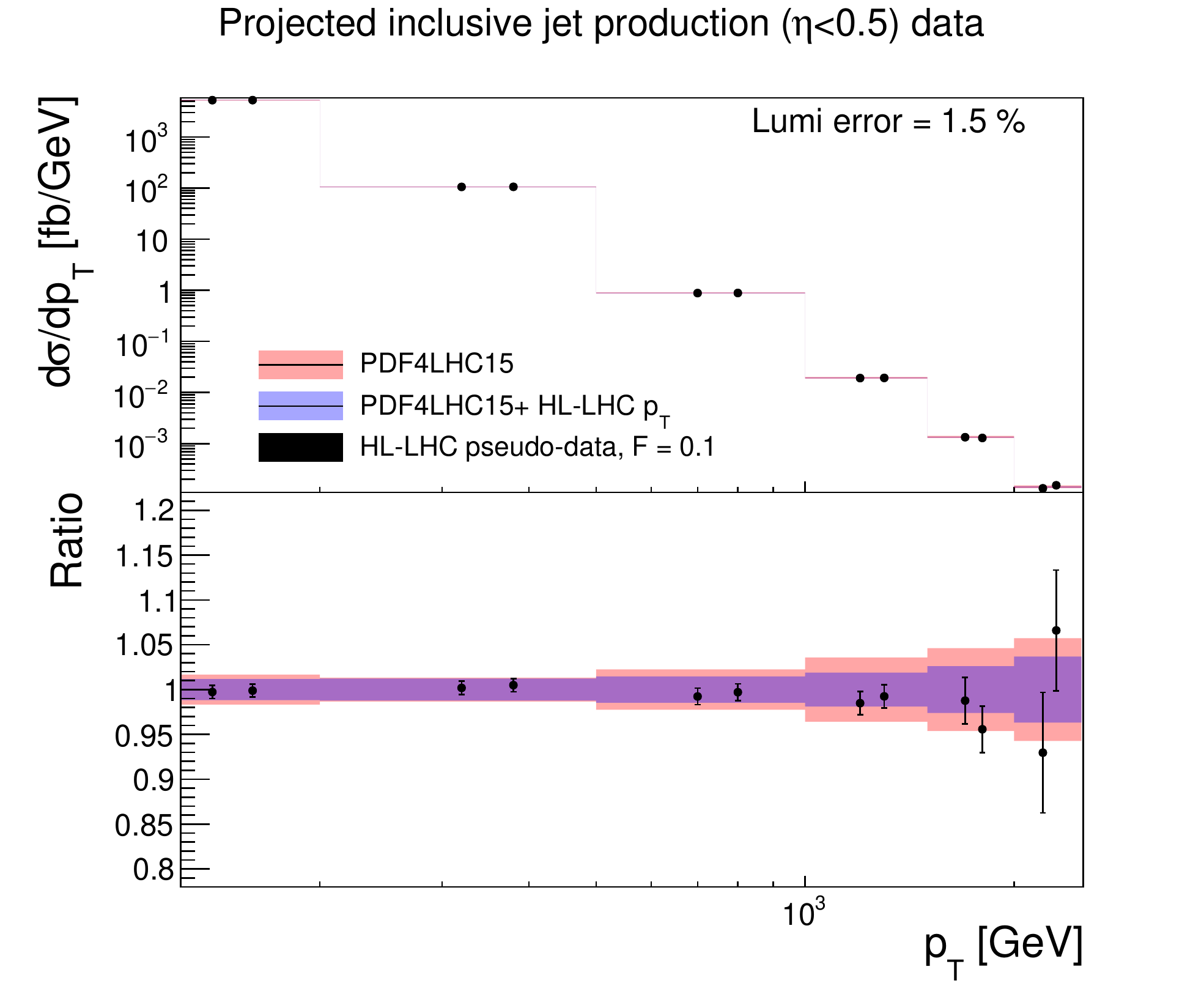}
\includegraphics[width=0.49\linewidth]{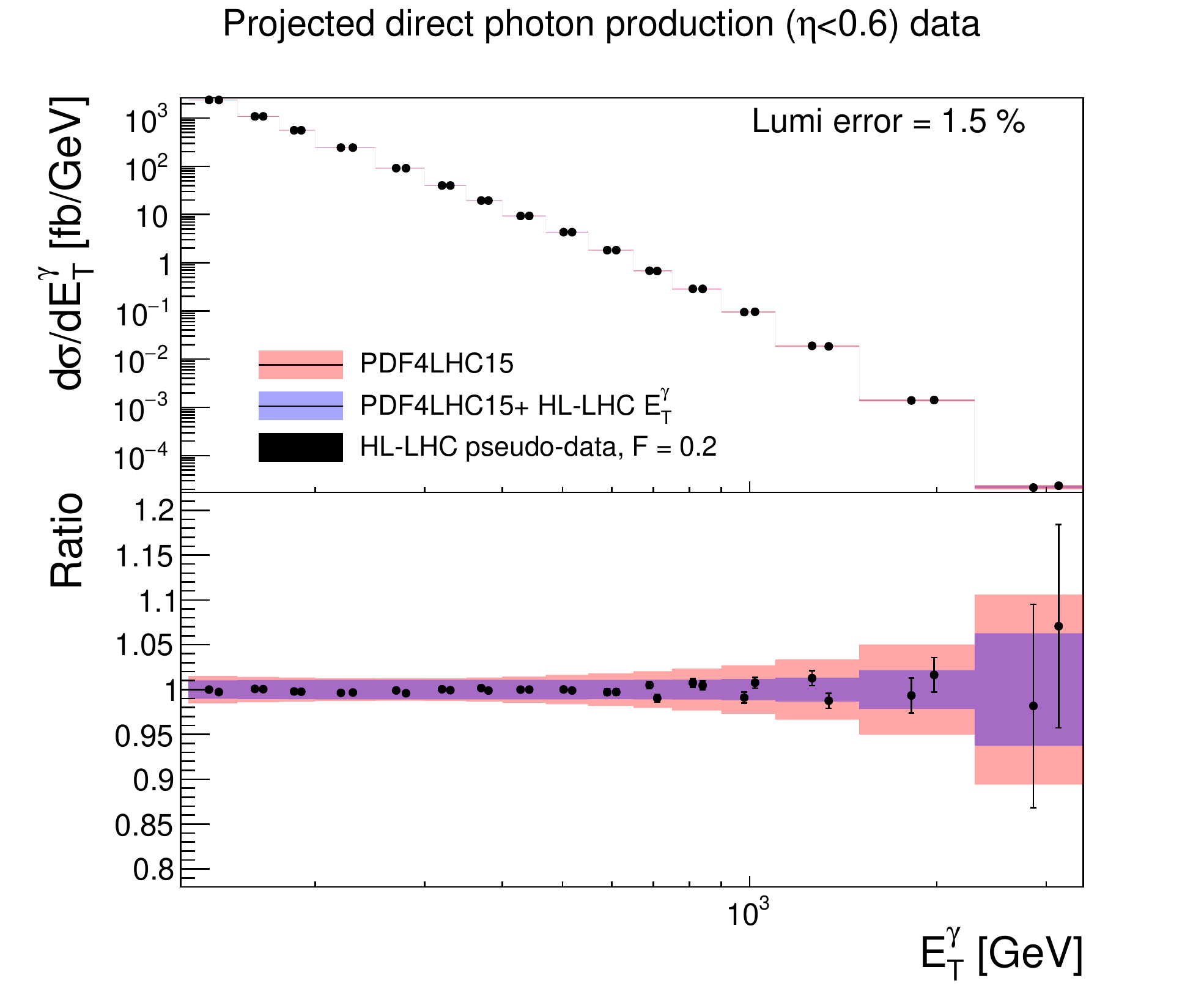}
\caption{\small As in Fig.~\ref{fig:datatheory_DY}, now for the central rapidity bins
  of the inclusive jet production (left plot) and the direct photon production
  (right plot) HL--LHC pseudo--data.
     \label{fig:datath_jetsphotons} }
  \end{center}
\end{figure}

The corresponding comparison between the theory predictions
and the HL--LHC pseudo--data, before and after adding the latter in the fit,
is shown in Fig.~\ref{fig:datath_jetsphotons}.
Note that while we show the results only for central rapidity bins,
the PDF fits include the constraints from all the available rapidity bins.
It is interesting to observe the excellent coverage that the HL--LHC will offer in the TeV region
for these two processes.
In the case of inclusive jet (direct photon) production, we expect to measure the differential
cross sections up to $p_T^{\rm jet}\simeq 2-3$ TeV ($E_T^\gamma \simeq 3$ TeV), a marked
improvement in comparison to current coverage.
Given the back--to--back topology at Born level for these two processes, we see that they can probe
scales up to $Q\simeq 6$ TeV.
We find that the effect of adding the HL--LHC pseudo--data is to reduce
the PDF uncertainties in a range of $p_T^{\rm jet}$ and $E_T^\gamma$.
In the case of direct photon production, the effects are seen across most of the $E_T^\gamma$ range,
while in the case of inclusive jet production they are more localised around the
$p_T^{\rm jet}\sim 1$ TeV region.

\begin{figure}[t]
  \begin{center}
\includegraphics[width=0.49\linewidth]{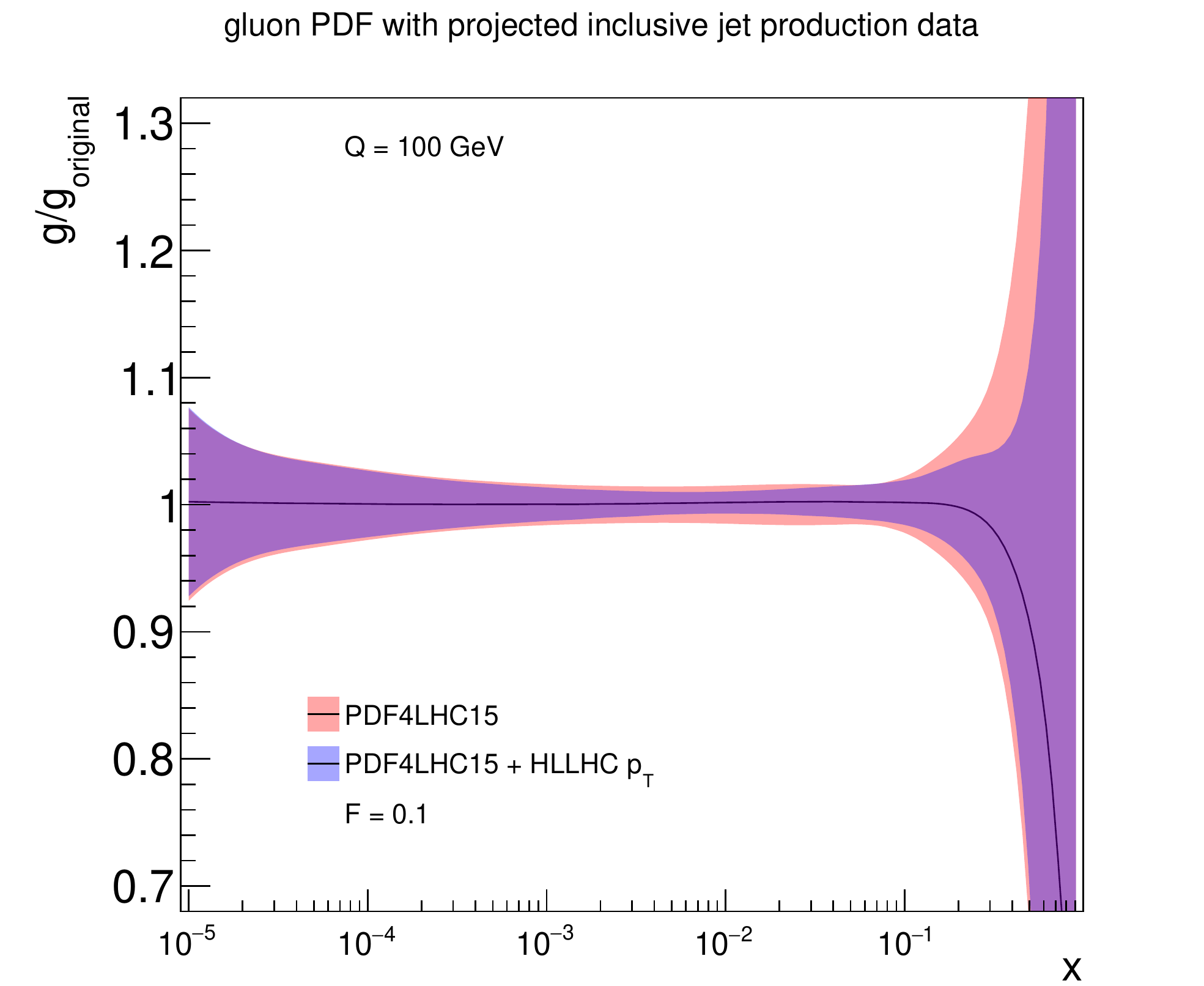}
\includegraphics[width=0.49\linewidth]{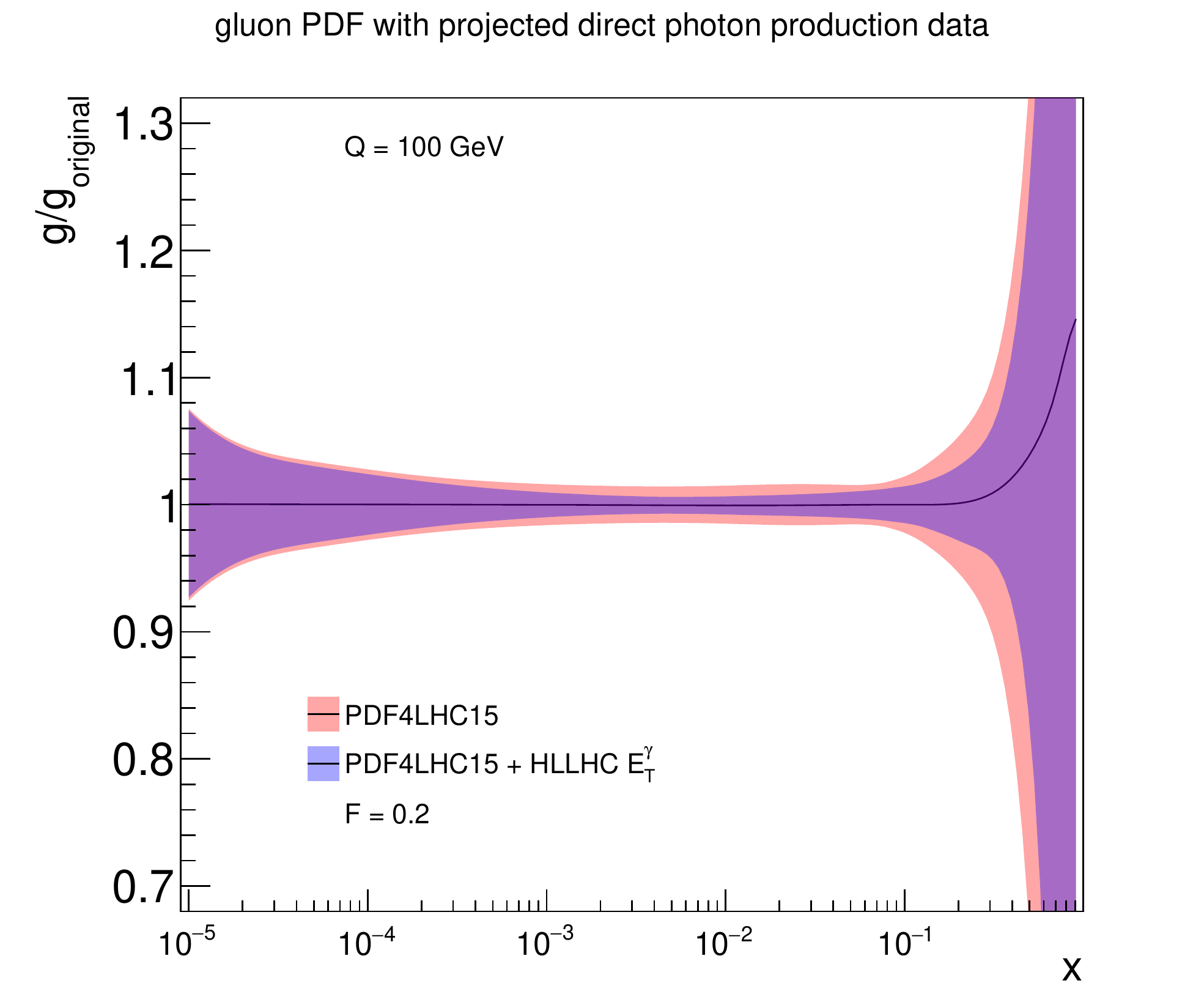}
\caption{\small As in Fig.~\ref{fig:PDFimpact_DY}, now for the gluon
  PDF after including the HL--LHC pseudo--data on inclusive
  jet production (left plot) and on direct photon production (right plot).
 \label{fig:pdf_jetsphotons} }
  \end{center}
\end{figure}

Concerning the impact of these two types of HL--LHC pseudo--data on the PDFs, shown
in Fig.~\ref{fig:pdf_jetsphotons},  one sees that in both cases there
is a visible reduction on the gluon uncertainties at both intermediate and large
values of $x$, of comparable size at high $x$, while at intermediate $x$ the isolated photon data is somewhat more constraining.
Note again that as expected the shift in the central values of the PDFs after profiling are much
smaller than the PDF uncertainties themselves.
Taking into account the results found when adding top quark pair production data
into the fit, see Fig.~\ref{fig:mtt_output_F_0_1}, a clear picture
emerges showing that the HL--LHC measurements will provide
particularly stringent constraints on the large--$x$ gluon PDF, in addition
to those on the quarks.

\subsection{$W$ production in association with charm quarks}

\begin{figure}[t]
  \begin{center}
    \includegraphics[width=0.49\linewidth]{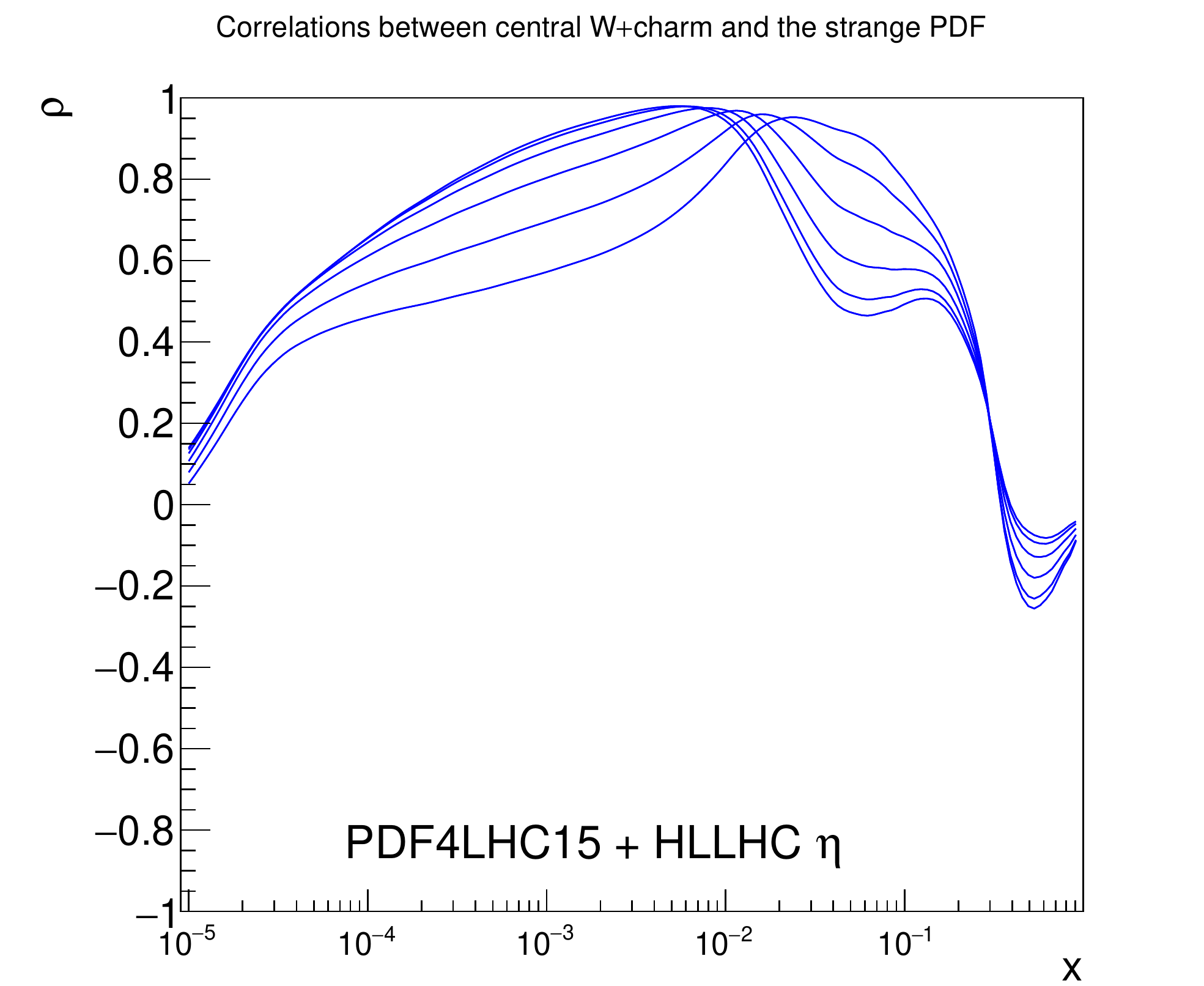}
    \includegraphics[width=0.49\linewidth]{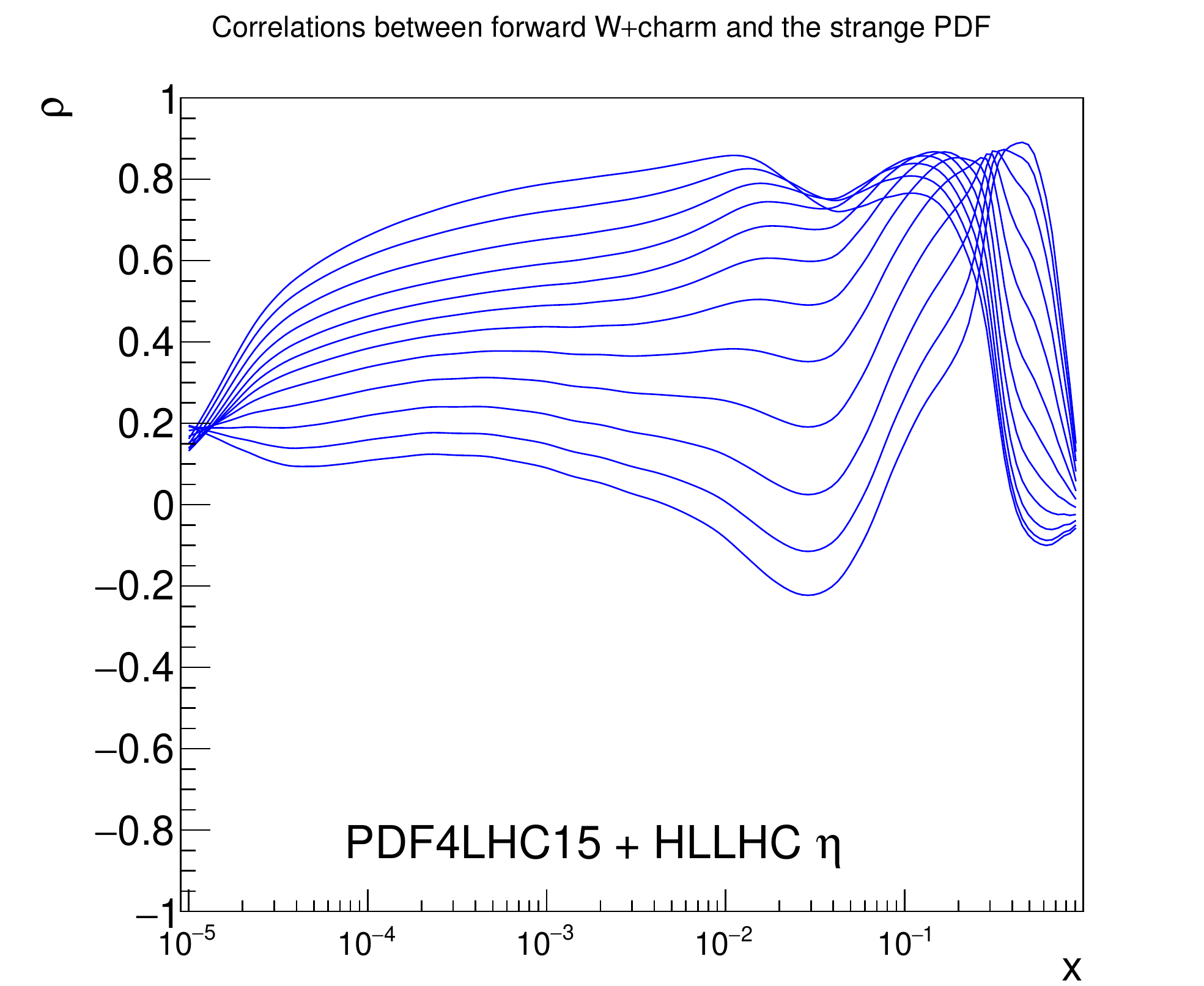}
\caption{\small As in Fig.~\ref{fig:correlations_DY}, now for
  the correlation coefficient between the strange PDF and the lepton rapidity
  distributions in $W$+charm production pseudo--data in the central rapidity region
  (left) and in the forward region (right plot).
  \label{fig:correlations_Wc} }
  \end{center}
\end{figure}

We now consider the impact of $W$ production
in association with charm quarks, which provides direct information on the strange content
of the proton.
As explained in Sect.~\ref{sec:pdfsensprocs} we have generated pseudo--data both
in the central rapidity region, relevant for ATLAS and CMS, and in the forward rapidity
region, relevant for LHCb.
In Fig.~\ref{fig:correlations_Wc} we show the correlation coefficient between the strange PDF and the lepton rapidity
distributions in $W$+charm production pseudo--data both
for the central and the forward rapidity regions.
We can see that indeed there is a large correlation between the strange PDF and the
$W$+charm production pseudo--data in a broad range of $x$ values.
For the case of central production, we find $\rho \ge 0.9$ in the range
of $10^{-3}\le x \le 0.1$, while for forward production the correlation coefficient
$\rho$ peaks at a somewhat smaller value, and covers a broader range in $x$,
with in particular a coverage of the small and large--$x$ regions that is
complementary to the central production pseudo--data.

\begin{figure}[t]
  \begin{center}
\includegraphics[width=0.49\linewidth]{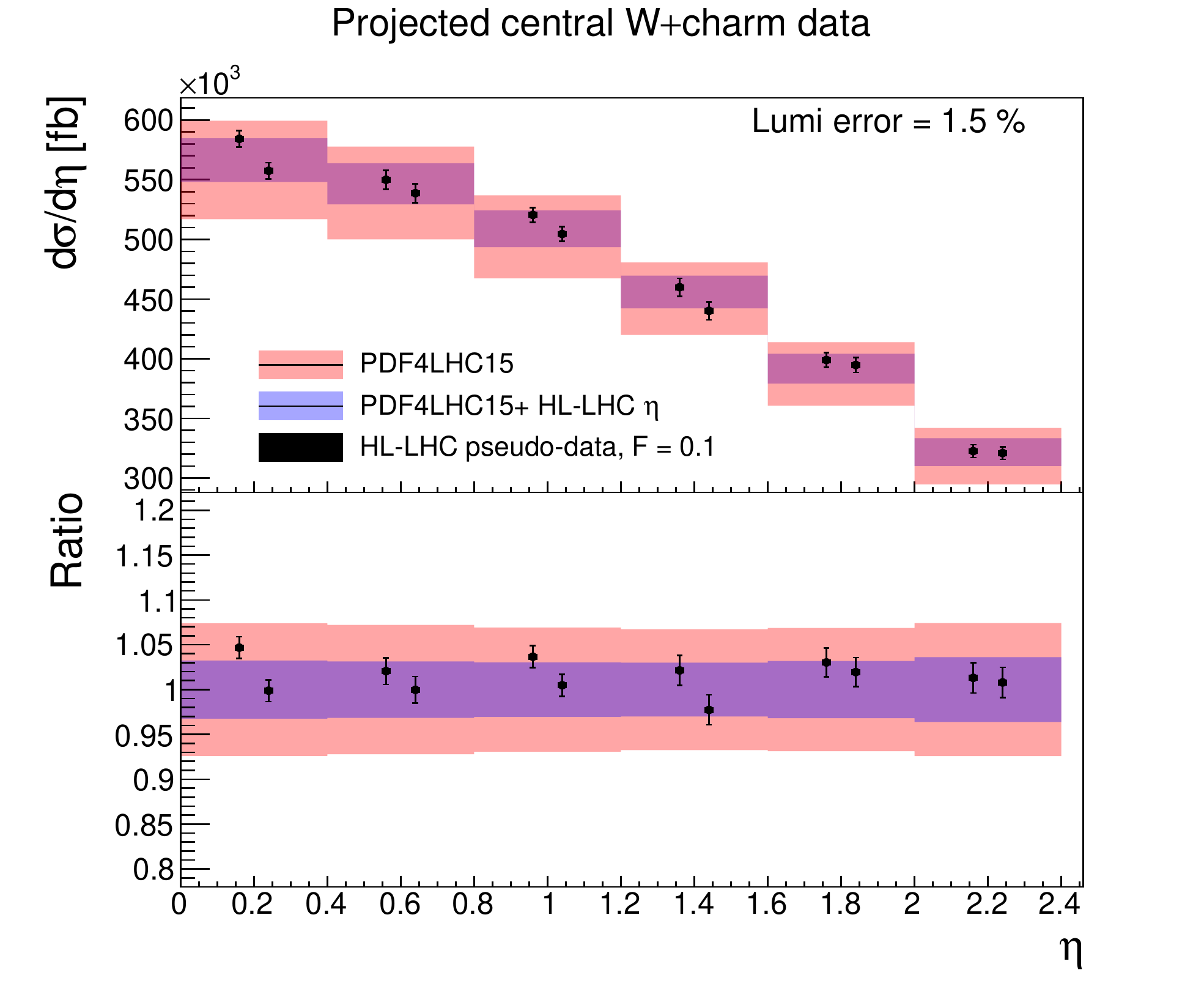}
\includegraphics[width=0.49\linewidth]{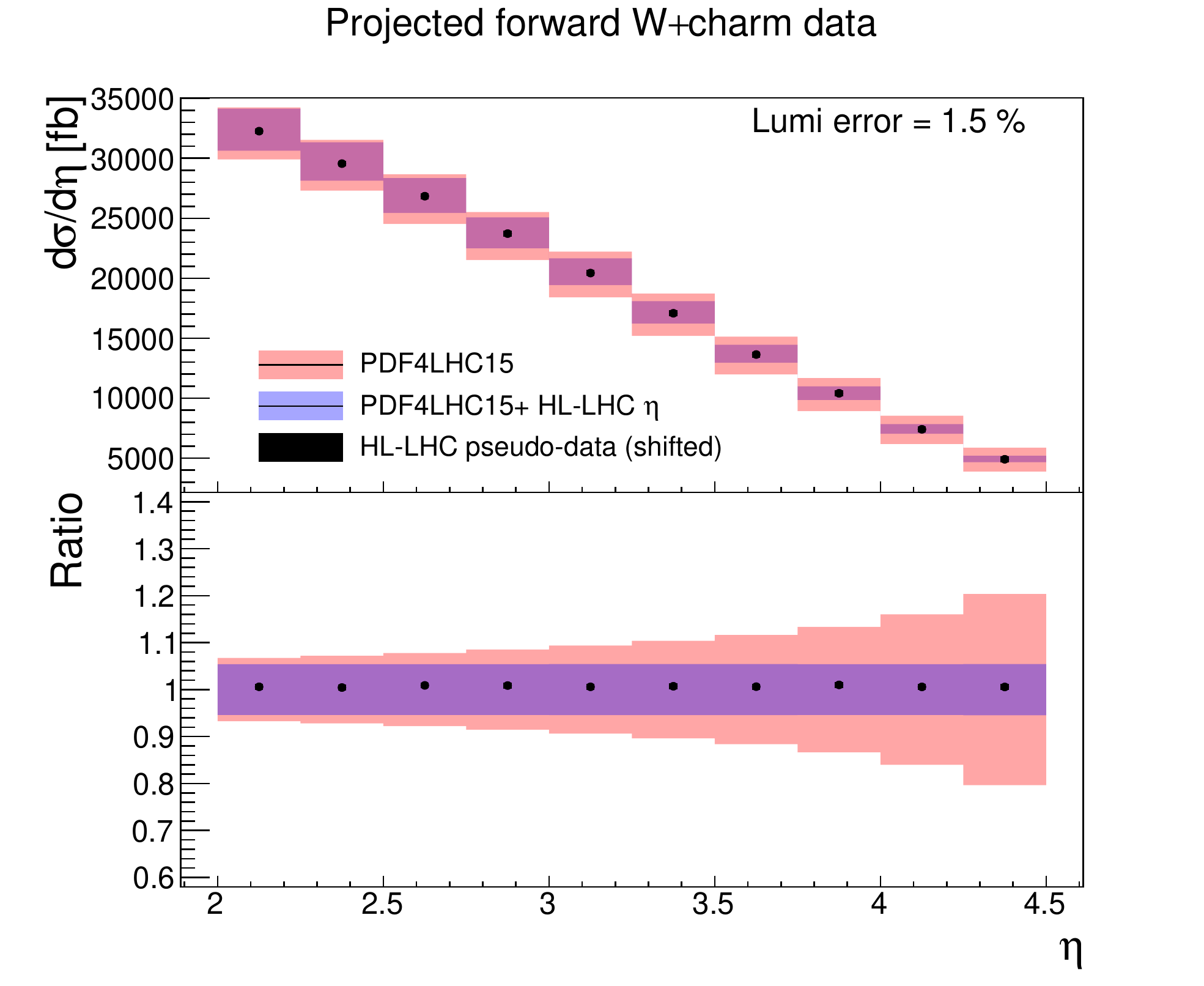}
\caption{\small As in Fig.~\ref{fig:datatheory_DY}, now for $W$+charm
  production in the central (left plot) and forward (right plot)
  rapidity regions. In the right plot only the statistical errors are shown, 
  while the data have been shifted by the dominant correlated source of uncertainty,
  namely the 5\% normalization uncertainty.
     \label{fig:datath_Wc} }
  \end{center}
\end{figure}

The comparison between the HL--LHC pseudo--data and the corresponding
theoretical predictions for $W$+charm production both in the central and
forward regions are collected in Fig.~\ref{fig:datath_Wc}.
In the central region, we see a clear reduction of the PDF uncertainties
after including the pseudo--data into the fit, by around a factor two.
This reduction of uncertainty is approximately constant  as a function
of the lepton rapidity.
At forward rapidities instead, we find that before adding the pseudo--data the
PDF uncertainties grow very fast with rapidity, reaching up to
30\% for $\eta_l \simeq 4.5$, while after including it they are markedly reduced
and become more or less constant with rapidity as in the central region.
Taking into account the correlation coefficients shown in
Fig.~\ref{fig:correlations_Wc}, these results indicates that $W$+charm production
in the forward region provides valuable constraints on the large--$x$ strangeness,
which is currently affected by large uncertainties.

\begin{figure}[t]
  \begin{center}
\includegraphics[width=0.49\linewidth]{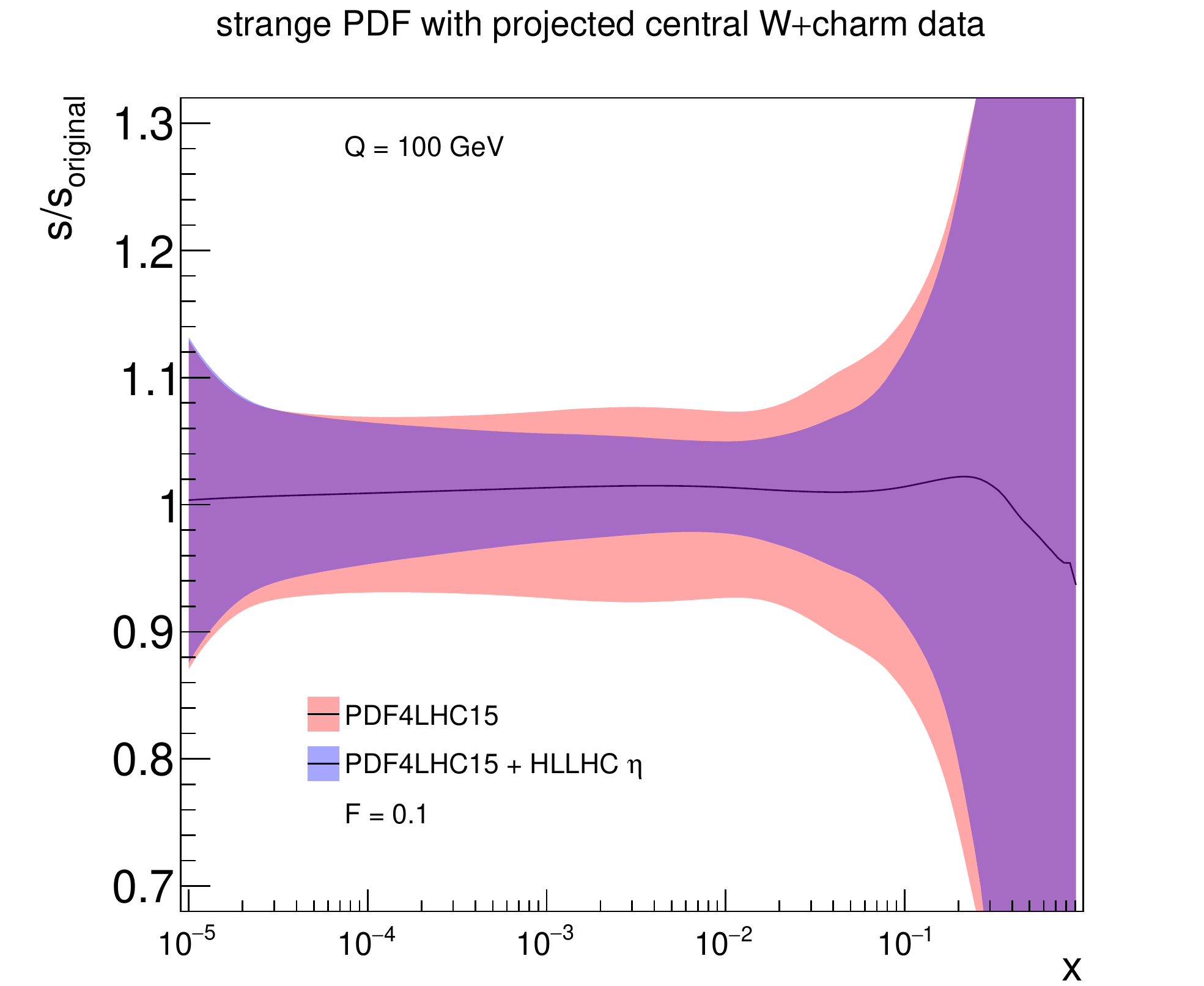}
\includegraphics[width=0.49\linewidth]{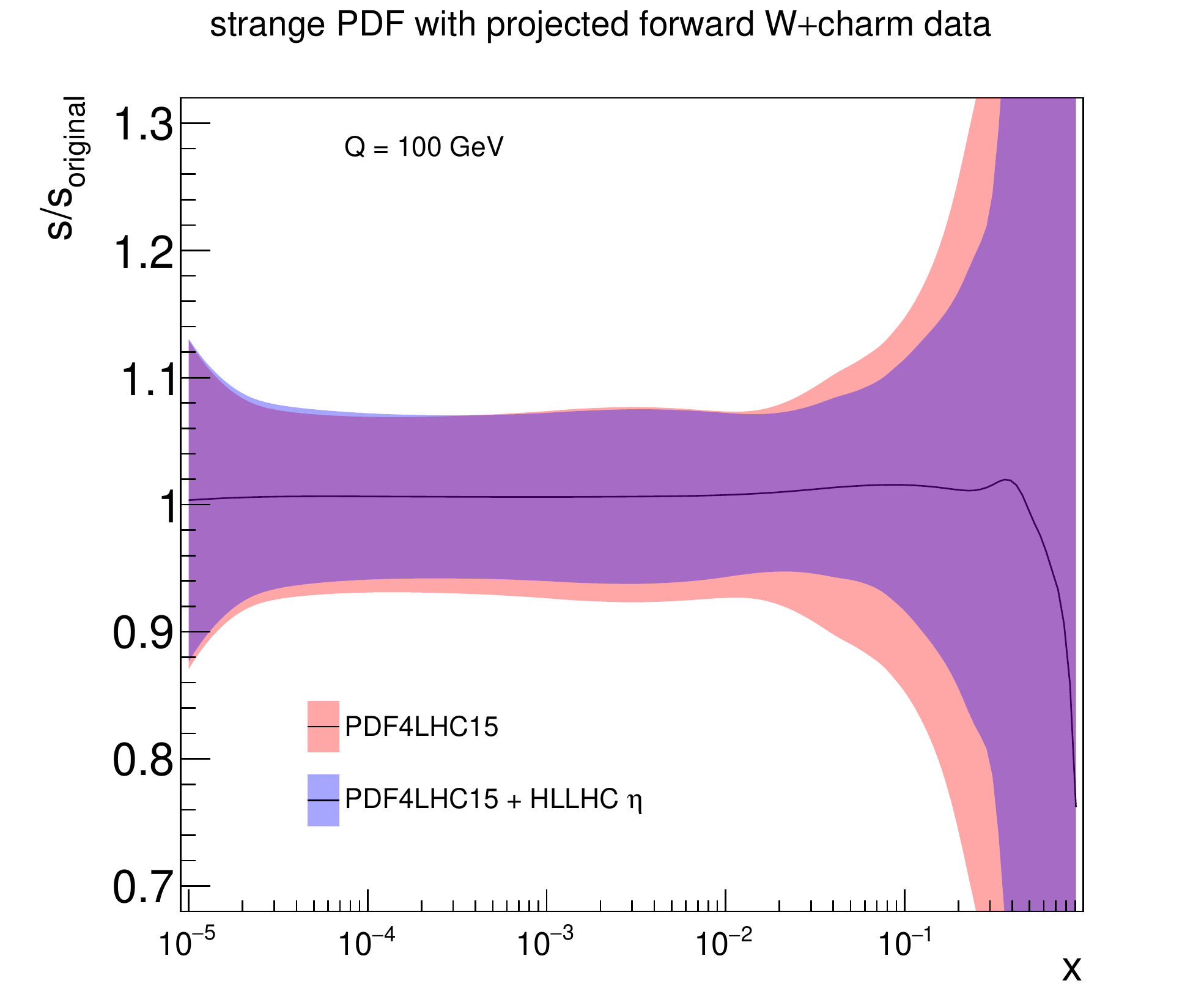}
\caption{\small As in Fig.~\ref{fig:PDFimpact_DY}, now
  for the strange quark PDF
  including the HL--LHC pseudo--data on $W$+charm production
  in the central (left plot) and in the forward region (right plot).
 \label{fig:pdf_Wc} }
  \end{center}
\end{figure}

This PDF uncertainty reduction on strangeness upon the addition of the $W+$charm pseudo--data
is quantified in Fig.~\ref{fig:pdf_Wc}.
For central production, we find that indeed most of the PDF uncertainty reduction
is concentrated in the region $10^{-3}\lsim x \lsim 0.1$, while the large--$x$ region
is affected only moderately.
For the pseudo--data in the forward region instead, there is a superior reduction
of the PDF uncertainties in the large--$x$ region.
The nice complementarity seen from Fig.~\ref{fig:pdf_Wc} illustrates the usefulness
of combine PDF--sensitive measurements in the central rapidity region with those
of the forward region.

\subsection{The transverse momentum of $Z$ bosons}

\begin{figure}[t]
  \begin{center}
\includegraphics[width=0.49\linewidth]{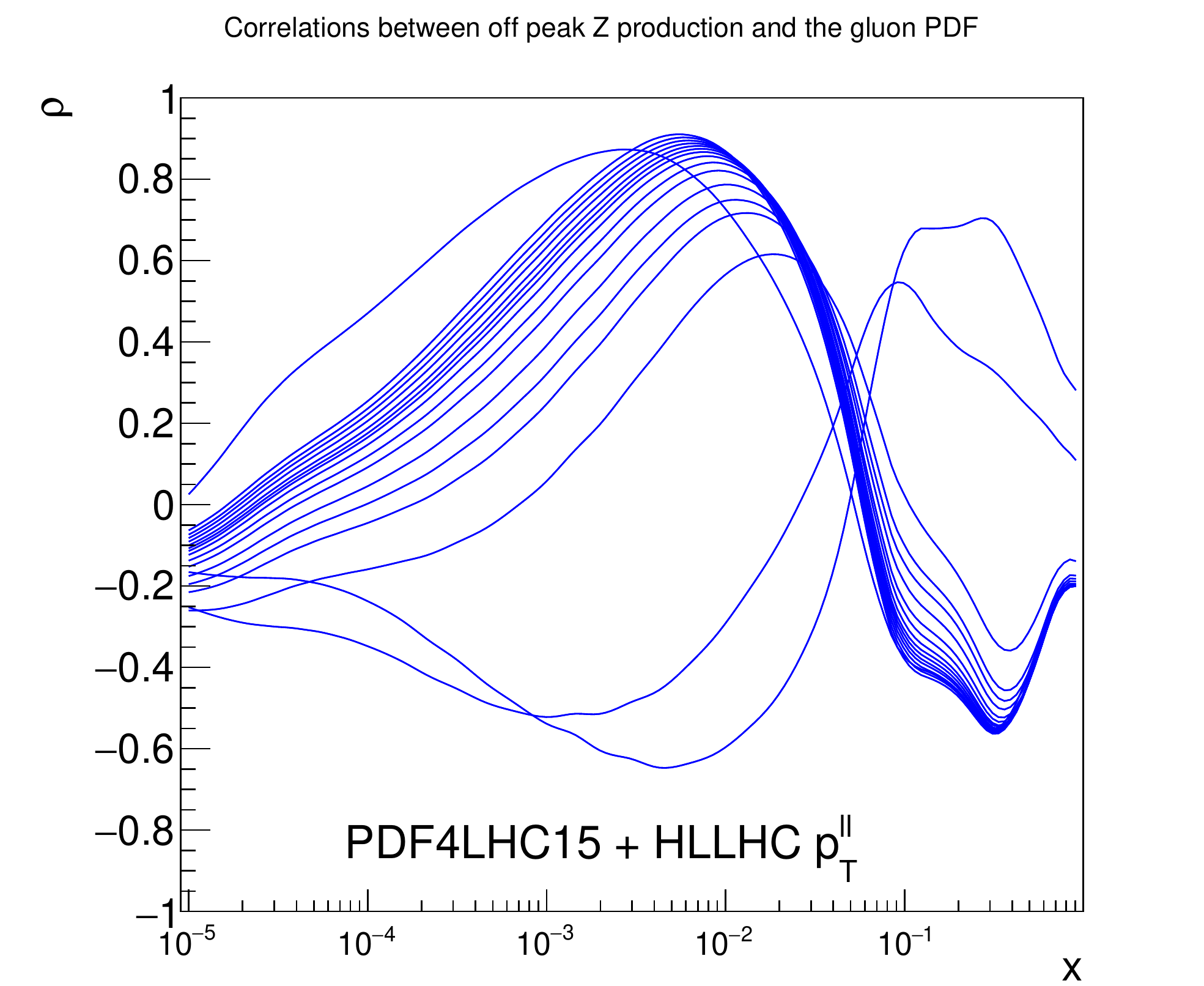}
\includegraphics[width=0.49\linewidth]{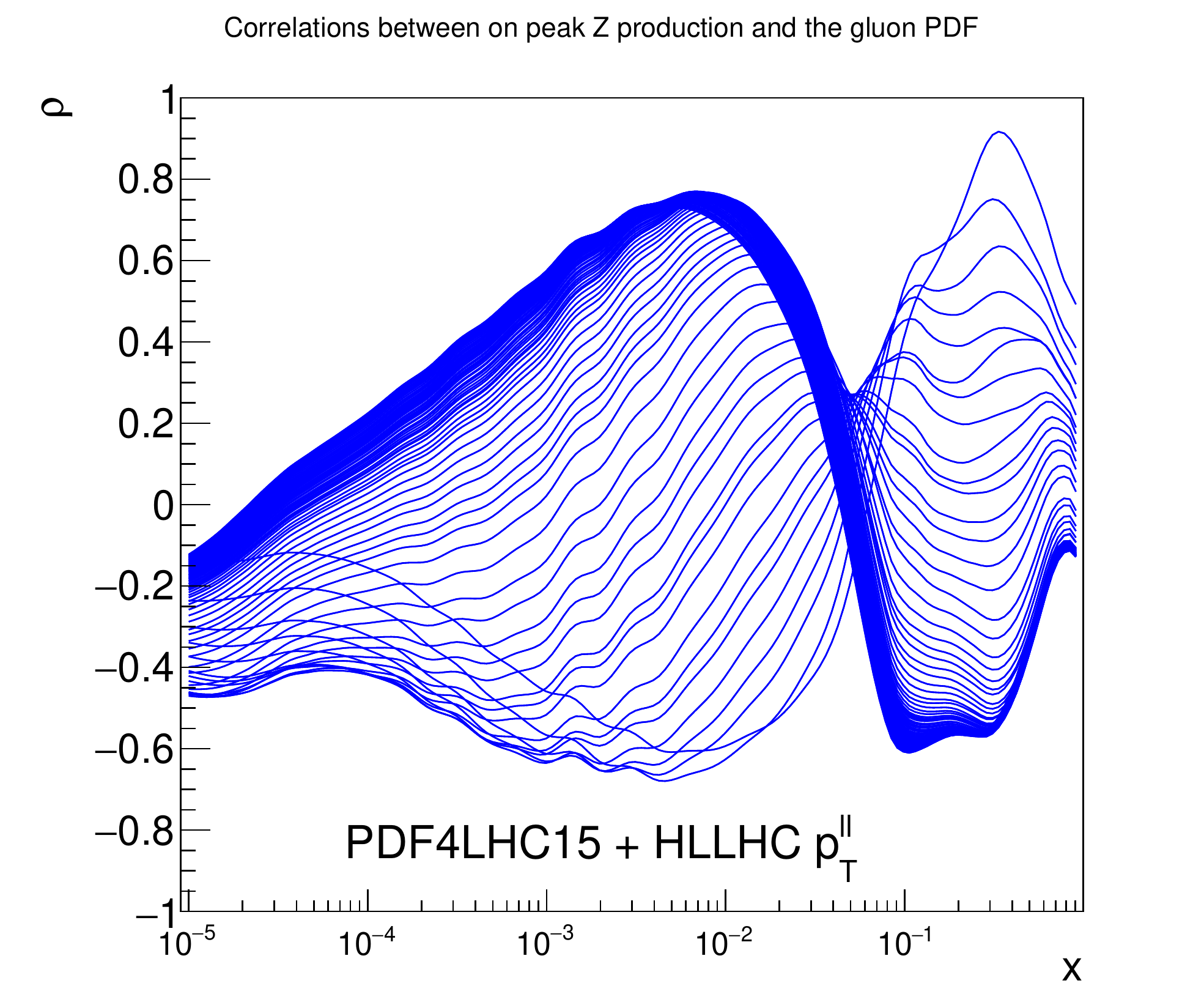}
\caption{\small As in Fig.~\ref{fig:correlations_DY}, now for
  the correlation coefficients between the gluon PDF and the $Z$ transverse
  momentum distributions in the central rapidity region, for
  the dilepton invariant mass $10\,{\rm GeV}\le m_{ll}\le 20$ GeV (left plot)
  and $66\,{\rm GeV}\le m_{ll}\le 116$ GeV (right plot).
  \label{fig:correlations_ZpT} }
  \end{center}
\end{figure}

We complete the study of the impact of individual HL--LHC pseudo--data on the PDFs
with the transverse momentum of $Z$ bosons, a process which is known
to influence the gluon and the quark PDFs at intermediate values of $x$.
We will show the data vs. theory comparison
only for the dominant dilepton invariant mass bin,
$66\,{\rm GeV}\le m_{ll} \le 116$ GeV, although in the PDF profiling the effects
of all the six $m_{ll}$ bins are being taken into account.
To begin with, in Fig.~\ref{fig:correlations_ZpT} we show
the correlation coefficients between the gluon PDF and the $Z$ transverse
  momentum distributions in the central rapidity region, for
  the dilepton invariant mass $10\,{\rm GeV}\le m_{ll}\le 20$ GeV 
  and $66\,{\rm GeV}\le m_{ll}\le 116$ GeV.
  As we can see from this comparison, the pseudo--data on the
  $Z$ $p_T$ provides information on the gluon PDF, being mostly
  sensitive at $x\simeq 10^{-2}$, as well as around $x\simeq 0.3$
  in the case of the on--peak dilepton invariant mass bin.

\begin{figure}[t]
  \begin{center}
\includegraphics[width=0.49\linewidth]{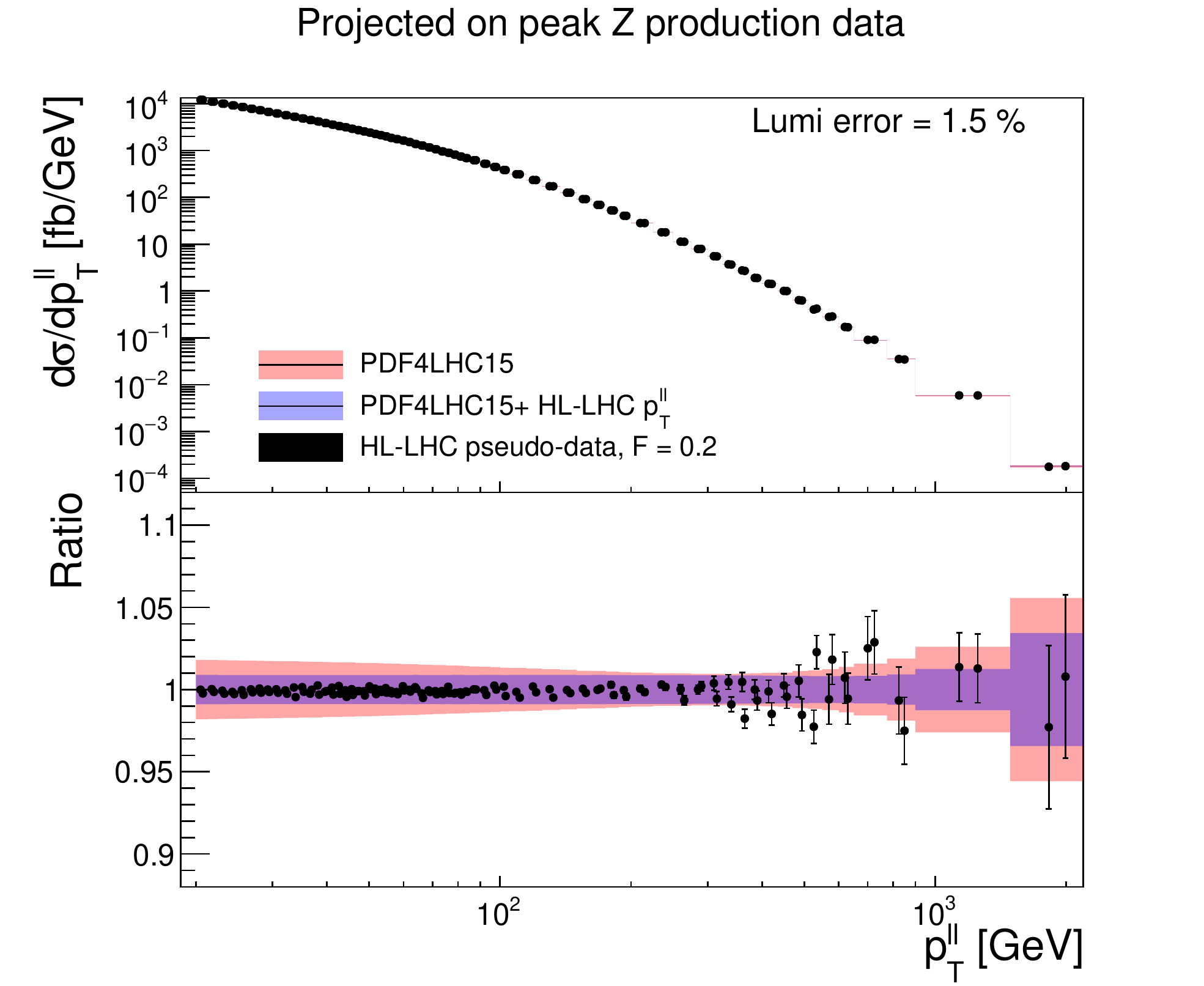}
\includegraphics[width=0.49\linewidth]{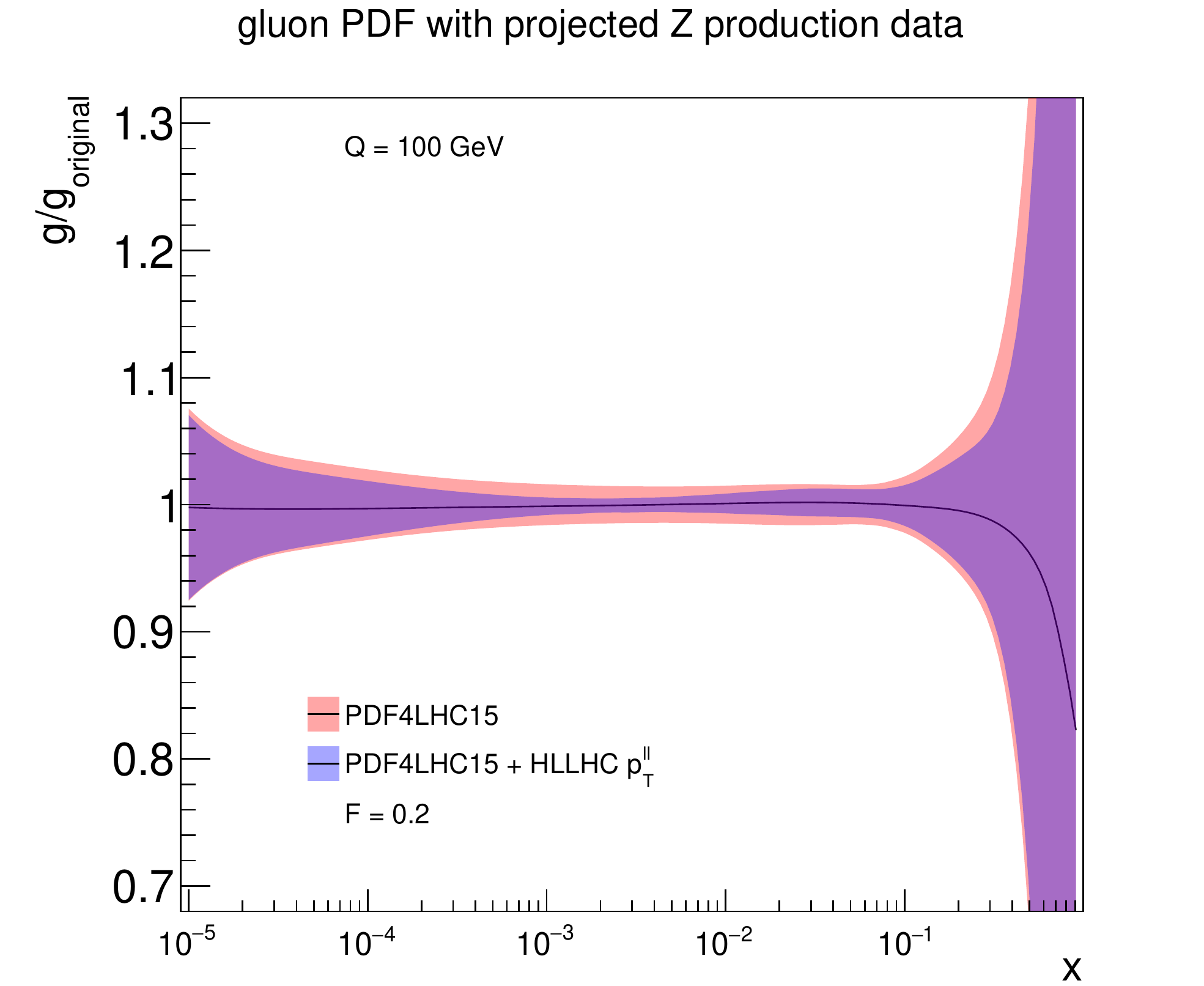}
\caption{\small
  Left: as in Fig.~\ref{fig:datatheory_DY}, now
  for the $p_T$ distribution of $Z$ bosons in the dilepton final state,
  in the on--peak bin defined by $66~{\rm GeV}\le m_{ll} \le 116~{\rm GeV}$.
  Right: as in Fig.~\ref{fig:PDFimpact_DY}, now for the gluon
  PDF after including the HL--LHC $Z$ transverse momentum pseudo--data
  in the fit.
  \label{fig:data_pdf_ZpT} }
  \end{center}
\end{figure}

The comparison between HL--LHC pseudo--data and theoretical predictions
in the on--peak bin defined by $66~{\rm GeV}\le m_{ll} \le 116~{\rm GeV}$
is shown in Fig.~\ref{fig:data_pdf_ZpT}, where we can see that coverage up
to $p_T^{ll}\simeq 3$ TeV is expected, similar as in the case of
direct photon production.
We find a moderate reduction in the PDF uncertainties once the HL--LHC
pseudo--data is added to the fit by means of Hessian profiling.
Concerning its effects on the gluon, we see that the $Z$ $p_T$ measurements
provide valuable information in the intermediate $x$ region
between $10^{-3}$ and $10^{-2}$ with a clear reduction of PDF uncertainties
even if in this region these were quite small to begin with.

%% file: sec-ultimatepdfs.tex
\section{Ultimate PDFs with HL--LHC pseudo--data}
\label{sec:ultimatepdfs}

In this section we  combine the complete set of HL--LHC pseudo--data
listed in Table~\ref{tab:PseudoData} to produce
the final profiled PDF sets, which
quantify
the impact of future HL--LHC measurements
on our knowledge of the quark and gluon structure of the proton.

In Table~\ref{tab:Scenarios} we list the
three scenarios for the systematic uncertainties of the HL--LHC pseudo--data
    that we assume in the present analysis.
    These scenarios, ranging from more conservative to more optimistic, differ among them in
    the reduction factor $f_{\rm red}$, Eq.~(\ref{eq:totalExpError}),
    applied to the systematic errors of the reference
    8 TeV or 13 TeV measurements.
    In particular, in the optimistic scenario we assume a reduction
    of the systematic errors by a factor 2.5 (5) as compared to the
    reference 8 TeV (13 TeV) measurements, while for the conservative scenario we assume no reduction in systematic
    errors with respect to 8 TeV reference.
    We also indicate in each case the name of the corresponding {\tt LHAPDF} grid.
    Reassuringly, as we show below, the qualitative results of our
    study depend only mildly in the specific assumption for
    the values of $f_{\rm red}$.

\begin{table}[h!]
  \centering
  \renewcommand{\arraystretch}{1.20}
  \begin{tabular}{c|c|c|c|c}
    Scenario    &   $f_{\rm red}$ (8 TeV)  & $f_{\rm red}$ (13 TeV) &   {\tt LHAPDF} set  &
    Comments \\
    \toprule
     A          &   1   &  0.5   & {\tt PDF4LHC\_nnlo\_hllhc\_scen1}  & Conservative \\
    \midrule
    B         &   0.7   &  0.36  & {\tt PDF4LHC\_nnlo\_hllhc\_scen2}  & Intermediate \\
     \midrule
      C          &   0.4   &  0.2  &  {\tt PDF4LHC\_nnlo\_hllhc\_scen3}  & Optimistic \\
 \bottomrule
  \end{tabular}
  \vspace{0.3cm}
  \caption{\small \label{tab:Scenarios}
    The three scenarios for the systematic uncertainties of the HL--LHC pseudo--data
    that we assume in the present study.
    These scenarios, ranging from conservative to optimistic, differ among them in
    the reduction factor $f_{\rm red}$, Eq.~(\ref{eq:totalExpError}),
    applied to the systematic errors of the reference
    8 TeV or 13 TeV measurements.
    We also indicate in each case the name of the corresponding {\tt LHAPDF} grid.
  }
\end{table}

In this section, we study how the HL--LHC pseudo--data constraints
the parton distributions and the PDF luminosities for
proton--proton collisions at $\sqrt{s}=14$ TeV.
Then we present an initial study with some representative
implications of the ultimate PDFs for LHC phenomenology.

\subsection{Parton distributions}
\label{sec:pdfcomparisons}

\begin{figure}[t]
  \begin{center}
\includegraphics[width=0.49\linewidth]{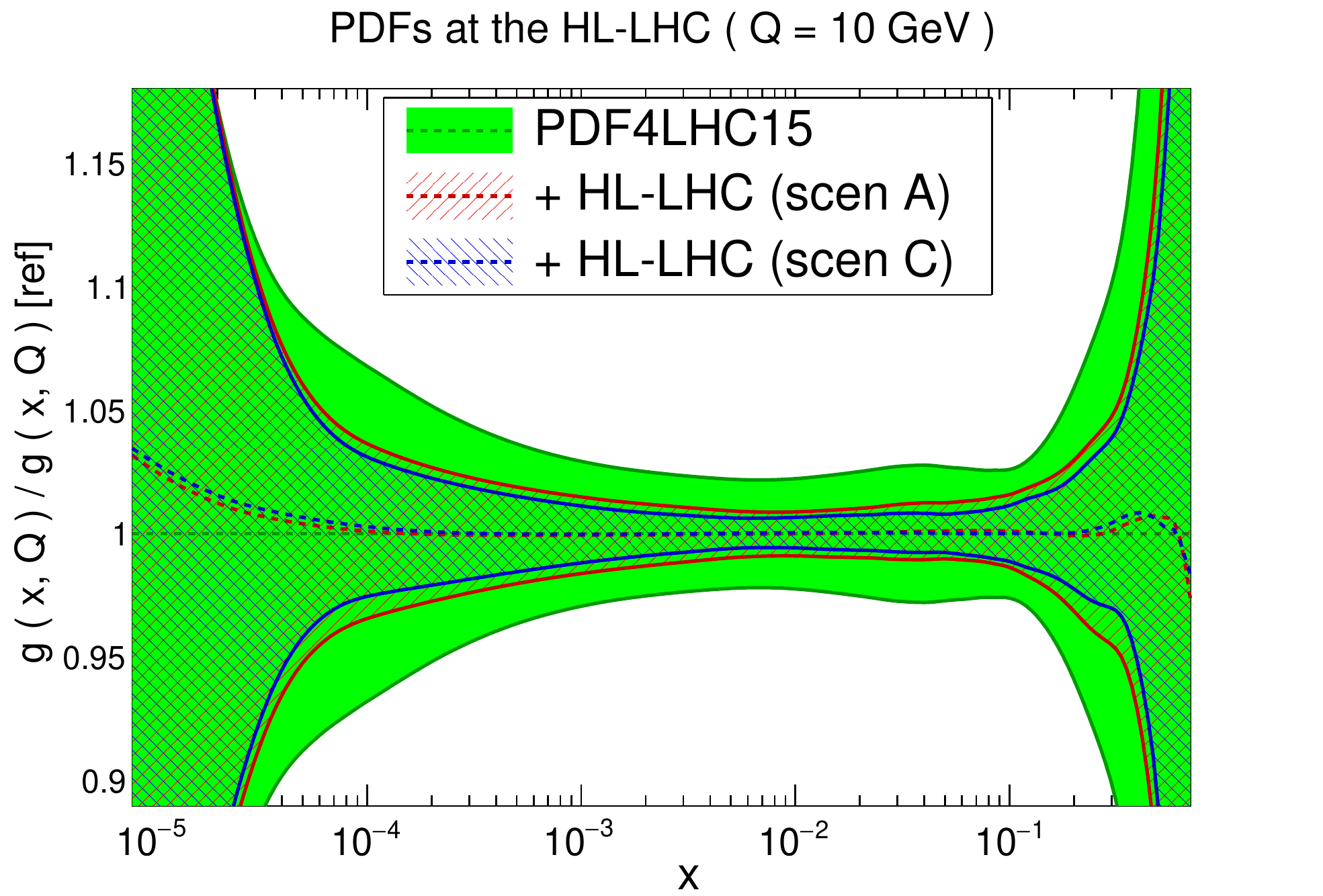}
\includegraphics[width=0.49\linewidth]{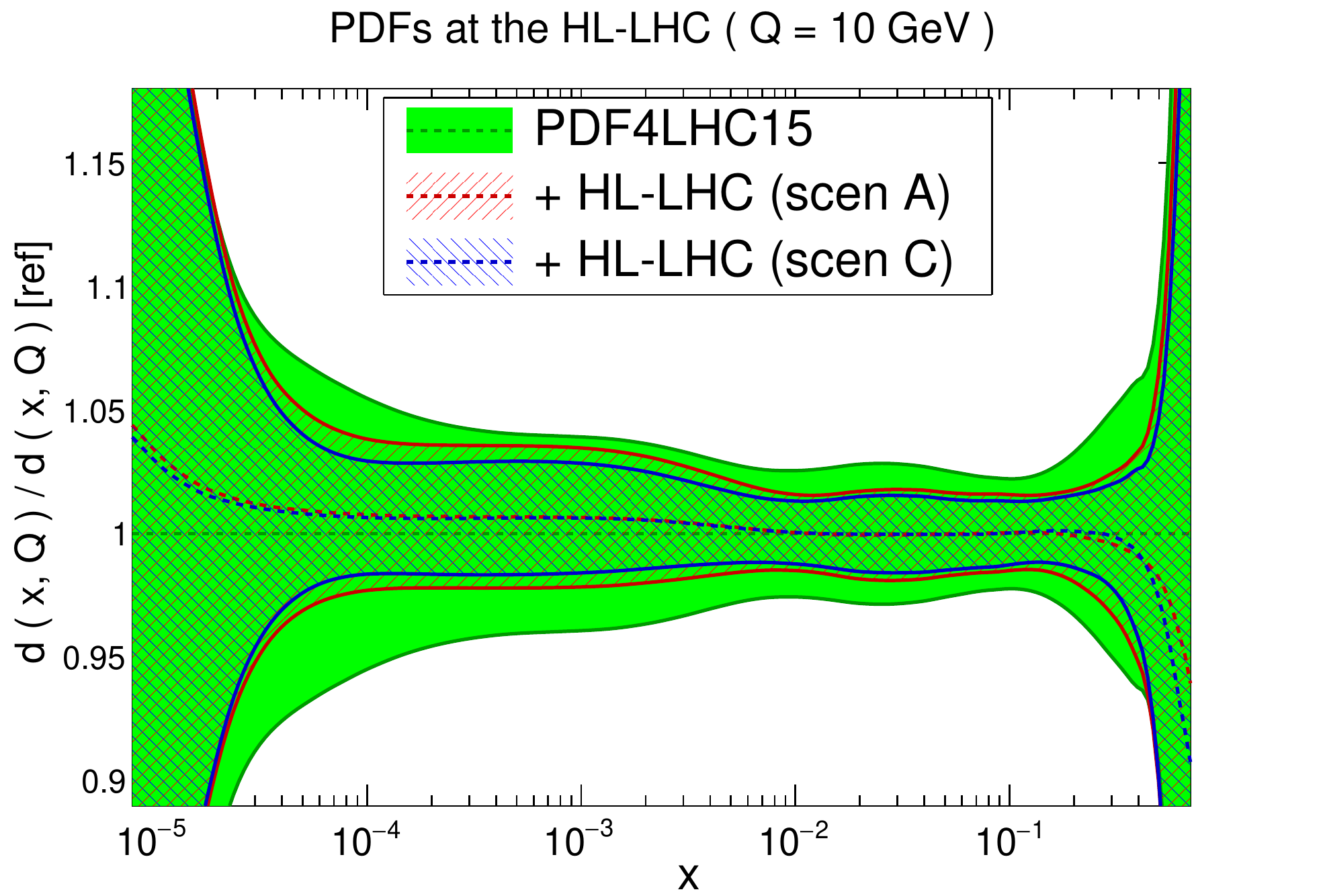}
\includegraphics[width=0.49\linewidth]{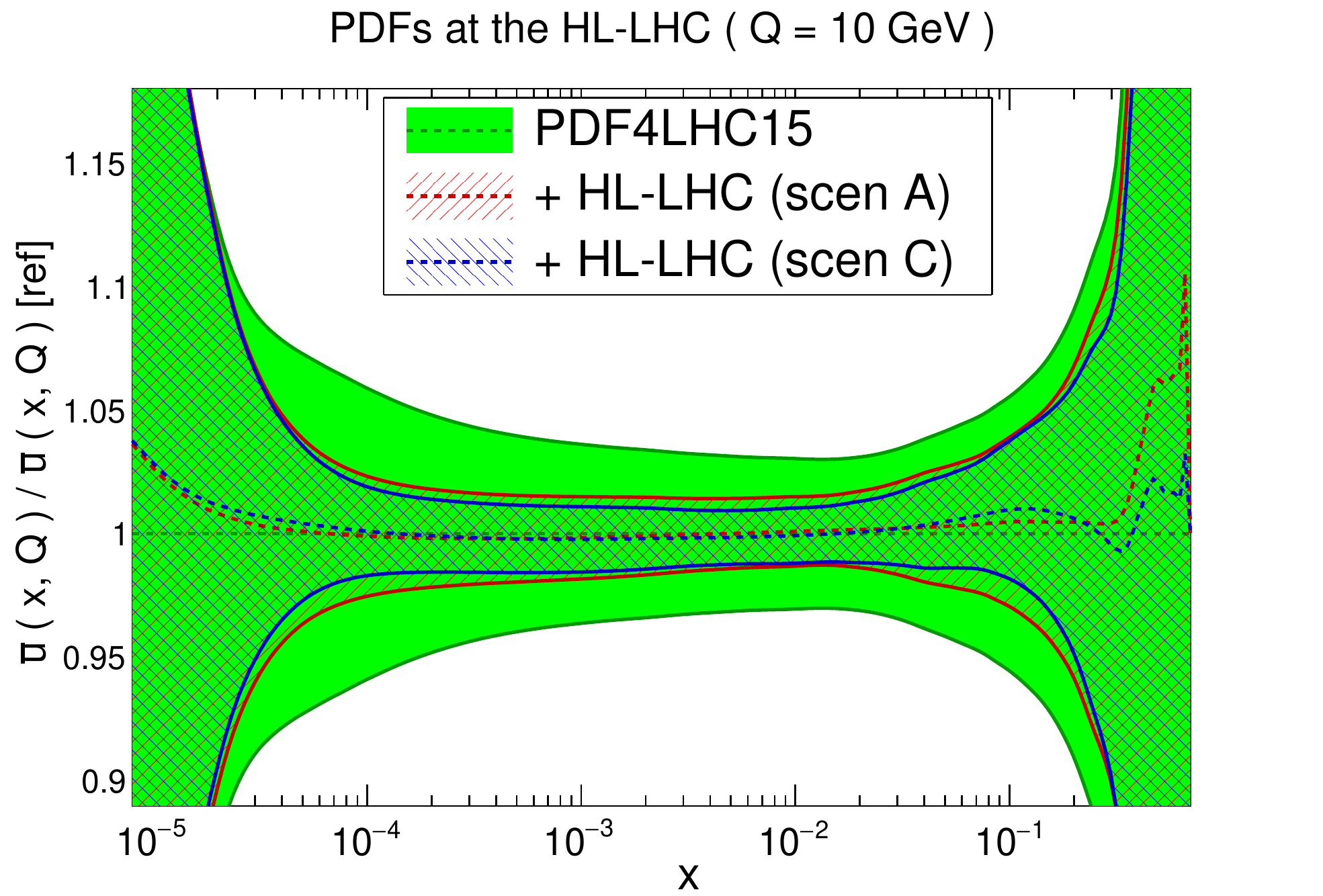}
\includegraphics[width=0.49\linewidth]{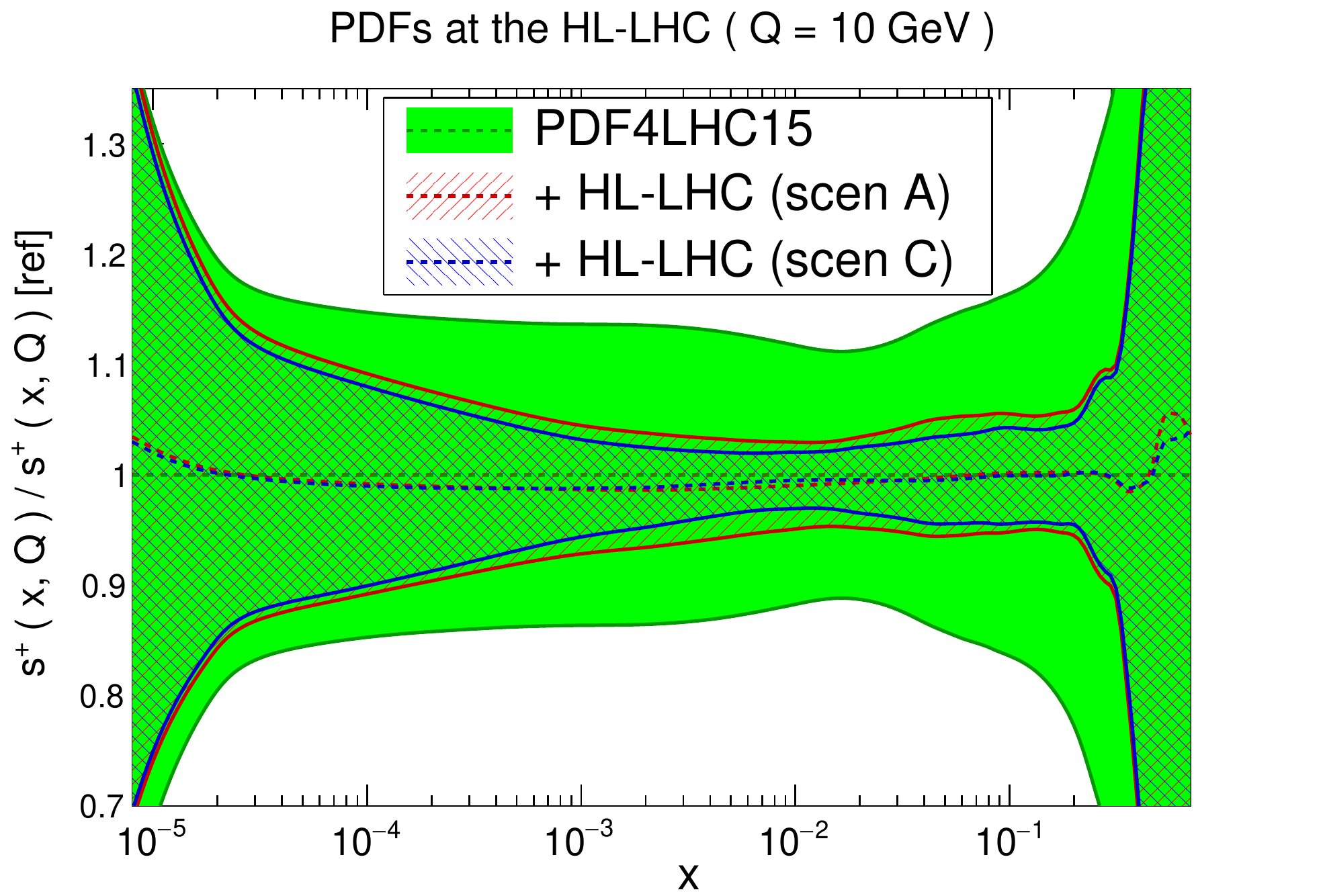}
\caption{\small Comparison of the PDF4LHC15 set with the 
  HL--LHC profiled set in scenarios A and C, defined
  in Table~\ref{tab:Scenarios}.
  We show the gluon, down quark, up anti--quark, and total
  strangeness at $Q=10$ GeV, normalized to the central value of
  the PDF4LHC15 baseline.
  The bands correspond to the one--sigma PDF uncertainties.
     \label{fig:PDFratios} }
  \end{center}
\end{figure}

In Fig.~\ref{fig:PDFratios} we present a comparison of
the baseline PDF4LHC15 set with the profiled sets
based on HL--LHC pseudo--data from scenarios A (conservative)
and C (optimistic) as defined
in Table~\ref{tab:Scenarios}.
  Specifically, we show the gluon, down quark, up anti--quark, and total
  strangeness at $Q=10$ GeV, normalized to the central value of
  the PDF4LHC15 baseline.
  In this comparison,
  the bands correspond to the one--sigma PDF uncertainties.

  First of all, we observe that the
  impact of the HL--LHC pseudo--data is reasonably similar
  in the conservative and optimistic scenarios.
This is not so surprising, as we have
  explicitly chosen those datasets which will benefit from a significant improvement in statistics, and these tend to lie in kinematic regions where the PDFs themselves are generally less
  well determined, see the discussion in Sect.~\ref{sec:pseudodata}.
  Therefore, the dominant reason for the observed reduction
  of PDF uncertainties
  is the increased statistics and the corresponding extended kinematic reach
  that becomes available at the HL--LHC, rather than the specific assumptions about the systematic uncertainties.
  This demonstrates that our results are robust against the
  details of the projections
  of how the experimental systematic
  uncertainties will be reduced in the HL--LHC era.

  From Fig.~\ref{fig:PDFratios} we observe
  a marked reduction of the PDF uncertainties in all cases.
  This 
  is particularly significant for the gluon and the sea quarks,
  for the reason that these are currently affected by larger
  uncertainties than in the case of the valence quarks.
  In the case of the gluon PDF, there is an improvement of
  uncertainties across a very broad range of $x$.
  This is a direct consequence of the fact that
  we have included several HL--LHC processes
  that have direct sensitivity to the gluon
  content of the proton, namely jet, direct photon, and top quark pair
  production, as well as the transverse momentum of $Z$ bosons.

  Another striking feature of Fig.~\ref{fig:PDFratios} concerns
  the strange PDF.
  In this case, the PDF uncertainties are reduced by almost
  a factor 4, from around 15\% to a few percent, in a wide
  region of $x$.
  This result highlights the importance of the $W$+charm measurements
  at the HL--LHC, specially those in the forward region by LHCb,
  see Fig.~\ref{fig:correlations_Wc}, which represent a unique
  handle on the poorly known strange content of the proton.
  In turn, such an improved understanding of the strange PDF
  will feed into a reduction of theory uncertainties in crucial
  HL--LHC measurements such as those of $M_W$ or $\sin^2\theta_W$.

\subsection{Partonic luminosities}
\label{sec:luminosities}

\begin{figure}[t]
  \begin{center}
\includegraphics[width=0.49\linewidth]{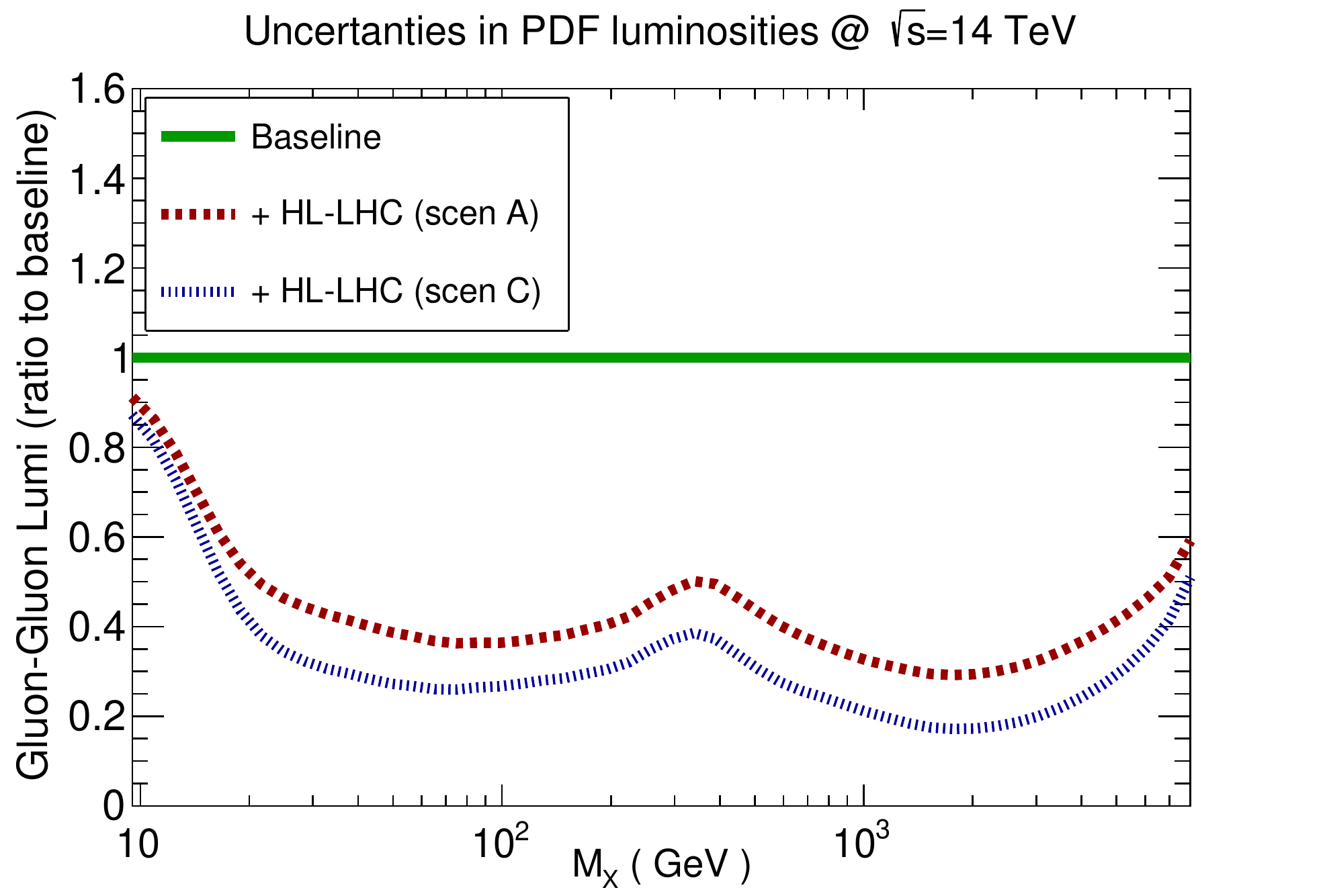}
\includegraphics[width=0.49\linewidth]{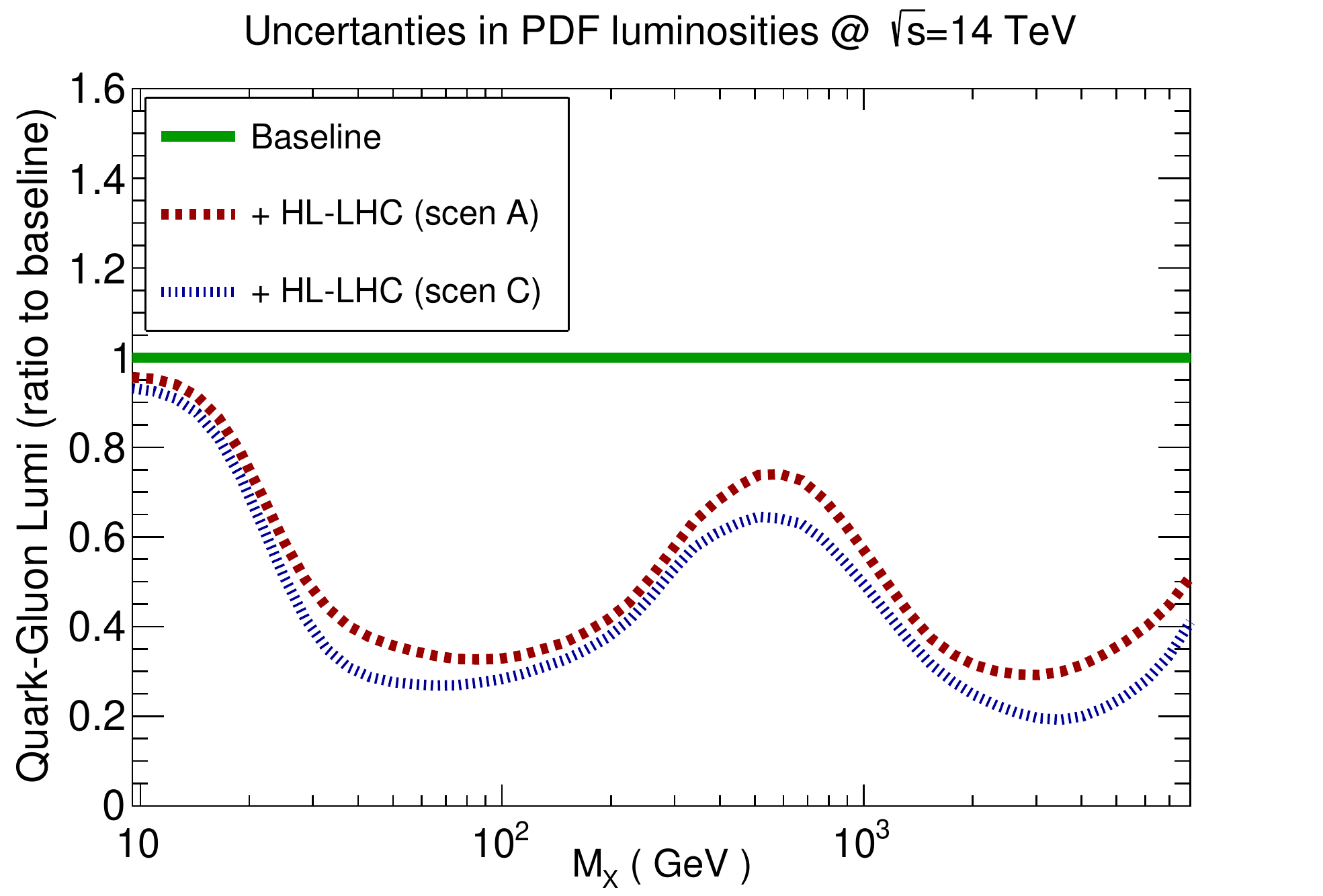}
\includegraphics[width=0.49\linewidth]{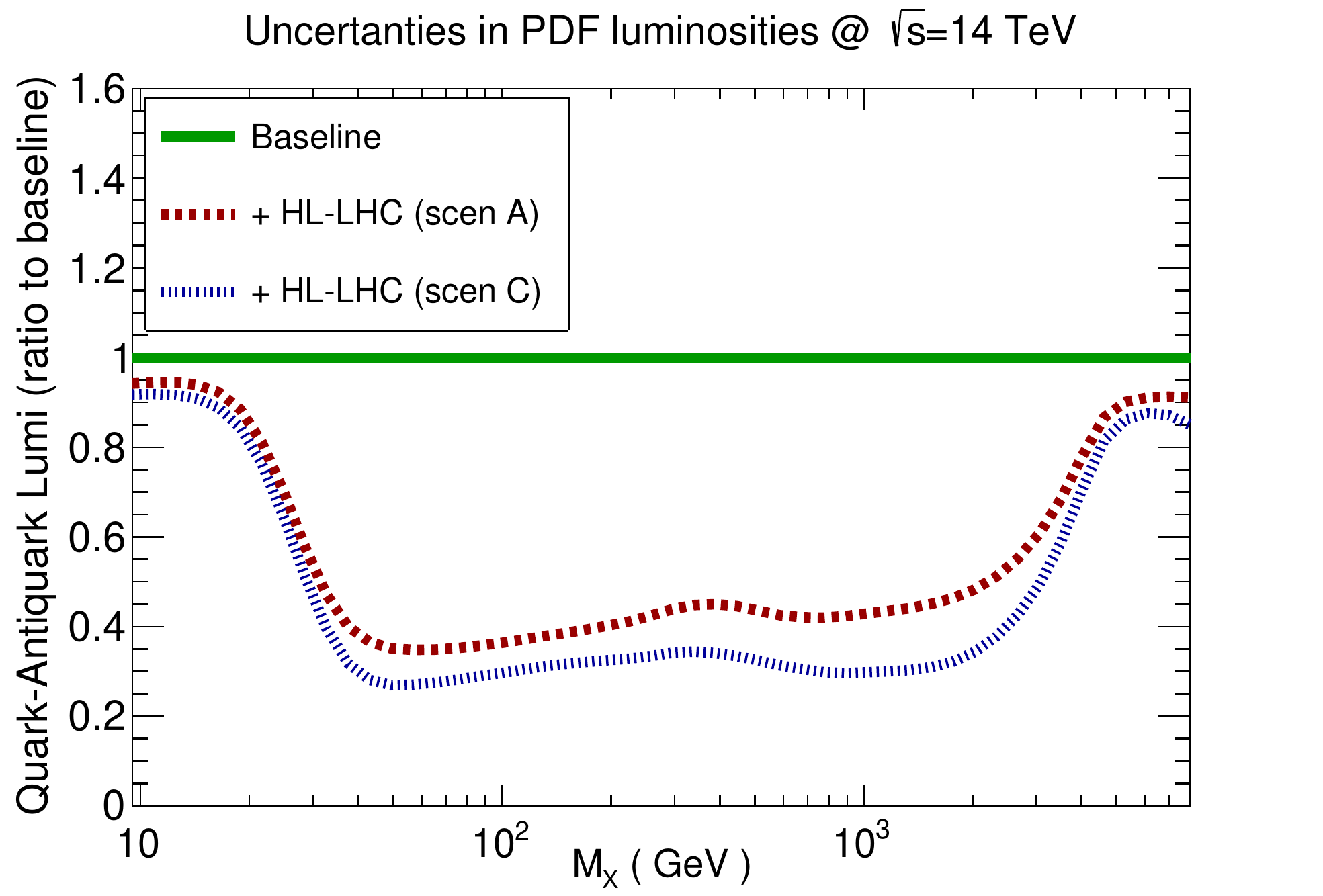}
\includegraphics[width=0.49\linewidth]{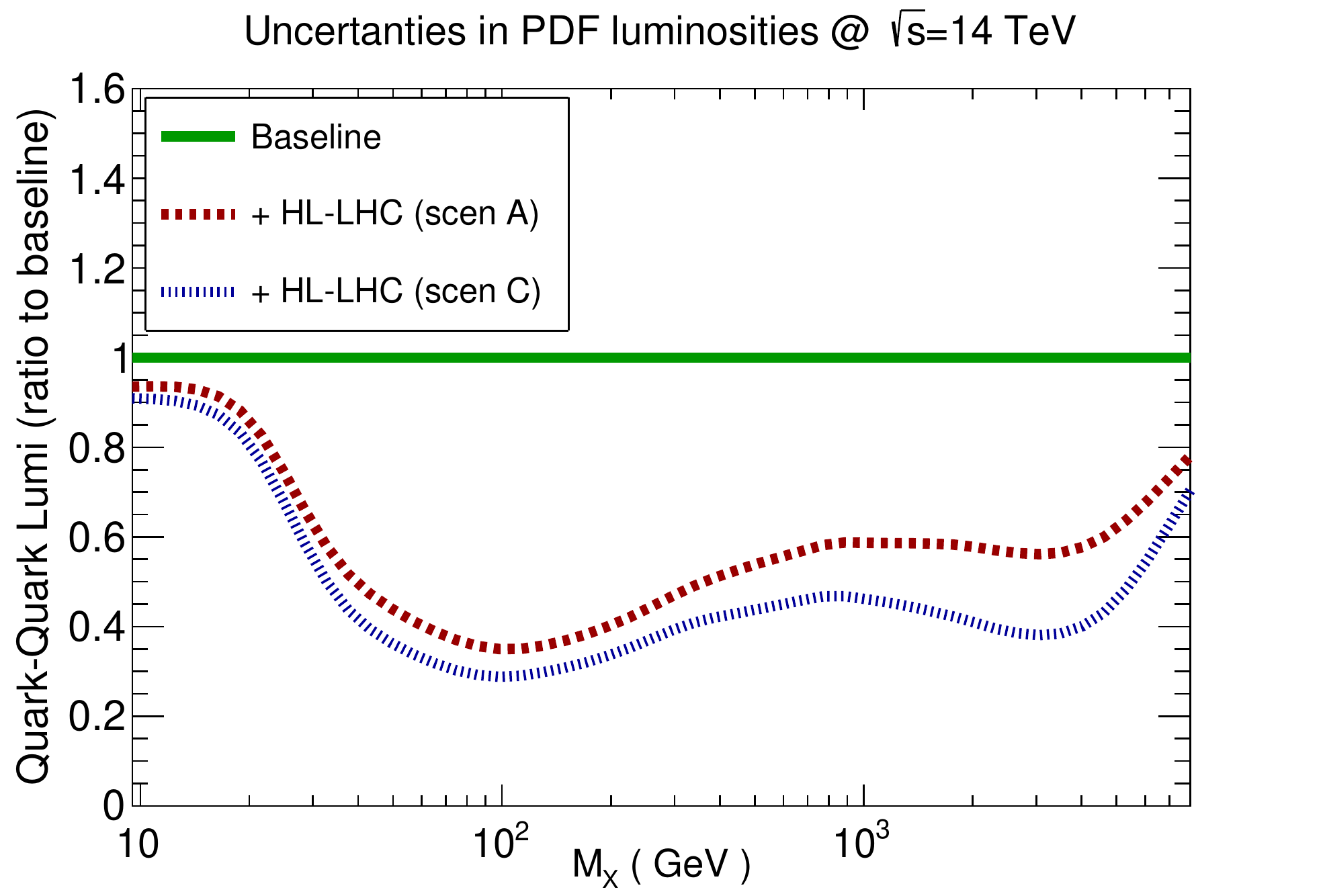}
\includegraphics[width=0.49\linewidth]{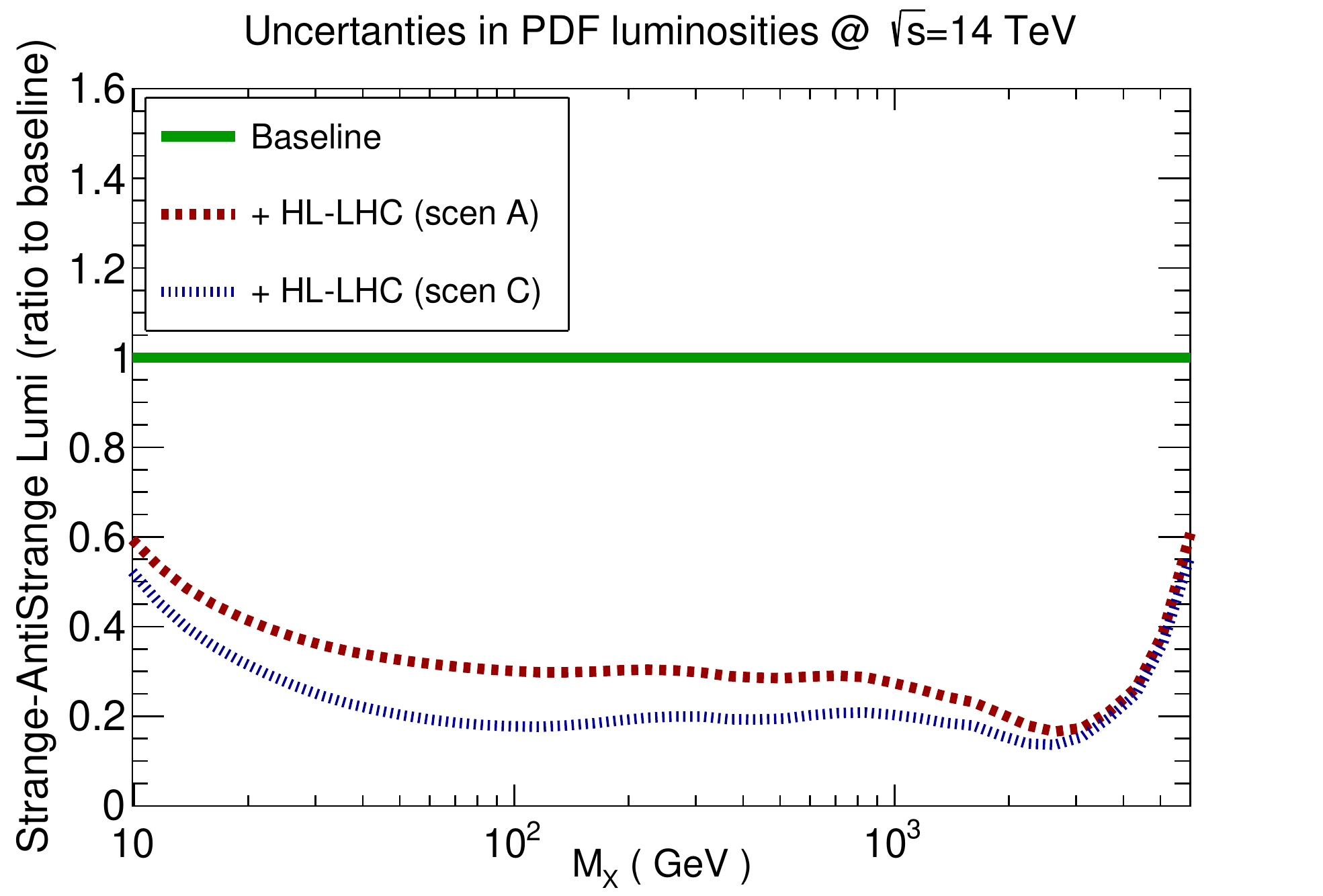}
\includegraphics[width=0.49\linewidth]{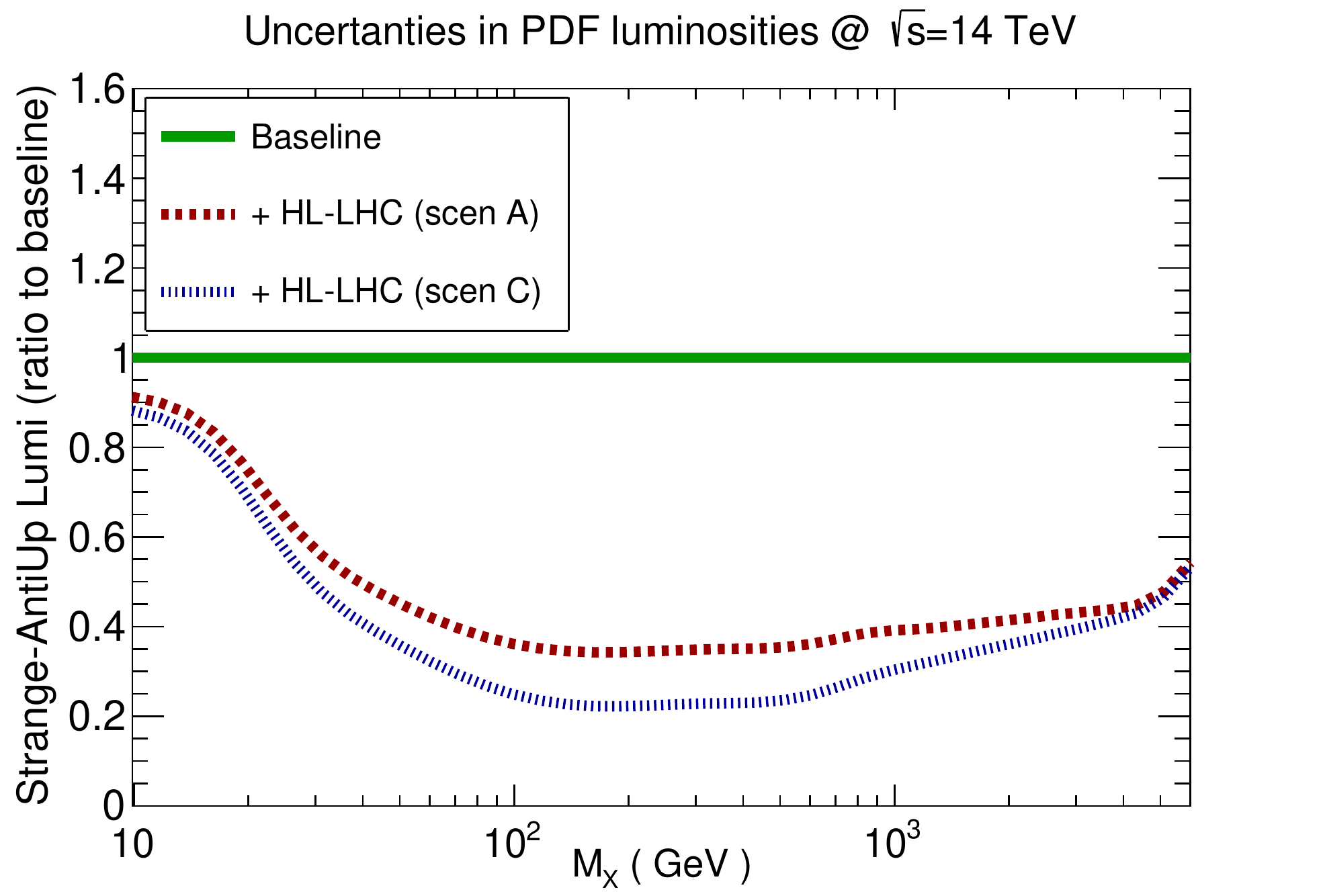}
\caption{\small The reduction of the uncertainties in the PDF luminosities
 at $\sqrt{s}=14$ TeV once the HL--LHC pseudo--data has been included,
 compared to the PDF4LHC15 baseline.
 We show the $gg$, $qg$, $q\bar{q}$,
$qq$, $s\bar{s}$, and $s\bar{u}$ luminosities for the conservative (A)
and optimistic (C) scenarios.
The average values for the PDF uncertainty reduction in different bins
of $M_X$ is also reported in Fig.~\ref{fig:PDFs-HL--LHC-summaryTable}.
     \label{fig:PDFluminosities} }
  \end{center}
\end{figure}

Next we take a look at the partonic luminosities, to
quantify the improvement in the PDF uncertainties in different
initial--state partonic combinations from the HL--LHC pseudo--data.
In Fig.~\ref{fig:PDFluminosities} we show the reduction of PDF uncertainties
in the $gg$, $qg$, $q\bar{q}$,
  and $qq$, $s\bar{s}$, and $s\bar{u}$
  luminosities at $\sqrt{s}=14$ TeV that can be expected
  as a consequence of adding the HL--LHC pseudo--data
on top of
 the PDF4LHC15 baseline.
 Note that a value of 1 in these plots corresponds
 to no uncertainty reduction.
  As in the case of the PDF comparisons, results are shown
  both for the conservative (A) and optimistic (C) scenarios
  for our projections of the experimental systematic
  uncertainties.

In addition, in Table~\ref{fig:PDFs-HL--LHC-summaryTable} we also report the
 average values of these PDF uncertainty reductions for three different invariant mass bins.
 In particular, we consider
low invariant masses, $10~{\rm GeV}\le M_X\le 40~{\rm GeV}$, relevant for
instance for Monte Carlo tuning and QCD studies; intermediate masses,
$40~{\rm GeV}\le M_X\le 1~{\rm TeV}$, relevant for electroweak, top,
and Higgs measurements; and large invariance masses, $1~{\rm TeV}\le M_X\le 6~{\rm TeV}$,
relevant for searches of new heavy particles.
These averages are
  computed from 10 points per mass bin, logarithmically spaced in $M_X$.
  In Table~\ref{fig:PDFs-HL--LHC-summaryTable},
  the values shown outside (inside) the brackets correspond to the
  optimistic (conservative) scenario.

\begin{table}[t]
  \begin{center}
  \small
   \renewcommand{\arraystretch}{1.70}
\begin{tabular}{c|c|c|c}
\toprule
Ratio to baseline  & $10~{\rm GeV}\le M_X\le 40~{\rm GeV}$   &
 $40~{\rm GeV}\le M_X\le 1~{\rm TeV}$&
  $1~{\rm TeV}\le M_X\le 6~{\rm TeV}$\\
\midrule
gluon--gluon & 0.50 (0.60)  & 0.28 (0.40)  & 0.22 (0.34)     \\
gluon--quark & 0.66 (0.72) & 0.42 (0.45) &   0.28 (0.37)     \\
quark--quark &  0.74 (0.79) & 0.37 (0.46) &  0.43 (0.59)     \\
quark--antiquark & 0.71 (0.76) & 0.31 (0.40)  &  0.50 (0.60)     \\
\midrule
strange--antistrange & 0.34 (0.44)  & 0.19 (0.30)  &  0.23 (0.27)    \\
strange--antiup & 0.67 (0.73)  & 0.27 (0.38)  & 0.38 (0.43)     \\
\bottomrule
\end{tabular}
\vspace{+0.4cm}
\caption{\small The reduction of the PDF uncertainties
 compared to the PDF4LHC15 baseline for
different initial partonic combinations (that is, a value
of 1 corresponds to no reduction at all).
  Results are presented for three different bins
of the invariant mass $M_X$ of the produced system,
averaging over 10 points logarithmically spaced within each bin.
  The values shown outside (inside) the brackets correspond to the
  optimistic (conservative) scenario.
  The corresponding results differential in $M_X$ are
  presented in Fig.~\ref{fig:PDFluminosities}.
     \label{fig:PDFs-HL--LHC-summaryTable}
 }
  \end{center}
\end{table}

From the comparisons in Fig.~\ref{fig:PDFluminosities} and
Table~\ref{fig:PDFs-HL--LHC-summaryTable}, we
observe again that the reduction in the uncertainties
of the PDF luminosities is rather robust with respect to
the assumed projections
for the experimental systematic uncertainties.
For instance, for intermediate values of the final--state
invariant mass, $40~{\rm GeV}\le M_X\le 1~{\rm TeV}$,
we find that for all the partonic initial states the reduction
factor varies between 0.28 and 0.40 (0.42 and 0.45, 0.31 and 0.40)
in the optimistic and conservative scenario for the gluon--gluon
(gluon--quark, quark--antiquark) luminosities.
These results again reinforce our conclusion that the results
of this study are only mildly sensitive to the details
of the projected pseudo--data.

We find that in the intermediate $M_X$ bin
the reduction of PDF uncertainties ranges approximately between a factor
2 and a factor 5,
depending on the specific
partonic channel and the scenario for the systematic errors.
For example, for the gluon--gluon
luminosity in the range relevant for Higgs production in gluon fusion,
one finds a reduction by almost
a factor 4 in the optimistic scenario.
The improvement in the strange--initiated processes is also remarkable,
for example the PDF uncertainties  in the $s\bar{s}$ luminosity
are expected to be reduced by a factor 5 (3) in the optimistic
(conservative) scenario.
Recall that strange--initiated processes are important
for a variety of LHC analysis, from measurements of
$M_W$ and $\sin^2\theta_W$ to searches for BSM $W'$ bosons.
We also find that the uncertainties in
quark--antiquark luminosities, relevant for
example for precision electroweak measurements, are expected
to be reduced by up to a factor 3 in this invariant mass range.

Similar improvements in the
PDF luminosities
are found in the high mass region, $M_X\ge 1$ TeV, directly
relevant for BSM searches.
For instance, in
the optimistic scenario, the PDF error reduction at higher masses is expected to be as large as a factor 5
for the gluon--gluon luminosity.
Again this is a consequence of the inclusion in the profiling
of gluon--dominated processes such as $t\bar{t}$ and inclusive jets
that at the HL--LHC, which cover the region up to 6 TeV,
see Fig.~\ref{fig:kinHLLHC}.
The impact of the HL--LHC pseudo--data is less marked
for the quark--quark and quark--antiquark luminosities
in this high--mass region, due to the fact that of the data points
included in the profiling only a fraction of them
are both quark--initiated and cover the large--$x$ region.

It is worth emphasizing again here that
the list of processes studied in this work
and summarised in Table~\ref{fig:PDFs-HL--LHC-summaryTable} are just a subset of
those HL--LHC measurements with PDF--constraining potential.
Therefore, it is conceivable that the actual reduction of PDF errors
presented in Table~\ref{fig:PDFs-HL--LHC-summaryTable} would actually
be more significant than our estimates here.

\subsection{Implications for HL--LHC phenomenology}
\label{sec:pheno}

We now turn to present some representative results
of the phenomenological implications that these ``ultimate'' PDFs
will have at the HL--LHC, both for processes within the SM and beyond it.
It is beyond the scope of this work to carry out a comprehensive
phenomenological study, and we refer the reader to the upcoming
Yellow Report~\cite{yellowreport} describing the physics case
of the HL--LHC, where more detailed projections and analyses will be presented.

Let us begin by assessing the PDF impact of HL--LHC measurements
on representative Standard Model processes.
In particular, we consider diphoton production, dijet production,
and Higgs production in gluon fusion, both inclusive and in association
with a hard jet.
In the following all cross sections have been computed at $\sqrt{s}=14$ TeV
using leading order (LO) matrix elements
with {\tt MCFMv8.2}~\cite{Boughezal:2016wmq} and
applying the standard
ATLAS/CMS central acceptance cuts.
Since the comparison is restricted to ratios of cross sections, the LO
calculation is sufficient to illustrate the impact of the improvement
in the PDF uncertainties in each of these processes.
Indeed, we are only interested here in
illustrating the relative impact of the PDF error reduction, rather
than providing state--of--the--art predictions for the rates, which
will be presented elsewhere in the Yellow Report~\cite{yellowreport}.

\begin{figure}[t]
  \begin{center}
    \includegraphics[width=0.49\linewidth]{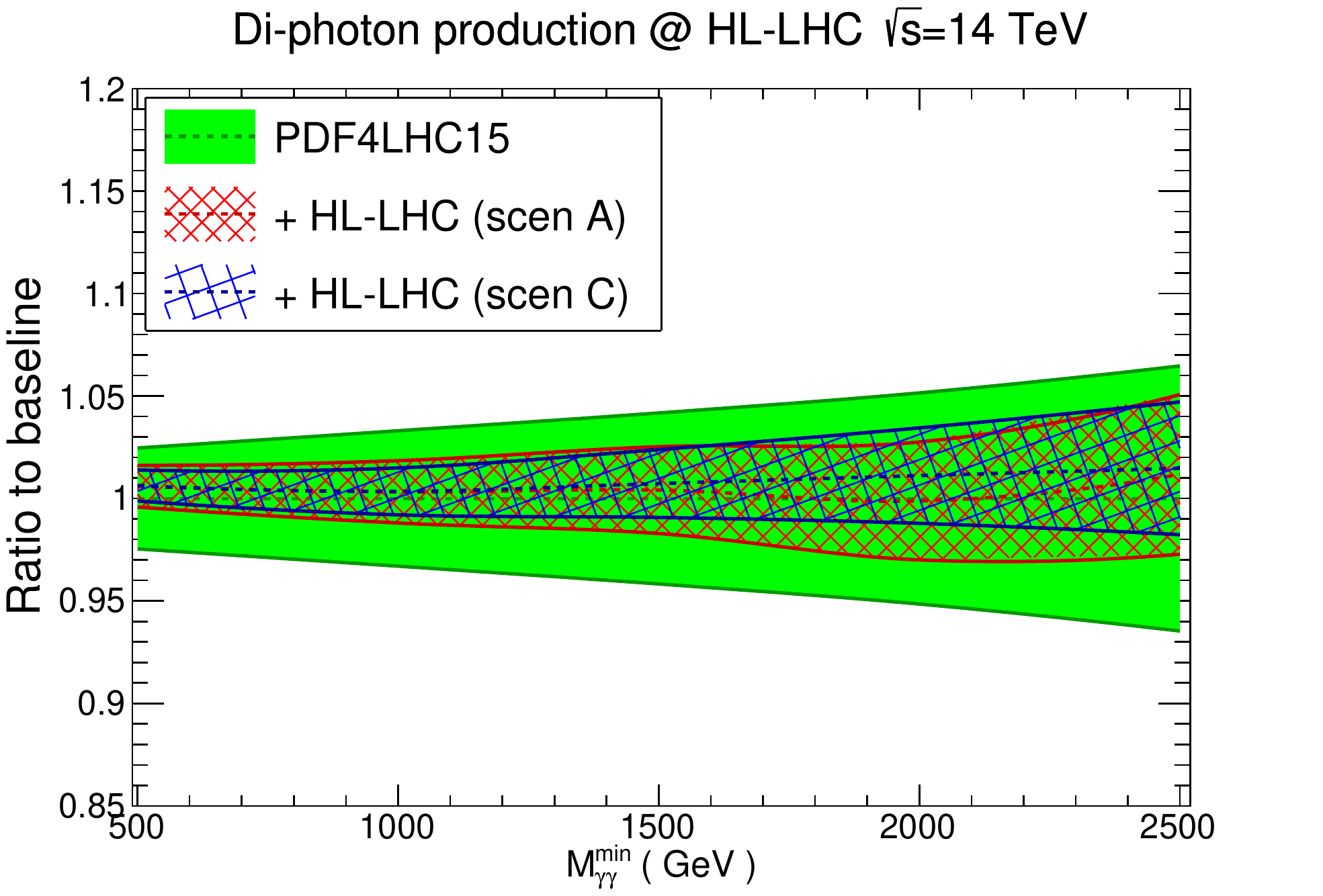}
    \includegraphics[width=0.49\linewidth]{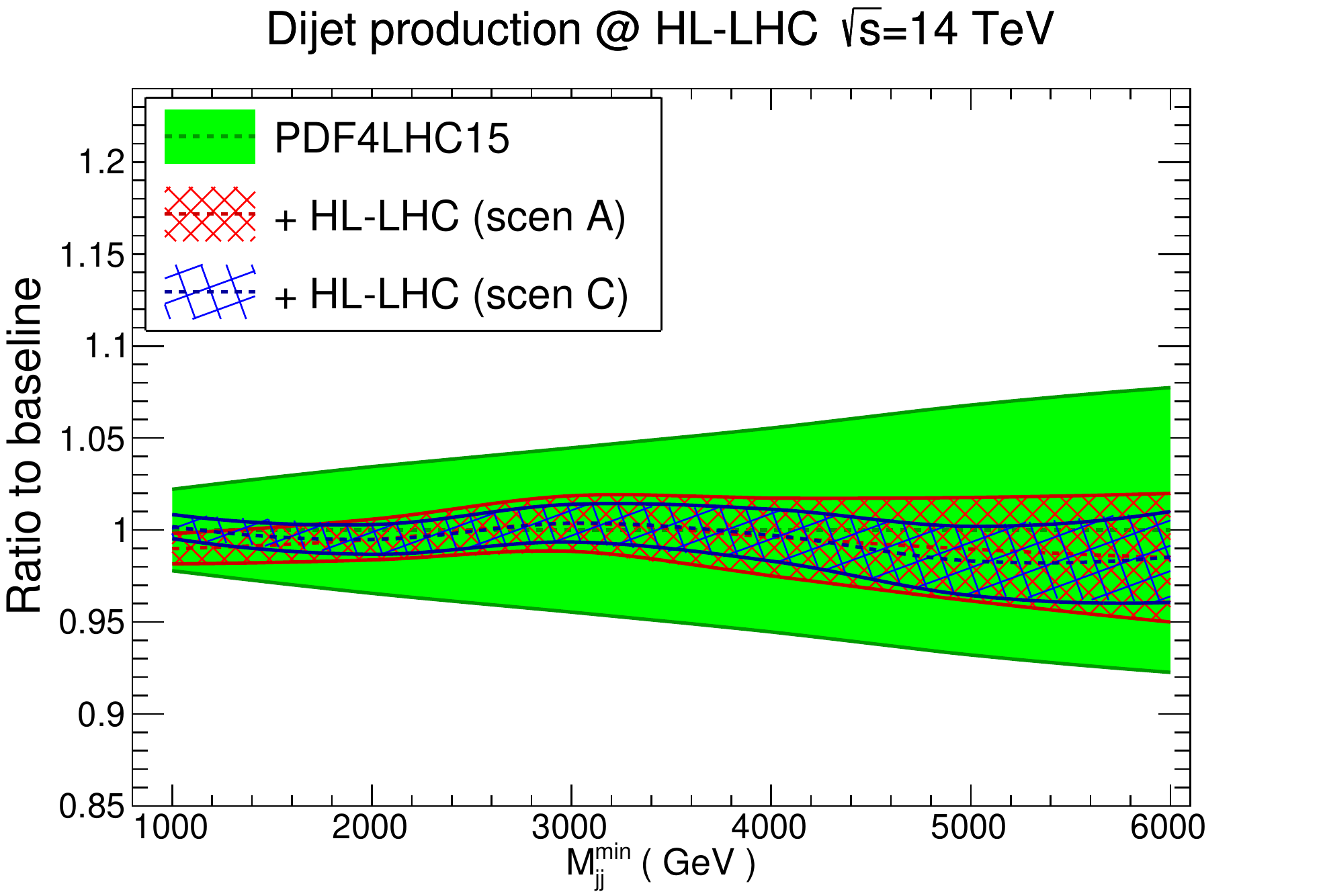}
    \includegraphics[width=0.49\linewidth]{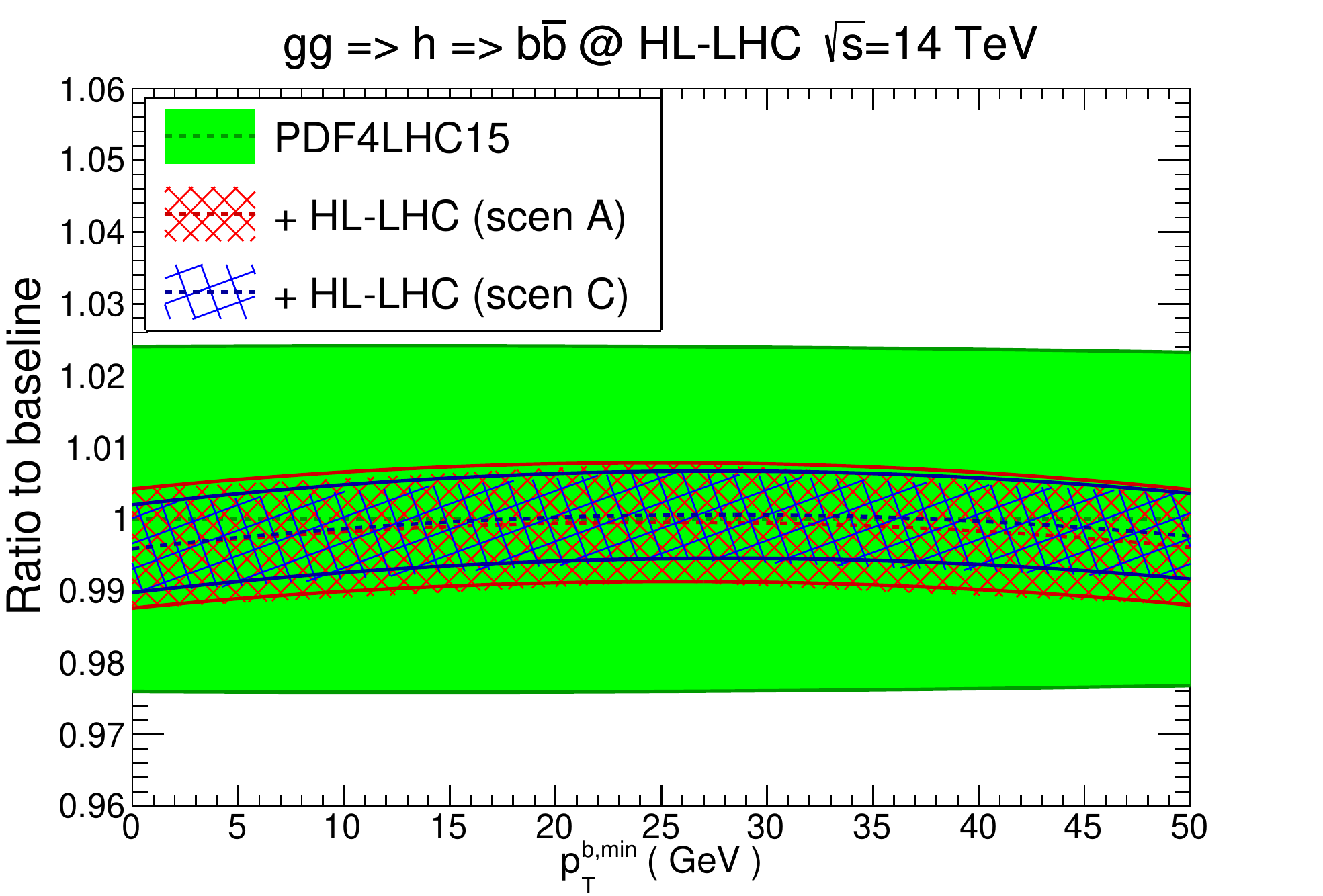}
    \includegraphics[width=0.49\linewidth]{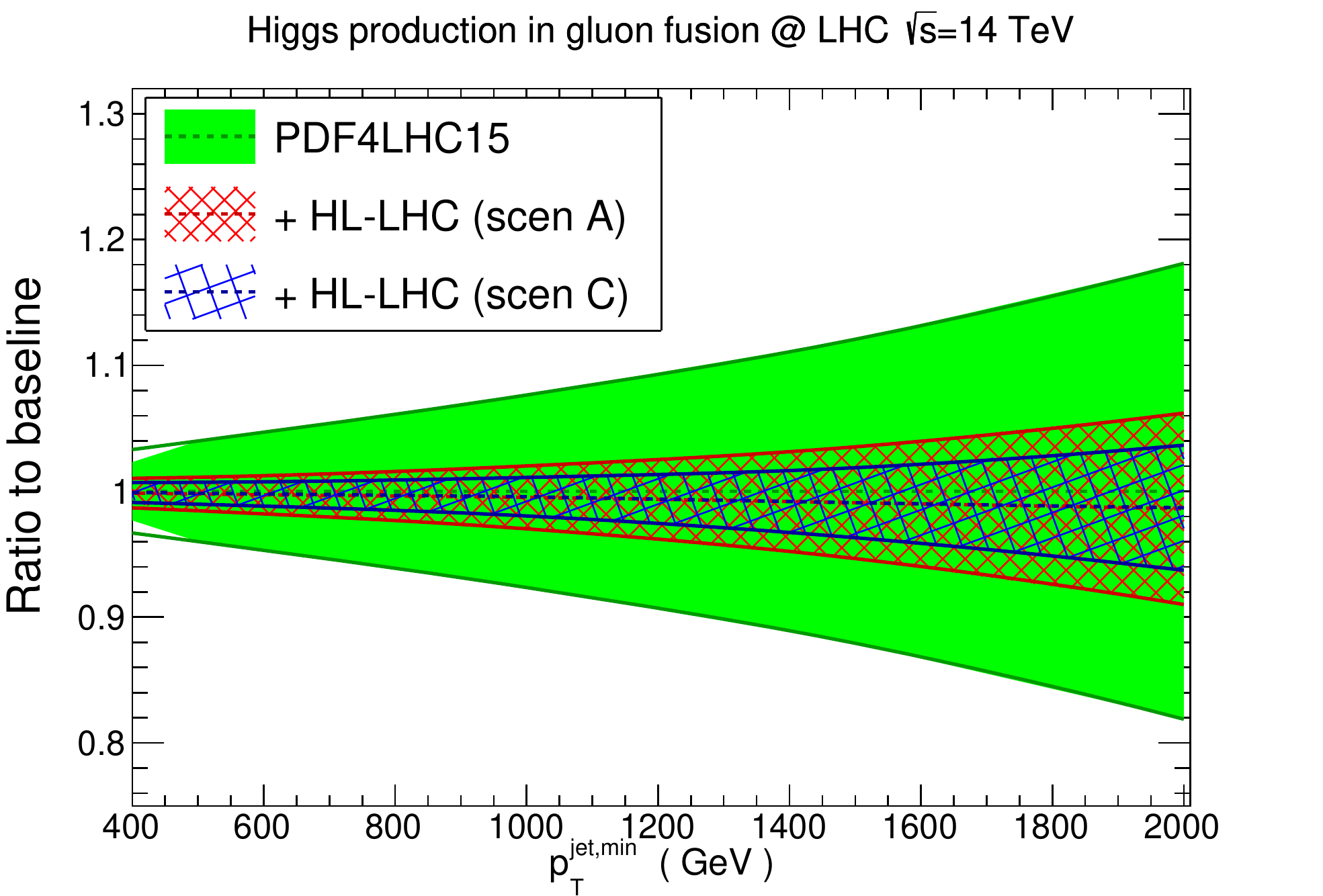}
    \caption{\small Comparison of the predictions for representative
    SM cross sections at $\sqrt{s}=14$ TeV between the PDF4LHC15
baseline and the HL--LHC profiled sets in
the conservative (A) and optimistic (C) scenarios.
Results are shown normalised to the
central value of PDF4LHC15.
      The upper plots show the diphoton (left) and dijet (right plot) production
      cross sections as a function of
      the minimum invariant masses of the final state, $M_{\gamma\gamma}^{\rm min}$
      and $M_{jj}^{\rm min}$ respectively.
      The bottom plots show Higgs boson production in gluon fusion with heavy top
      quark effective theory, both inclusive
      and decaying into $b\bar{b}$ as a function of $p_T^{b,\rm min}$ (left), and
      then in association with a hard jet as a function its transverse
      momentum $p_{\rm T}^{jet,\rm min}$ (right plot).
      The calculations have been performed using {\tt MCFM8.2} with leading--order
      matrix elements.
     \label{fig:MCFMxsects} }
  \end{center}
\end{figure}

First of all, we show the production cross sections of pairs of
photons (left) and of jets (right)
in the upper panels of Fig.~\ref{fig:MCFMxsects}.
We compare the PDF4LHC15
baseline with the HL--LHC profiled PDF sets in
the conservative (A) and optimistic (C) scenarios of Table~\ref{tab:Scenarios},
normalised  to the central value of PDF4LHC15.
In the considered kinematic regions, these two processes are mostly sensitive to the quark--antiquark initial state, 
and to the quark--gluon and quark--(anti)quark initial states, respectively.
The cross sections are presented  as a function of
      the minimum invariant mass of the final state, $M_{\gamma\gamma}^{\rm min}$
      and $M_{jj}^{\rm min}$ respectively, in order to facilitate their
      comparison with the corresponding PDF luminosities shown in
      Fig.~\ref{fig:PDFluminosities}.
     
      From this comparison, we find again
that both the optimistic and conservative scenarios lead to similar results
in terms of the expected reduction of the
PDF uncertainties in the entire kinematical range accessible
at the HL--LHC.
In the case of dijet production, we find that PDF uncertainties could
be reduced down to $\simeq$2$\%$ even for invariant masses as large
as $M_{jj}=6$ TeV.
The resulting improved theory predictions for dijet production
should also become relevant at the HL--LHC 
for BSM searches, {\it e.g.} for quark compositeness~\cite{Gao:2012qpa,Khachatryan:2014cja}.
Note that since the initial partonic states are the same, similar improvements are
expected
for inclusive jet production, see also Fig.~\ref{fig:datath_jetsphotons}, as well
as for multijet production.
Similar considerations apply for diphoton production, where the expected
PDF error reduction is a bit less marked since it is driven by the quark--antiquark
luminosity, see also the comparisons in Table~\ref{fig:PDFs-HL--LHC-summaryTable}.

In the two lower plots of Fig.~\ref{fig:MCFMxsects}, we present the corresponding
comparisons for the case of Higgs boson production via gluon fusion, using heavy
top quark effective theory.
In the case of inclusive production with decay into bottom quarks (left plot),
we find that the constraints from HL--LHC measurements are expected
to reduce PDF uncertainties down to the $1\%$ level.
Needless to say, this will directly benefit the characterisation of the Higgs
sector at the HL--LHC, where a few percent is the typical uncertainty target
for the determination of its couplings.
In the case of Higgs boson production in association with a hard jet (right plot), also there
we find a marked error reduction, indicating that PDF uncertainties
in the Higgs transverse momentum distribution could be reduced down to the $\simeq$2\% level
 in the entire  kinematical range relevant at the HL--LHC.
We recall that the large Higgs transverse momentum region is sensitive to
new heavy particles running in the loops as well as to BSM effects such as partial
Higgs compositeness~\cite{Grojean:2013nya}.

As we have discussed above in Sects.~\ref{sec:pdfcomparisons} and~\ref{sec:luminosities},
the impact of the HL--LHC pseudo--data is also significant in the large--$x$ region,
which in turn corresponds to large invariant masses for the PDF luminosities.
This is of course an important region for the searches of BSM heavy particles, where
PDF uncertainties often represent the dominant source of theoretical uncertainty.
With this motivation,
to illustrate the benefits that HL--LHC measurements will provide for BSM
searches we consider here high--mass supersymmetric (SUSY) particle production at
$\sqrt{s}=14$ TeV, where the HL--LHC reach extends to
sparticles masses up to around $M\simeq 3$ TeV.
While we use  SUSY production as a benchmark process, our results also apply
to the production of other heavy particles predicted in different BSM scenarios.

\begin{figure}[t]
  \begin{center}
    \includegraphics[width=0.49\linewidth]{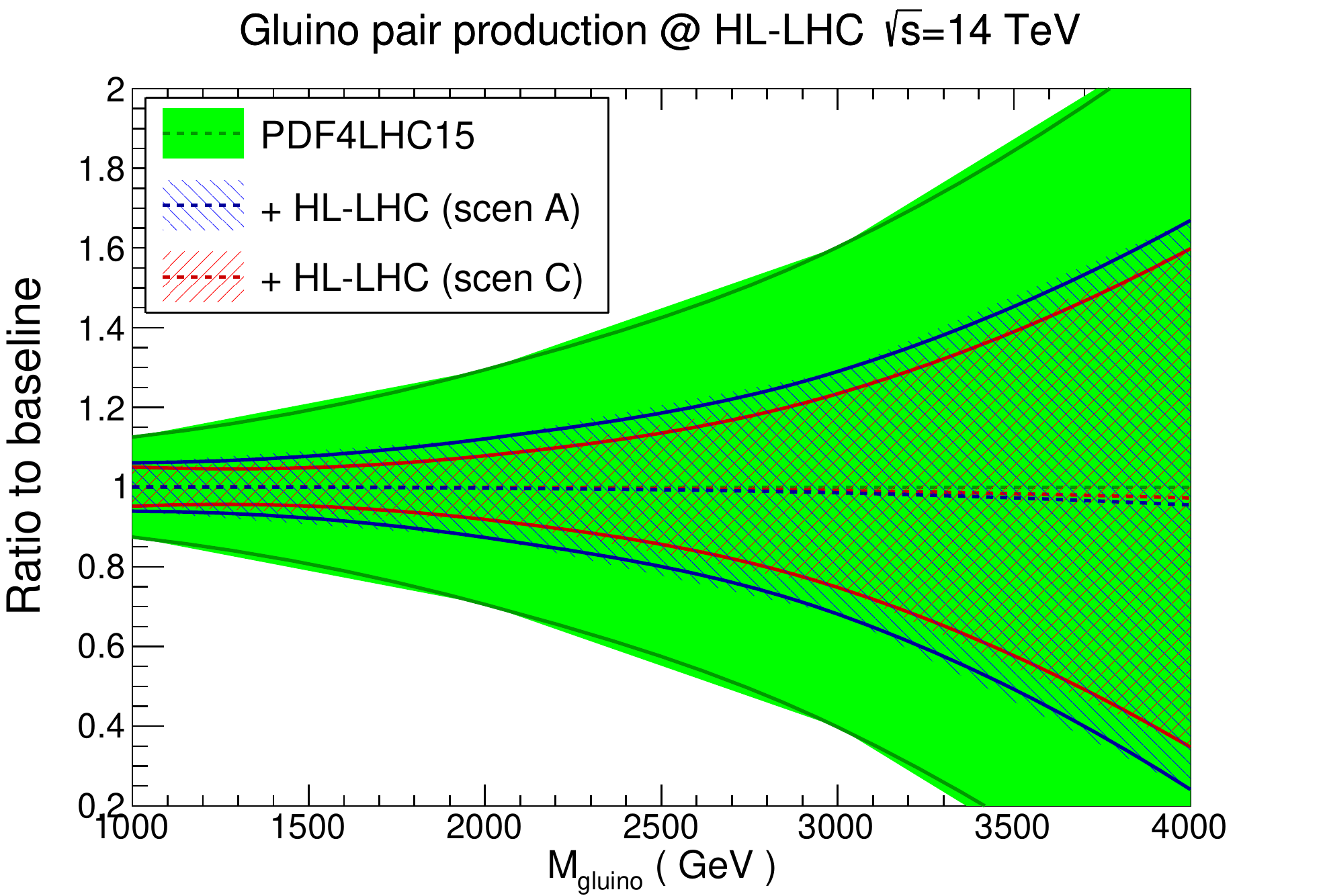}
    \includegraphics[width=0.49\linewidth]{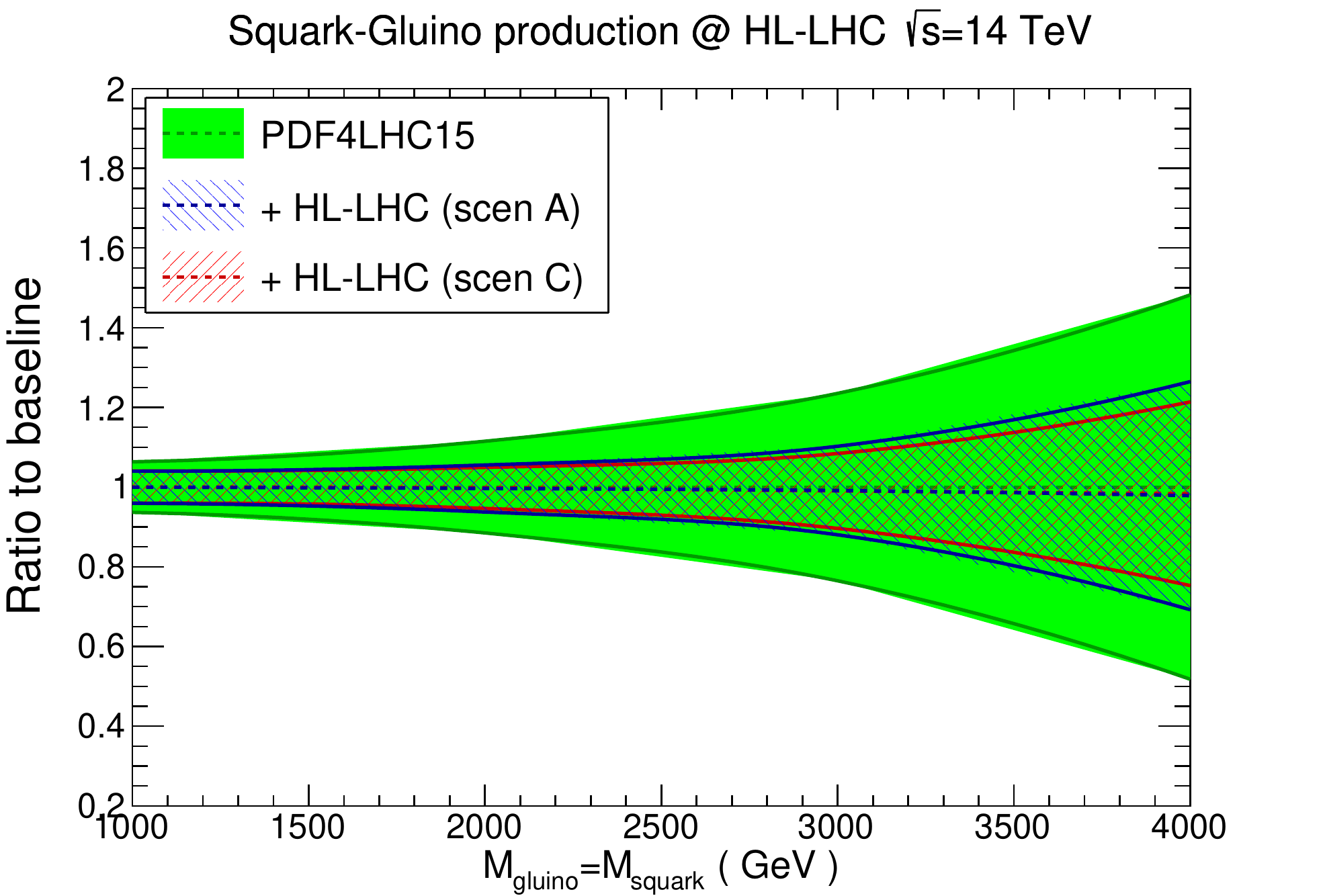}
\caption{\small The cross sections for high--mass supersymmetric particle production
at $\sqrt{s}=14$ TeV, comparing the predictions of the PDF4LHC15
baseline with those of the HL--LHC PDF sets in
the conservative (A) and optimistic (C) scenarios, normalised to the
central value of PDF4LHC15.
We show the results corresponding to gluino pair production
(left) and squark--gluino production (right).
The cross sections have been
evaluated with {\tt Pythia8.235} using leading--order
matrix elements and the SLHA2 benchmark point as model input.
     \label{fig:susyxsects}
}
  \end{center}
\end{figure}

In Fig.~\ref{fig:susyxsects} we show the comparison between
the  PDF4LHC15 predictions
with the corresponding results from the profiled PDF sets
with HL--LHC pseudo--data,
normalised to the central value of the PDF4LHC15 baseline.
As in Fig.~\ref{fig:MCFMxsects}, we provide results for scenarios
A and C, the conservative and optimistic ones respectively.
Specifically, we show the cross sections
for gluino--gluino and squark--gluino production at $\sqrt{s}=14$ TeV -- similar
conclusions are derived from squark--squark and squark--antisquark production.
 The theoretical calculations have been obtained using leading order (LO)
  matrix elements with
  {\tt Pythia8.235}~\cite{Sjostrand:2007gs}
  and assuming the SLHA2 benchmark point~\cite{Allanach:2008qq}, for a range
  of sparticle masses within the HL--LHC reach.
  For simplicity, underlying event and multiple interactions have been switched off
  in the calculation.
  Again, we are not interested here in providing state--of--the--art predictions
  for the event rates, which can be found elsewhere~\cite{Beenakker:2016lwe}.

From the comparisons in Fig.~\ref{fig:susyxsects}, we can
see that the constraints
on the PDFs expected from the HL--LHC data permit
a significant reduction of the uncertainties in the high--mass SUSY cross sections.
The size of this reduction is  consistent with the corresponding results at the level
of luminosities, reported in Fig.~\ref{fig:PDFluminosities} and Table~\ref{fig:PDFs-HL--LHC-summaryTable},
recalling that gluino--gluino and gluino--squark production are driven by the
gluon--gluon and gluon--quark initial states respectively~\cite{Beenakker:2015rna}.
For instance, for gluino pair--production with $M_{\widetilde{g}}=3$ TeV,
the PDF uncertainties are reduced from $\simeq 60\%$ to $\simeq 20\%$ in
the optimistic scenario.
A somewhat milder reduction is found for the squark--gluino cross sections.
For squark--squark and squark--antisquark production, driven by the quark--quark
and quark--antiquark initial states respectively, a PDF uncertainty
reduction by around a factor two at
high masses is found, consistently with Table~\ref{fig:PDFs-HL--LHC-summaryTable}.

To summarise, the initial phenomenological study presented in this section
nicely illustrates the internal coherence of the HL--LHC physics
program: high precision SM measurements will lead to a much improved understanding of
the quark and gluon structure of protons, which in turn will benefit many other important analyses,
from the characterisation of the Higgs sector to the searches of new heavy particles.

%% file: sec-summary.tex
\section{Summary}
\label{sec:summary}

 In this study, we have quantified the expected constraints that precision
 HL--LHC measurements will impose on the quark and gluon structure of the proton.
 To achieve this goal,
 we have assessed the impact of a range of relevant PDF--sensitive processes, from weak gauge
 boson and jet production to top quark and photon production.
 Moreover, we have studied the robustness
 of our results with respect to different projections for the experimental
 systematic uncertainties, from a more conservative one, where systematics
 are assumed to have the same size as in
 current measurements, to a more optimistic one,
 where they are markedly reduced.
 
 Our main finding is that HL--LHC data has the potential to significantly
 reduce the PDF uncertainties in a wide kinematic range and for
 all relevant partonic final states.
 This is true both for the region of intermediate invariant masses, relevant
 for precision Higgs, electroweak, and top quark measurements, as well
 as in the TeV region relevant for searches of new heavy particles.
 Even in the most conservative scenario, in the region $M_X\gsim 40$ GeV
 we find that HL--LHC measurements can reduce PDF uncertainties by at least a factor
 between 2 and 3 as compared to the current PDF4LHC15 baseline.
 The PDF constraining information from the HL--LHC
 is expected to be specially significant
 for gluon-- and for strange--initiated processes.
 We also find that the quark--antiquark luminosity at the electroweak scale,
 a central input for legacy LHC measurements
 such as $M_W$ and $\sin^2\theta_W$, could be improved
 by more than a factor 3 in the optimistic scenario.

 This improved knowledge of the quark and gluon structure of the proton
 which will become possible at the HL--LHC will directly benefit
 a number of phenomenologically important process, due to the reduction
 of the associated theoretical errors.
 For instance, the PDF uncertainties in Higgs production in gluon fusion can
 be reduced down to $\lsim 2\%$ for the entire range of Higgs transverse momenta
 accessible at the HL--LHC.
 Likewise, PDF uncertainties in high--mass supersymmetric particle production
 can be decreased by up to a factor 3, with a similar impact expected for other BSM scenarios.
 This improvement should strengthen the bounds derived in the case of null searches,
 or facilitate their
 characterisation in the case of an eventual discovery.
 Similar improvements are found for Standard Model process, for example
 dijet production, which provides a unique opportunity
 to measurement the running of the strong coupling constant at the
 TeV scale.
 More detailed studies of the phenomenological implications of our study
 will be presented in the upcoming HL--LHC Yellow Report.
 
 Two caveats are relevant at this point.
 First, it should be emphasised again that in
 this study we have only considered
a subset of all possible measurements of relevance for PDF fits.
There are certainly
processes for which data is and will be available, such as
multijet production and single top production, that
 we have not considered here.
 Moreover, we can also reasonably expect that various new processes may be 
 added to the PDF toolbox on the rather long timescales we consider here.
 Thus, we may certainly expect further
 constraints to become available for PDF studies
 by the end of HL--LHC running.

 Second, in this study we have
 ignored any possible issues such as data incompatibilities,
 limitations of the theoretical calculations,
 or issues affecting the data correlation models.
 These are common in PDF fits, and indeed
 have already been found when comparing theory calculations
 against existing LHC data from Runs I and II.
 Such potential problems may eventually
 limit the PDF constraining power, in comparison to the estimates presented
 in this work, when the actual global fit with
 real HL--LHC data is performed.
 Clearly, such questions can only be tackled once the HL--LHC measurements
 are carried out, and indeed doing so will present an important programme of experimental and theoretical PDF--related work on its own.
 We cannot anticipate such
 work in our present study, which instead represents our best quantitative projections using our current knowledge.

 The results of this study are made
 publicly available in the {\tt LHAPDF6} format~\cite{Buckley:2014ana},
 with the grid names listed in Table~\ref{tab:Scenarios} for the three
 scenarios that have been considered.
 These three grid files can be downloaded from:
 \begin{center}
{\tt  https://data.nnpdf.science/HLLHC\_YR/PDF4LHC15\_nnlo\_hllhc\_scen1.tgz}\\
{\tt https://data.nnpdf.science/HLLHC\_YR/PDF4LHC15\_nnlo\_hllhc\_scen2.tgz}\\
{\tt https://data.nnpdf.science/HLLHC\_YR/PDF4LHC15\_nnlo\_hllhc\_scen3.tgz}
\end{center}
The ``ultimate'' PDFs produced in this exercise
 can then be straightforwardly applied to other physics projections
 of HL--LHC processes, taking into account our improved knowledge
 of the partonic structure of the proton which is expected by then.
We believe that the results of this work represent
 an important ingredient towards sharpening as much as possible
 the physics reach of the LHC in its upcoming high--luminosity era.

 \subsection*{Acknowledgements}
 We are grateful to W.~Barter, M.~Campanelli, C.~Gwenlan, S.~Farry, and
 K. Lipka for discussion about the projections of future HL--LHC measurements
 at ATLAS, CMS, and LHCb.
 We thank P.~Starovoitov for providing the {\tt APPLgrids} for
 the inclusive jet measurements at the HL--LHC.
 R.~A.~K. and J.~R. are
 supported by the European Research Council (ERC) Starting Grant ``PDF4BSM'' and by the Dutch
 Organization for Scientific Research (NWO).
 The work of J.~G. is sponsored by Shanghai Pujiang Program and by the National
 Natural Science Fundation of China under the Grant No. 11875189.
 S.~B. is supported by the Science and Technology Facilities Council (STFC).
 L.~H.~L thanks the Science and Technology Facilities Council (STFC)
 for support via grant award ST/L000377/1.

%% file: HLLHC_PDFs.bbl
\providecommand{\href}[2]{#2}\begingroup\raggedright\begin{thebibliography}{10}

\bibitem{Gao:2017yyd}
J.~Gao, L.~Harland-Lang, and J.~Rojo, {\it {The Structure of the Proton in the
  LHC Precision Era}},  {\em Phys. Rept.} {\bf 742} (2018) 1--121,
  [\href{http://arxiv.org/abs/1709.04922}{{\tt arXiv:1709.04922}}].

\bibitem{Rojo:2015acz}
J.~Rojo et~al., {\it {The PDF4LHC report on PDFs and LHC data: Results from Run
  I and preparation for Run II}},  {\em J. Phys.} {\bf G42} (2015) 103103,
  [\href{http://arxiv.org/abs/1507.00556}{{\tt arXiv:1507.00556}}].

\bibitem{Forte:2013wc}
S.~Forte and G.~Watt, {\it {Progress in the Determination of the Partonic
  Structure of the Proton}},  {\em Ann.Rev.Nucl.Part.Sci.} {\bf 63} (2013) 291,
  [\href{http://arxiv.org/abs/1301.6754}{{\tt arXiv:1301.6754}}].

\bibitem{deFlorian:2016spz}
{\bf LHC Higgs Cross Section Working Group} Collaboration, D.~de~Florian
  et~al., {\it {Handbook of LHC Higgs Cross Sections: 4. Deciphering the Nature
  of the Higgs Sector}},  \href{http://arxiv.org/abs/1610.07922}{{\tt
  arXiv:1610.07922}}.

\bibitem{Beenakker:2015rna}
W.~Beenakker, C.~Borschensky, M.~Kramer, A.~Kulesza, E.~Laenen, S.~Marzani, and
  J.~Rojo, {\it {NLO+NLL squark and gluino production cross-sections with
  threshold-improved parton distributions}},  {\em Eur. Phys. J.} {\bf C76}
  (2016), no.~2 53, [\href{http://arxiv.org/abs/1510.00375}{{\tt
  arXiv:1510.00375}}].

\bibitem{Alioli:2017jdo}
S.~Alioli, M.~Farina, D.~Pappadopulo, and J.~T. Ruderman, {\it {Precision
  Probes of QCD at High Energies}},  {\em JHEP} {\bf 07} (2017) 097,
  [\href{http://arxiv.org/abs/1706.03068}{{\tt arXiv:1706.03068}}].

\bibitem{Aaboud:2017svj}
{\bf ATLAS} Collaboration, M.~Aaboud et~al., {\it {Measurement of the $W$-boson
  mass in pp collisions at $\sqrt{s}=7$ TeV with the ATLAS detector}},  {\em
  Eur. Phys. J.} {\bf C78} (2018), no.~2 110,
  [\href{http://arxiv.org/abs/1701.07240}{{\tt arXiv:1701.07240}}].

\bibitem{Aaltonen:2018dxj}
{\bf CDF, D0} Collaboration, T.~A. Aaltonen et~al., {\it {Tevatron Run II
  combination of the effective leptonic electroweak mixing angle}},  {\em Phys.
  Rev.} {\bf D97} (2018), no.~11 112007,
  [\href{http://arxiv.org/abs/1801.06283}{{\tt arXiv:1801.06283}}].

\bibitem{Ball:2018iqk}
{\bf NNPDF} Collaboration, R.~D. Ball, S.~Carrazza, L.~Del~Debbio, S.~Forte,
  Z.~Kassabov, J.~Rojo, E.~Slade, and M.~Ubiali, {\it {Precision determination
  of the strong coupling constant within a global PDF analysis}},  {\em Eur.
  Phys. J.} {\bf C78} (2018), no.~5 408,
  [\href{http://arxiv.org/abs/1802.03398}{{\tt arXiv:1802.03398}}].

\bibitem{Becciolini:2014lya}
D.~Becciolini, M.~Gillioz, M.~Nardecchia, F.~Sannino, and M.~Spannowsky, {\it
  {Constraining new colored matter from the ratio of 3 to 2 jets cross sections
  at the LHC}},  {\em Phys. Rev.} {\bf D91} (2015), no.~1 015010,
  [\href{http://arxiv.org/abs/1403.7411}{{\tt arXiv:1403.7411}}]. [Addendum:
  Phys. Rev.D92,no.7,079905(2015)].

\bibitem{Ball:2017nwa}
{\bf NNPDF} Collaboration, R.~D. Ball et~al., {\it {Parton distributions from
  high-precision collider data}},  {\em Eur. Phys. J.} {\bf C77} (2017), no.~10
  663, [\href{http://arxiv.org/abs/1706.00428}{{\tt arXiv:1706.00428}}].

\bibitem{Dulat:2015mca}
S.~Dulat, T.-J. Hou, J.~Gao, M.~Guzzi, J.~Huston, P.~Nadolsky, J.~Pumplin,
  C.~Schmidt, D.~Stump, and C.~P. Yuan, {\it {New parton distribution functions
  from a global analysis of quantum chromodynamics}},  {\em Phys. Rev.} {\bf
  D93} (2016), no.~3 033006, [\href{http://arxiv.org/abs/1506.07443}{{\tt
  arXiv:1506.07443}}].

\bibitem{Harland-Lang:2014zoa}
L.~A. Harland-Lang, A.~D. Martin, P.~Motylinski, and R.~S. Thorne, {\it {Parton
  distributions in the LHC era: MMHT 2014 PDFs}},  {\em Eur. Phys. J.} {\bf
  C75} (2015) 204, [\href{http://arxiv.org/abs/1412.3989}{{\tt
  arXiv:1412.3989}}].

\bibitem{Alekhin:2017kpj}
S.~Alekhin, J.~Bl{\"u}mlein, S.~Moch, and R.~Placakyte, {\it {Parton
  distribution functions, $\alpha_s$, and heavy-quark masses for LHC Run II}},
  {\em Phys. Rev.} {\bf D96} (2017), no.~1 014011,
  [\href{http://arxiv.org/abs/1701.05838}{{\tt arXiv:1701.05838}}].

\bibitem{Czakon:2016olj}
M.~Czakon, N.~P. Hartland, A.~Mitov, E.~R. Nocera, and J.~Rojo, {\it {Pinning
  down the large-x gluon with NNLO top-quark pair differential distributions}},
   {\em JHEP} {\bf 04} (2017) 044, [\href{http://arxiv.org/abs/1611.08609}{{\tt
  arXiv:1611.08609}}].

\bibitem{Guzzi:2014wia}
M.~Guzzi, K.~Lipka, and S.-O. Moch, {\it {Top-quark pair production at hadron
  colliders: differential cross section and phenomenological applications with
  DiffTop}},  {\em JHEP} {\bf 01} (2015) 082,
  [\href{http://arxiv.org/abs/1406.0386}{{\tt arXiv:1406.0386}}].

\bibitem{Boughezal:2017nla}
R.~Boughezal, A.~Guffanti, F.~Petriello, and M.~Ubiali, {\it {The impact of the
  LHC Z-boson transverse momentum data on PDF determinations}},  {\em JHEP}
  {\bf 07} (2017) 130, [\href{http://arxiv.org/abs/1705.00343}{{\tt
  arXiv:1705.00343}}].

\bibitem{d'Enterria:2012yj}
D.~d'Enterria and J.~Rojo, {\it {Quantitative constraints on the gluon
  distribution function in the proton from collider isolated-photon data}},
  {\em Nucl.Phys.} {\bf B860} (2012) 311--338,
  [\href{http://arxiv.org/abs/1202.1762}{{\tt arXiv:1202.1762}}].

\bibitem{Campbell:2018wfu}
J.~M. Campbell, J.~Rojo, E.~Slade, and C.~Williams, {\it {Direct photon
  production and PDF fits reloaded}},  {\em Eur. Phys. J.} {\bf C78} (2018),
  no.~6 470, [\href{http://arxiv.org/abs/1802.03021}{{\tt arXiv:1802.03021}}].

\bibitem{Zenaiev:2015rfa}
{\bf PROSA} Collaboration, O.~Zenaiev et~al., {\it {Impact of heavy-flavour
  production cross sections measured by the LHCb experiment on parton
  distribution functions at low x}},  {\em Eur. Phys. J.} {\bf C75} (2015),
  no.~8 396, [\href{http://arxiv.org/abs/1503.04581}{{\tt arXiv:1503.04581}}].

\bibitem{Gauld:2016kpd}
R.~Gauld and J.~Rojo, {\it {Precision determination of the small-$x$ gluon from
  charm production at LHCb}},  {\em Phys. Rev. Lett.} {\bf 118} (2017), no.~7
  072001, [\href{http://arxiv.org/abs/1610.09373}{{\tt arXiv:1610.09373}}].

\bibitem{Aad:2014xca}
{\bf ATLAS} Collaboration, G.~Aad et~al., {\it {Measurement of the production
  of a $W$ boson in association with a charm quark in $pp$ collisions at
  $\sqrt{s} =$ 7 TeV with the ATLAS detector}},  {\em JHEP} {\bf 1405} (2014)
  068, [\href{http://arxiv.org/abs/1402.6263}{{\tt arXiv:1402.6263}}].

\bibitem{Chatrchyan:2013mza}
{\bf CMS} Collaboration, S.~Chatrchyan et~al., {\it {Measurement of the muon
  charge asymmetry in inclusive pp to WX production at $\sqrt{s}$ = 7 TeV and
  an improved determination of light parton distribution functions}},  {\em
  Phys.Rev.} {\bf D90} (2014) 032004,
  [\href{http://arxiv.org/abs/1312.6283}{{\tt arXiv:1312.6283}}].

\bibitem{CMS-PAS-SMP-17-014}
{\bf CMS Collaboration} Collaboration, {\it {Measurement of associated
  production of W bosons with charm quarks in proton-proton collisions at
  $\sqrt{s}=13~\mathrm{TeV}$ with the CMS experiment at the LHC}},  Tech. Rep.
  CMS-PAS-SMP-17-014, CERN, Geneva, 2018.

\bibitem{Currie:2016bfm}
J.~Currie, E.~W.~N. Glover, and J.~Pires, {\it {Next-to-Next-to Leading Order
  QCD Predictions for Single Jet Inclusive Production at the LHC}},  {\em Phys.
  Rev. Lett.} {\bf 118} (2017), no.~7 072002,
  [\href{http://arxiv.org/abs/1611.01460}{{\tt arXiv:1611.01460}}].

\bibitem{Rojo:2014kta}
J.~Rojo, {\it {Constraints on parton distributions and the strong coupling from
  LHC jet data}},  {\em Int. J. Mod. Phys.} {\bf A30} (2015) 1546005,
  [\href{http://arxiv.org/abs/1410.7728}{{\tt arXiv:1410.7728}}].

\bibitem{yellowreport}
P.~Azzi et~al., {\it {The Physics at the HL/HE-LHC Yellow Report}}, .

\bibitem{Bediaga:2018lhg}
{\bf LHCb} Collaboration, I.~Bediaga et~al., {\it {Physics case for an LHCb
  Upgrade II - Opportunities in flavour physics, and beyond, in the HL-LHC
  era}},  \href{http://arxiv.org/abs/1808.08865}{{\tt arXiv:1808.08865}}.

\bibitem{AbelleiraFernandez:2012cc}
{\bf LHeC Study Group} Collaboration, J.~Abelleira~Fernandez et~al., {\it {A
  Large Hadron Electron Collider at CERN: Report on the Physics and Design
  Concepts for Machine and Detector}},  {\em J.Phys.} {\bf G39} (2012) 075001,
  [\href{http://arxiv.org/abs/1206.2913}{{\tt arXiv:1206.2913}}].

\bibitem{AbelleiraFernandez:2012ty}
{\bf LHeC Study Group} Collaboration, J.~L. Abelleira~Fernandez et~al., {\it
  {On the Relation of the LHeC and the LHC}},
  \href{http://arxiv.org/abs/1211.5102}{{\tt arXiv:1211.5102}}.

\bibitem{Paukkunen:2017phq}
{\bf LHeC study Group} Collaboration, H.~Paukkunen, {\it {An update on nuclear
  PDFs at the LHeC}},  {\em PoS} {\bf DIS2017} (2018) 109,
  [\href{http://arxiv.org/abs/1709.08342}{{\tt arXiv:1709.08342}}].

\bibitem{Ball:2013tyh}
{\bf NNPDF} Collaboration, R.~D. Ball, S.~Forte, A.~Guffanti, E.~R. Nocera,
  G.~Ridolfi, and J.~Rojo, {\it {Polarized Parton Distributions at an
  Electron-Ion Collider}},  {\em Phys. Lett.} {\bf B728} (2014) 524--531,
  [\href{http://arxiv.org/abs/1310.0461}{{\tt arXiv:1310.0461}}].

\bibitem{Marquet:2017bga}
C.~Marquet, M.~R. Moldes, and P.~Zurita, {\it {Unveiling saturation effects
  from nuclear structure function measurements at the EIC}},  {\em Phys. Lett.}
  {\bf B772} (2017) 607--614, [\href{http://arxiv.org/abs/1702.00839}{{\tt
  arXiv:1702.00839}}].

\bibitem{Aschenauer:2017oxs}
E.~C. Aschenauer, S.~Fazio, M.~A.~C. Lamont, H.~Paukkunen, and P.~Zurita, {\it
  {Nuclear Structure Functions at a Future Electron-Ion Collider}},  {\em Phys.
  Rev.} {\bf D96} (2017), no.~11 114005,
  [\href{http://arxiv.org/abs/1708.05654}{{\tt arXiv:1708.05654}}].

\bibitem{Aschenauer:2015ata}
E.~C. Aschenauer, R.~Sassot, and M.~Stratmann, {\it {Unveiling the Proton Spin
  Decomposition at a Future Electron-Ion Collider}},  {\em Phys. Rev.} {\bf
  D92} (2015), no.~9 094030, [\href{http://arxiv.org/abs/1509.06489}{{\tt
  arXiv:1509.06489}}].

\bibitem{Boer:2011fh}
D.~Boer, M.~Diehl, R.~Milner, R.~Venugopalan, W.~Vogelsang, et~al., {\it
  {Gluons and the quark sea at high energies: Distributions, polarization,
  tomography}},  \href{http://arxiv.org/abs/1108.1713}{{\tt arXiv:1108.1713}}.

\bibitem{Cooper-Sarkar:2016udp}
{\bf LHeC study Group} Collaboration, A.~M. Cooper-Sarkar, {\it {Improved
  measurement of parton distribution functions and $\alpha_s(M_Z)$ with the
  LHeC}},  {\em PoS} {\bf DIS2016} (2016) 274,
  [\href{http://arxiv.org/abs/1605.08579}{{\tt arXiv:1605.08579}}].

\bibitem{Aaboud:2016btc}
{\bf ATLAS} Collaboration, M.~Aaboud et~al., {\it {Precision measurement and
  interpretation of inclusive $W^+$ , $W^-$ and $Z/\gamma ^*$ production cross
  sections with the ATLAS detector}},  {\em Eur. Phys. J.} {\bf C77} (2017),
  no.~6 367, [\href{http://arxiv.org/abs/1612.03016}{{\tt arXiv:1612.03016}}].

\bibitem{Khachatryan:2016pev}
{\bf CMS} Collaboration, V.~Khachatryan et~al., {\it {Measurement of the
  differential cross section and charge asymmetry for inclusive $\mathrm
  {p}\mathrm {p}\rightarrow \mathrm {W}^{\pm }+X$ production at ${\sqrt{s}} =
  8$ TeV}},  {\em Eur. Phys. J.} {\bf C76} (2016), no.~8 469,
  [\href{http://arxiv.org/abs/1603.01803}{{\tt arXiv:1603.01803}}].

\bibitem{Butterworth:2015oua}
J.~Butterworth et~al., {\it {PDF4LHC recommendations for LHC Run II}},  {\em J.
  Phys.} {\bf G43} (2016) 023001, [\href{http://arxiv.org/abs/1510.03865}{{\tt
  arXiv:1510.03865}}].

\bibitem{Gao:2013bia}
J.~Gao and P.~Nadolsky, {\it {A meta-analysis of parton distribution
  functions}},  {\em JHEP} {\bf 1407} (2014) 035,
  [\href{http://arxiv.org/abs/1401.0013}{{\tt arXiv:1401.0013}}].

\bibitem{Carrazza:2015hva}
S.~Carrazza, J.~I. Latorre, J.~Rojo, and G.~Watt, {\it {A compression algorithm
  for the combination of PDF sets}},  {\em Eur. Phys. J.} {\bf C75} (2015) 474,
  [\href{http://arxiv.org/abs/1504.06469}{{\tt arXiv:1504.06469}}].

\bibitem{Carrazza:2015aoa}
S.~Carrazza, S.~Forte, Z.~Kassabov, J.~I. Latorre, and J.~Rojo, {\it {An
  Unbiased Hessian Representation for Monte Carlo PDFs}},  {\em Eur. Phys. J.}
  {\bf C75} (2015), no.~8 369, [\href{http://arxiv.org/abs/1505.06736}{{\tt
  arXiv:1505.06736}}].

\bibitem{Paukkunen:2014zia}
H.~Paukkunen and P.~Zurita, {\it {PDF reweighting in the Hessian matrix
  approach}},  {\em JHEP} {\bf 12} (2014) 100,
  [\href{http://arxiv.org/abs/1402.6623}{{\tt arXiv:1402.6623}}].

\bibitem{Schmidt:2018hvu}
C.~Schmidt, J.~Pumplin, C.~P. Yuan, and P.~Yuan, {\it {Updating and Optimizing
  Error PDFs in the Hessian Approach}},
  \href{http://arxiv.org/abs/1806.07950}{{\tt arXiv:1806.07950}}.

\bibitem{Buckley:2014ana}
A.~Buckley, J.~Ferrando, S.~Lloyd, K.~NordstrÃ¶m, B.~Page, et~al., {\it
  {LHAPDF6: parton density access in the LHC precision era}},  {\em
  Eur.Phys.J.} {\bf C75} (2015) 132,
  [\href{http://arxiv.org/abs/1412.7420}{{\tt arXiv:1412.7420}}].

\bibitem{Aad:2016zzw}
{\bf ATLAS} Collaboration, G.~Aad et~al., {\it {Measurement of the
  double-differential high-mass Drell-Yan cross section in pp collisions at $
  \sqrt{s}=8 $ TeV with the ATLAS detector}},  {\em JHEP} {\bf 08} (2016) 009,
  [\href{http://arxiv.org/abs/1606.01736}{{\tt arXiv:1606.01736}}].

\bibitem{Rojo:2017xpe}
{\bf NNPDF} Collaboration, J.~Rojo, {\it {Improving quark flavor separation
  with forward W and Z production at LHCb}},  {\em PoS} {\bf DIS2017} (2018)
  198, [\href{http://arxiv.org/abs/1705.04468}{{\tt arXiv:1705.04468}}].

\bibitem{Aaij:2015zlq}
{\bf LHCb} Collaboration, R.~Aaij et~al., {\it {Measurement of forward W and Z
  boson production in $pp$ collisions at $ \sqrt{s}=8 $ TeV}},  {\em JHEP} {\bf
  01} (2016) 155, [\href{http://arxiv.org/abs/1511.08039}{{\tt
  arXiv:1511.08039}}].

\bibitem{Aad:2015mbv}
{\bf ATLAS} Collaboration, G.~Aad et~al., {\it {Measurements of top-quark pair
  differential cross-sections in the lepton+jets channel in $pp$ collisions at
  $\sqrt{s}=8$ TeV using the ATLAS detector}},  {\em Eur. Phys. J.} {\bf C76}
  (2016), no.~10 538, [\href{http://arxiv.org/abs/1511.04716}{{\tt
  arXiv:1511.04716}}].

\bibitem{ATL-PHYS-PUB-2018-017}
{\bf ATLAS Collaboration} Collaboration, {\it {Determination of the parton
  distribution functions of the proton from ATLAS measurements of differential
  $W$ and $Z/\gamma^*$ and $t\bar{t}$ cross sections}},  Tech. Rep.
  ATL-PHYS-PUB-2018-017, CERN, Geneva, Aug, 2018.

\bibitem{Aad:2015auj}
{\bf ATLAS} Collaboration, G.~Aad et~al., {\it {Measurement of the transverse
  momentum and $\phi ^*_{\eta }$ distributions of Drell-Yan lepton pairs in
  proton-proton collisions at $\sqrt{s}=8$ TeV with the ATLAS detector}},  {\em
  Eur. Phys. J.} {\bf C76} (2016), no.~5 291,
  [\href{http://arxiv.org/abs/1512.02192}{{\tt arXiv:1512.02192}}].

\bibitem{Stirling:2012vh}
W.~Stirling and E.~Vryonidou, {\it {Charm production in association with an
  electroweak gauge boson at the LHC}},  {\em Phys.Rev.Lett.} {\bf 109} (2012)
  082002, [\href{http://arxiv.org/abs/1203.6781}{{\tt arXiv:1203.6781}}].

\bibitem{LHCbprivcomm}
Will Barter, Stephen Farry, private communication.

\bibitem{Aaboud:2017cbm}
{\bf ATLAS} Collaboration, M.~Aaboud et~al., {\it {Measurement of the cross
  section for inclusive isolated-photon production in $pp$ collisions at $\sqrt
  s=13$ TeV using the ATLAS detector}},  {\em Phys. Lett.} {\bf B770} (2017)
  473--493, [\href{http://arxiv.org/abs/1701.06882}{{\tt arXiv:1701.06882}}].

\bibitem{Currie:2017eqf}
J.~Currie, A.~Gehrmann-De~Ridder, T.~Gehrmann, E.~W.~N. Glover, A.~Huss, and
  J.~Pires, {\it {Precise predictions for dijet production at the LHC}},  {\em
  Phys. Rev. Lett.} {\bf 119} (2017), no.~15 152001,
  [\href{http://arxiv.org/abs/1705.10271}{{\tt arXiv:1705.10271}}].

\bibitem{Berger:2017zof}
E.~L. Berger, J.~Gao, and H.~X. Zhu, {\it {Differential Distributions for
  t-channel Single Top-Quark Production and Decay at Next-to-Next-to-Leading
  Order in QCD}},  {\em JHEP} {\bf 11} (2017) 158,
  [\href{http://arxiv.org/abs/1708.09405}{{\tt arXiv:1708.09405}}].

\bibitem{Boughezal:2016wmq}
R.~Boughezal, J.~M. Campbell, R.~K. Ellis, C.~Focke, W.~Giele, X.~Liu,
  F.~Petriello, and C.~Williams, {\it {Color singlet production at NNLO in
  MCFM}},  {\em Eur. Phys. J.} {\bf C77} (2017), no.~1 7,
  [\href{http://arxiv.org/abs/1605.08011}{{\tt arXiv:1605.08011}}].

\bibitem{Carli:2010rw}
T.~Carli et~al., {\it {A posteriori inclusion of parton density functions in
  NLO QCD final-state calculations at hadron colliders: The APPLGRID Project}},
   {\em Eur.Phys.J.} {\bf C66} (2010) 503,
  [\href{http://arxiv.org/abs/0911.2985}{{\tt arXiv:0911.2985}}].

\bibitem{Nagy:2003tz}
Z.~Nagy, {\it {Next-to-leading order calculation of three-jet observables in
  hadron hadron collision}},  {\em Phys. Rev.} {\bf D68} (2003) 094002,
  [\href{http://arxiv.org/abs/hep-ph/0307268}{{\tt hep-ph/0307268}}].

\bibitem{Aaboud:2017wsi}
{\bf ATLAS} Collaboration, M.~Aaboud et~al., {\it {Measurement of inclusive jet
  and dijet cross-sections in proton-proton collisions at $\sqrt{s}=13$ TeV
  with the ATLAS detector}},  {\em JHEP} {\bf 05} (2018) 195,
  [\href{http://arxiv.org/abs/1711.02692}{{\tt arXiv:1711.02692}}].

\bibitem{Ball:2014uwa}
{\bf NNPDF} Collaboration, R.~D. Ball et~al., {\it {Parton distributions for
  the LHC Run II}},  {\em JHEP} {\bf 04} (2015) 040,
  [\href{http://arxiv.org/abs/1410.8849}{{\tt arXiv:1410.8849}}].

\bibitem{Manohar:2017eqh}
A.~V. Manohar, P.~Nason, G.~P. Salam, and G.~Zanderighi, {\it {The Photon
  Content of the Proton}},  {\em JHEP} {\bf 12} (2017) 046,
  [\href{http://arxiv.org/abs/1708.01256}{{\tt arXiv:1708.01256}}].

\bibitem{Bertone:2017bme}
{\bf NNPDF} Collaboration, V.~Bertone, S.~Carrazza, N.~P. Hartland, and
  J.~Rojo, {\it {Illuminating the photon content of the proton within a global
  PDF analysis}},  {\em SciPost Phys.} {\bf 5} (2018) 008,
  [\href{http://arxiv.org/abs/1712.07053}{{\tt arXiv:1712.07053}}].

\bibitem{Nathvani:2018pys}
R.~Nathvani, L.~Harland-Lang, R.~Thorne, and A.~Martin, {\it {Ad Lucem: The
  Photon in the MMHT PDFs}},  in {\em {26th International Workshop on Deep
  Inelastic Scattering and Related Subjects (DIS 2018) Port Island, Kobe,
  Japan, April 16-20, 2018}}, 2018.
\newblock \href{http://arxiv.org/abs/1807.07846}{{\tt arXiv:1807.07846}}.

\bibitem{Harland-Lang:2017ytb}
L.~A. Harland-Lang, A.~D. Martin, and R.~S. Thorne, {\it {The Impact of LHC Jet
  Data on the MMHT PDF Fit at NNLO}},  {\em Eur. Phys. J.} {\bf C78} (2018),
  no.~3 248, [\href{http://arxiv.org/abs/1711.05757}{{\tt arXiv:1711.05757}}].

\bibitem{Khachatryan:2015oqa}
{\bf CMS} Collaboration, V.~Khachatryan et~al., {\it {Measurement of the
  differential cross section for top quark pair production in pp collisions at
  $\sqrt{s} = 8\,\text {TeV} $}},  {\em Eur. Phys. J.} {\bf C75} (2015), no.~11
  542, [\href{http://arxiv.org/abs/1505.04480}{{\tt arXiv:1505.04480}}].

\bibitem{Chatrchyan:2013uja}
{\bf CMS} Collaboration, S.~Chatrchyan et~al., {\it {Measurement of associated
  W + charm production in pp collisions at $\sqrt{s}$ = 7 TeV}},  {\em JHEP}
  {\bf 02} (2014) 013, [\href{http://arxiv.org/abs/1310.1138}{{\tt
  arXiv:1310.1138}}].

\bibitem{Ball:2011gg}
R.~D. Ball, V.~Bertone, F.~Cerutti, L.~Del~Debbio, S.~Forte, et~al., {\it
  {Reweighting and Unweighting of Parton Distributions and the LHC W lepton
  asymmetry data}},  {\em Nucl.Phys.} {\bf B855} (2012) 608--638,
  [\href{http://arxiv.org/abs/1108.1758}{{\tt arXiv:1108.1758}}].

\bibitem{Ball:2010gb}
{\bf The NNPDF} Collaboration, R.~D. Ball et~al., {\it {Reweighting NNPDFs: the
  W lepton asymmetry}},  {\em Nucl. Phys.} {\bf B849} (2011) 112--143,
  [\href{http://arxiv.org/abs/1012.0836}{{\tt arXiv:1012.0836}}].

\bibitem{Ball:2009qv}
{\bf The NNPDF} Collaboration, R.~D. Ball et~al., {\it {Fitting Parton
  Distribution Data with Multiplicative Normalization Uncertainties}},  {\em
  JHEP} {\bf 05} (2010) 075, [\href{http://arxiv.org/abs/0912.2276}{{\tt
  arXiv:0912.2276}}].

\bibitem{Ball:2012wy}
R.~D. Ball, S.~Carrazza, L.~Del~Debbio, S.~Forte, J.~Gao, et~al., {\it {Parton
  Distribution Benchmarking with LHC Data}},  {\em JHEP} {\bf 1304} (2013) 125,
  [\href{http://arxiv.org/abs/1211.5142}{{\tt arXiv:1211.5142}}].

\bibitem{Gao:2012qpa}
J.~Gao, C.~S. Li, and C.~P. Yuan, {\it {NLO QCD Corrections to dijet Production
  via Quark Contact Interactions}},  {\em JHEP} {\bf 07} (2012) 037,
  [\href{http://arxiv.org/abs/1204.4773}{{\tt arXiv:1204.4773}}].

\bibitem{Khachatryan:2014cja}
{\bf CMS} Collaboration, V.~Khachatryan et~al., {\it {Search for quark contact
  interactions and extra spatial dimensions using dijet angular distributions
  in proton–proton collisions at $\sqrt s =$ 8 TeV}},  {\em Phys. Lett.} {\bf
  B746} (2015) 79--99, [\href{http://arxiv.org/abs/1411.2646}{{\tt
  arXiv:1411.2646}}].

\bibitem{Grojean:2013nya}
C.~Grojean, E.~Salvioni, M.~Schlaffer, and A.~Weiler, {\it {Very boosted Higgs
  in gluon fusion}},  {\em JHEP} {\bf 05} (2014) 022,
  [\href{http://arxiv.org/abs/1312.3317}{{\tt arXiv:1312.3317}}].

\bibitem{Sjostrand:2007gs}
T.~Sjostrand, S.~Mrenna, and P.~Z. Skands, {\it {A Brief Introduction to PYTHIA
  8.1}},  {\em Comput. Phys. Commun.} {\bf 178} (2008) 852--867,
  [\href{http://arxiv.org/abs/0710.3820}{{\tt arXiv:0710.3820}}].

\bibitem{Allanach:2008qq}
B.~C. Allanach et~al., {\it {SUSY Les Houches Accord 2}},  {\em Comput. Phys.
  Commun.} {\bf 180} (2009) 8--25, [\href{http://arxiv.org/abs/0801.0045}{{\tt
  arXiv:0801.0045}}].

\bibitem{Beenakker:2016lwe}
W.~Beenakker, C.~Borschensky, M.~Krämer, A.~Kulesza, and E.~Laenen, {\it
  {NNLL-fast: predictions for coloured supersymmetric particle production at
  the LHC with threshold and Coulomb resummation}},  {\em JHEP} {\bf 12} (2016)
  133, [\href{http://arxiv.org/abs/1607.07741}{{\tt arXiv:1607.07741}}].

\end{thebibliography}\endgroup
